\documentclass[11pt]{article}

\usepackage{amssymb,amsmath,amsthm,amsfonts,titlesec}
\usepackage[utf8]{inputenc}
\usepackage[T1]{fontenc}
\usepackage[tt=false,type1=true]{libertine}
\usepackage[varqu]{zi4}
\usepackage[libertine]{newtxmath}
\usepackage[margin=1in]{geometry}
\usepackage{enumitem}
\setlist{nosep,topsep=2pt,leftmargin=*}

\usepackage{xcolor}
\usepackage{hyperref}
\hypersetup{colorlinks,allcolors=teal}

\usepackage[sortcites,sorting=nyt,style=alphabetic,maxbibnames=7]{biblatex}
\AtBeginBibliography{\setlength{\emergencystretch}{3em}}
\newcommand{\citet}[1]{\textcite{#1}}

\usepackage{tikz,xifthen,physics,xcolor,suffix,nicefrac,enumitem,subcaption,float,thm-restate,wrapfig}
\usepackage{graphicx}
\usepackage[ruled, vlined]{algorithm2e}
\usetikzlibrary{shapes, arrows, positioning}
\usepackage[bb=dsserif]{mathalpha}
\usepackage[capitalize]{cleveref}
\usepackage{wrapfig}
\usepackage{xspace}

\makeatletter
\@ifclassloaded{acmart}{}{%
  \newtheorem{theorem}{Theorem}[section]
  
  \newtheorem{proposition}[theorem]{Proposition}
  \newtheorem{lemma}[theorem]{Lemma}
  \newtheorem{corollary}[theorem]{Corollary}
  \theoremstyle{definition}
  \newtheorem{example}[theorem]{Example}
  \newtheorem{definition}[theorem]{Definition}
}
\makeatother

\AtBeginDocument{%
  \ifcsname th@acmdefinition\endcsname
    \theoremstyle{acmdefinition}%
  \else
    \theoremstyle{definition}%
  \fi
  \newtheorem{remark}[theorem]{Remark}%
  \ifcsname th@acmplain\endcsname
    \theoremstyle{acmplain}%
  \else
    \theoremstyle{plain}%
  \fi
}

\DeclareMathOperator*{\E}{\mathbb{E}}
\DeclareMathOperator*{\Var}{\mathrm{Var}}

\newcommand{\Unif}{\mathrm{Unif}}
\newcommand{\Bern}{\mathrm{Bern}}
\newcommand{\Binom}{\mathrm{Binom}}
\newcommand{\Geom}{\mathrm{Geom}}
\newcommand{\spec}{\texttt{SPEC}\xspace}
\newcommand{\pec}{\texttt{PEC}\xspace}

\let\hat\widehat
\let\tilde\widetilde
\renewcommand{\Pr}{\operatorname*{\mathbb{P}}}

\SetAlFnt{\small}
\SetAlCapFnt{\small}
\SetAlCapNameFnt{\small}
\SetAlCapHSkip{0pt}
\IncMargin{-\parindent}

\providecommand{\Description}[1]{}

\title{Non-Monetary Mechanisms for Stochastic Scheduling via Sampled Pairwise Externality Compensation}

\author{
David X. Lin%
\thanks{While at Cornell, supported by AFOSR grant FA9550-231-0068, NSF grant ECCS1847393 and through a SPROUT Award from Cornell Engineering.}\\
The University of Texas at Austin\footnote{Conducted while an M.S. student at Cornell University.}\\
\texttt{davidlin@cs.utexas.edu}
\and
Chido Onyeze%
\thanks{Supported by AFOSR grant FA9550-231-0068, NSF grant ECCS1847393 and through a SPROUT Award from Cornell Engineering.}\\
Cornell University\\
\texttt{chidoonyeze@cs.cornell.edu}
\and
Giannis Fikioris%
\thanks{While at Cornell, supported by the ONR MURI grant N000142412742.}\\
Google Research\footnote{Conducted while a Ph.D. student at Cornell University.}\\
\texttt{giannisf@google.com}
\and
Siddhartha Banerjee%
\thanks{Supported by AFOSR grant FA9550-231-0068, NSF grant ECCS1847393 and through a SPROUT Award from Cornell Engineering.}\\
Cornell University\\
\texttt{sbanerjee@cornell.edu}
\and
\'Eva Tardos%
\thanks{Supported in part by AFOSR grant FA9550-23-1-0410, AFOSR grant FA9550-231-0068, ONR MURI grant N000142412742, and through a SPROUT Award from Cornell Engineering.}\\
Cornell University\\
\texttt{eva.tardos@cornell.edu}
}
\date{}

\begin{document}
\maketitle
\thispagestyle{empty}

\begin{abstract}
How can a shared resource be allocated without money when agents have private, stochastic demands that can be queued for a while, and may behave strategically, or even adversarially? We introduce a paradigm for non-monetary online resource allocation called \emph{Sampled Pairwise Externality Compensator} (\spec). Our objective is to guarantee each agent a large fraction of its \emph{ideal utility}: an individual benchmark corresponding to serving the highest-valued fraction of its demand permitted by an equal share of capacity. We seek guarantees that hold regardless of other agents' behavior, including potentially adversarial and collusive behavior. Our approach builds on the idea of pairwise externality compensation, making it computationally efficient by replacing exact calculations of future externalities with sampled estimates. The central insight is that local mixing, exploited through coupled simulations, allows these estimates to be computed efficiently while preserving robust utility guarantees. In our main technical contribution, we instantiate \spec in a stochastic scheduling model in which agents receive jobs with private values, unserved jobs persist across periods subject to random expiration, and a single server processes at most one job per period. Our mechanism gives each agent a simple strategy that achieves the optimal worst-case fraction of ideal utility available to a fully informed planner, up to a sublinear additive loss in total utility. 
We also show that this fraction approaches $1$ in the long-horizon limit as the expiration probability vanishes: sufficiently persistent jobs allow each agent to secure nearly its full ideal utility.
The mechanism runs in $O(n^2)$ expected amortized time per period, where $n$ is the number of agents; significantly, the expected amortized computational overhead per period is independent of the time horizon $T$.
\end{abstract}

\clearpage
\setcounter{page}{1}
\section{Introduction}
Consider the problem of repeatedly allocating a resource among multiple agents: a growing line of work shows how to do so in fair and efficient ways without the use of money
\cite{guo2009competitive,gorokh2021remarkable,banerjee2023robust,onyeze2025allocating,fikioris2025beyond,lin2025online,lin2026robust,lin2026itcs}. 
A canonical model in this literature assumes that a request loses its value if it is not served immediately; this is partly for modeling simplicity, but also meaningful in applications like allocating time on a telescope, where a team aiming to observe an event in the sky needs the equipment when the event happens. A more general model of intertemporal dependence is one where unserved jobs can remain available for future service for a while, but then expire. 
An increasingly relevant example is 
ML training: for example, Google's internal Quota Marketplace uses dynamic pricing to allocate shared training resources~\cite{sivan2026quota}. 
These workloads can have different values and delay tolerances (i.e., some may still be useful to do tomorrow).
This motivates the core model we study in this work.\footnote{AI and LLM tools assisted with computational techniques, proof drafting, and error checking; \cref{sec:ai_disclosure} details the scope and extent of this use.}

Consider an organization (university/company) controlling access to a computing cluster shared between multiple competing agents (labs/teams). This setting exhibits several features. 
\begin{itemize}[nosep,leftmargin=*]
\item The computing cluster is a limited resource. In general, the demand for it exceeds its capacity.
\item Labs have uncertain and varying demand for the cluster over time. Sometimes, labs have a lot of important jobs that they wish to run and other times, they have few.
\item If a lab has an important job to run but the cluster is busy with other labs' jobs, they'll still likely want to run the job later when the cluster becomes available. However, if a lab does not get to run its job for too long a time, the lab becomes no longer interested in processing that job. For example, perhaps the conference deadline for the job passed.
\item The allocation procedure should be efficient and fair without charging the labs money. Furthermore, it is unclear from the department's perspective how to evaluate which labs should be given access to the cluster at each time. For example, asking each lab to rate each of their jobs' importance on a scale~1--10 and giving access to the jobs that have the highest ratings has several flaws: different labs may have different interpretations for how important a particular rating like~7 may be, or labs may act strategically and always report~10.
\item The allocation procedure should be simple; it should be easy for agents to understand, and participate in a way to be able to obtain access to the cluster when they need it most.
\end{itemize}

We consider a simple model for settings like these as follows. There is a fixed set of $n$ agents for a long time horizon $T$. Each agent maintains a queue of jobs. In each time period $t\in\{1,2,\dots,T\}$, each agent $i$ receives a job in their queue with value drawn from some distribution $\mathcal F_i$ i.i.d. across time. There is a single server that can process at most one job in each time period. Unserved jobs expire with probability $\delta$; unexpired jobs remain in their corresponding queues to the next time period. A principal decides which agent gets access to the server at each time period via some mechanism without knowledge of the agents' queue states, including the job values and how many jobs are in each agent's queue at any time. Each agent seeks to maximize the sum of the values of served jobs.

Previous work has introduced the benchmark of \textit{ideal utility} to quantify the efficiency and fairness of a mechanism; we continue to use this in our more general setting. 
Roughly speaking, the ideal utility of agent $i$ is the utility they would receive if their highest-valued $T/n$ jobs were processed. This is a reasonable upper bound for what an agent should receive in a ``fair'' world: at most $T$ total jobs can get processed, so an agent should only hope to process their highest-valued $T/n$ jobs. Notably, this benchmark is easy to compute: it depends only on the agent's own value distribution and the number of players, and does not require an analysis of other agents' demands or strategies.
Note that the ideal utility does not depend on $\delta$: it simulates an ideal world where the agent does not have to queue jobs at all, immediately serving their choice of request, limited only by their fair share.
However, it is not possible to guarantee each agent their ideal utility even if the scheduler has full information: for example, in the extreme case that jobs cannot be delayed ($\delta=1$), if each agent has $\Bern(1/n)$ value distribution, then in $(1-1/n)^nT$ time periods (in expectation), all agents' jobs have zero value. This implies that it is impossible to guarantee every agent more than a $(1-(1-1/n)^n)\approx 1-1/e\approx 0.63$ fraction of their ideal utility in expectation. When jobs do not have to be processed immediately, a higher fraction of ideal utility is achievable with central coordination. Let $ \lambda(\delta,n,T)\in (1-1/e, 1]$ denote the \emph{optimal fraction of ideal utility} that a central planner with full information of the agents' queue states and job values can guarantee each agent under worst-case value distributions $\mathcal F_i$. 
For each fixed $n$, we show that
$
\lim_{\delta\to 0}\lim_{T\to\infty}\lambda(\delta,n,T)=1$ (see \cref{fig:lambda_vs_delta} for how $\lambda(\delta,n,T)$ behaves for fixed $n$ and large $T$).
Thus, sufficiently persistent jobs allow nearly the full ideal-utility
benchmark to be attained over a sufficiently long horizon.
Our mechanism achieves this benefit even when agents' queue states
are private and other agents behave arbitrarily.

\subsection{Our Results and Techniques}

We introduce the \emph{Sampled Pairwise Externality Compensator}
(\spec) mechanism, a non-monetary scheduling mechanism for our setting which is both efficient and gives optimal ideal utility guarantees.
We develop it in two stages. First, we define \emph{Pairwise Externality Compensation} (\pec), a mechanism for our model using exact externalities; its pairwise accounting scheme draws on ideas in~\citet{safronov2018coalition} and its dynamic development in the Guaranteed Utility Mechanism (GUM) of \citet{csoka2023collusion}. Second, we obtain \spec by replacing exact externalities with sampled estimates. Coupled simulations and local mixing make this computation efficient while preserving robust individual utility guarantees. In particular, we obtain the following utility guarantee for \spec.

\begin{theorem}[First part of \cref{thm:main_result_one}, rephrased]
\label{thm:introbrief}
For fixed $\delta>0$, \spec gives each agent $i$ a strategy
that guarantees
$
\mathbb{E}[U_i]\geq\lambda(\delta,n,T)\,v_i^\star T-\widetilde{O}(\sqrt{nT})$,
regardless of the behavior of the other agents, where $U_i$ is
agent $i$'s total utility and $v_i^\star$ is its ideal utility per
period.
\end{theorem}
Thus, each agent can secure this expected-utility guarantee even
if all other agents collude to harm it. The leading factor
$\lambda(\delta,n,T)$ is optimal: even a fully informed central
planner cannot guarantee every agent a larger fraction of ideal
utility uniformly over value distributions. Our mechanism attains
this factor up to an additive loss of
$\widetilde{O}(\sqrt{nT})$ in total utility, which is sublinear in
$T$ for fixed $n$ and $\delta>0$.
We focus on the $\delta > 0$ case since for $\delta = 0$ jobs never disappear, making the problem trivial (in particular, round-robin is asymptotically optimal).

\noindent\textbf{Why persistent jobs create a computational challenge:}
When $0<\delta<1$, allocating the server changes both current utility
and future competition. Serving one agent leaves other agents'
jobs waiting, changing their future queues and service opportunities.
An agent's report can therefore affect other agents over many
subsequent periods. 
Our mechanism proposes a simple allocation rule that tracks these effects via pairwise externality compensation. However, evaluating them directly requires reasoning about the evolution of the queues over the remaining horizon. Our challenge is to estimate these future effects efficiently while preserving utility guarantees against arbitrary reporting strategies.

Now, let us give a brief overview of how our mechanism works. At each time $t$, each agent makes a report of how many ``high-valued'' jobs they currently have. The principal gives access to an agent uniformly at random among agents who report a nonzero number. One can show that if each agent uses the simple strategy of reporting the number of jobs in their queue of value in the top $1/n$-quantile of their value distribution, and processes the highest-valued job when given the opportunity, each agent obtains at least a $\lambda(\delta,n,T)$ fraction of their ideal utility in expectation.
This choice of threshold also simplifies the mechanism's information requirements. A newly arriving job is high-valued with probability exactly $1/n$, independently across agents and periods. Under the prescribed
strategy, the process of high-valued job counts depends
only on $n$, expiration probability $\delta$, and allocation rule, rather than on the value distributions $\mathcal F_i$.
This means the principal can \emph{simulate the job-count process} without knowing or learning the job value distributions; each agent only needs to identify its own highest-valued job when served.
This reduction also characterizes the optimal benchmark factor: under the target policy, $\lambda(\delta,n,T)T$ is the expected number of high-valued jobs served, and the symmetric instance $\mathcal F_i=\Bern(1/n)$ makes this tight --- every positive-valued job has value $1$, and no policy can serve more such jobs in expectation than a policy that serves one whenever one is available.

Of course, agents need not follow this strategy. Nevertheless, \pec allows an agent to secure this guarantee, up to an $\tilde O(\sqrt{n T})$ additive loss, \emph{even when other agents use arbitrary strategies}.
The relevant rules of \pec can be summarized through its pairwise virtual accounts:
for each ordered pair of agents, the mechanism maintains a virtual account recording the cumulative externality of one agent's reports on the other's anticipated service. These accounts involve no monetary transfers. If an agent's cumulative negative externality on another agent falls below a prescribed threshold, the former is temporarily required to report zero current jobs. Under the prescribed strategy, the externality increments have conditional mean zero, allowing concentration bounds to control the probability of such timeouts. An agent following the prescribed strategy avoids being timed out with high probability. Controlling both the externalities imposed by other agents and the loss from these timeouts yields the
expected-utility guarantee in~\cref{thm:introbrief}, regardless of the other agents' strategies.

The prescribed strategy is simple to implement: an agent reports the number of jobs in its queue whose
values lie in the top $1/n$ quantile of its own value distribution,
and serves its highest-valued job when selected. Securing the
guarantee requires neither observing other agents' reports nor
predicting their future behavior. In particular, the guarantee
does not depend on other agents following the prescribed strategy
or on play converging to an equilibrium.

While participating in the \pec mechanism is simple from the perspective of the agents, the principal has a more difficult task.
In particular, \pec requires the principal to compute the expected externality imposed on each other agent, which requires evaluating the effect of an agent's report for the rest of the horizon. The technically challenging part of our work is in showing how coupled simulations and local mixing can be used to estimate these efficiently while preserving the utility guarantee. This leads to the \spec mechanism, with the following runtime guarantee.

\begin{theorem}[Second part of \cref{thm:main_result_one}, rephrased] \label{thm:runtime_result_introduction}
For fixed $\delta>0$, \spec runs in $O(n^2\log T)$ time per period in the worst case, $O(n^2\log\log T)$ in expectation, and $O(n^2)$ amortized in expectation. These bounds hold under arbitrary reporting strategies.
\end{theorem}
In particular, the expected amortized computation per period is
independent of the horizon $T$, even though the externalities being
estimated include effects on future allocations.

\noindent\textbf{Pairwise externality compensation and utility guarantees:}
To explain how we obtain this result, we first describe the pairwise
externality accounting underlying the \pec mechanism. Given the reports so far, consider the \emph{expected} number of
high-valued jobs served for each agent over the horizon, assuming that
all agents follow the prescribed strategy from then on; call these the \emph{anticipated payoffs}. In each period, reports are considered sequentially: agent $i$'s \emph{externality} on agent $j$ is the change in $j$'s anticipated payoff when $i$'s report is incorporated. The mechanism records these externalities in pairwise virtual accounts. If an agent's cumulative externality on any other agent goes below a threshold, the former is temporarily required to report zero.

The core idea from~\citet{csoka2023collusion} is that this accounting can protect an agent against the effects of other agents' reports. In our setting, when an agent follows the prescribed strategy, its reports have zero expected externality conditional on the preceding history. Concentration bounds then show that it rarely reaches a virtual debt limit. At the same time, the limits restrict the cumulative negative externalities imposed on it by other agents. Together, these observations yield the robust utility guarantee; we give the formal argument in~\Cref{sec:gum}.

\noindent\textbf{Efficient estimation through coupling and local mixing:}
It is not obvious how the principal could run this mechanism efficiently. In particular, the anticipated payoffs count the expected number of high-valued jobs served over the entire horizon, and when an agent $i$ makes a report in a given period, their report affects other agents' opportunities in future periods as well. Computing these anticipated payoffs directly at every timestep could therefore be expensive.

The key idea that allows us to obtain our efficient algorithm is that, at each timestep, the mechanism only needs estimates
of \emph{externalities} (i.e., changes in anticipated payoffs) rather than the anticipated
payoffs themselves. We estimate these differences using coupled simulations. Specifically, when agent $i$ makes a report, to estimate its externality on agent $j$, we couple two simulated versions of the future process. The first incorporates $i$'s actual report in the current period; the second replaces it with a report sampled according to the prescribed strategy. In both versions, we assume agents follow the prescribed strategy in future periods. We maximally couple the two chains (i.e., use the same randomness wherever possible) and stop simulating once they sync (i.e., have the same state), since their subsequent service outcomes are henceforth the same. The difference in the number of high-valued jobs served for agent $j$ in the two simulations is thus an unbiased estimate of the externality.
We show these two Markov chains sync in time $O(\log T)$ in expectation, giving us an $O_n(\log T)$ expected runtime for this basic estimator.

To obtain the improved runtimes as in \cref{thm:runtime_result_introduction}, we observe that the $O(\log T)$ bound is pessimistic for queue sizes typically reported by agents following the prescribed strategy. We therefore add per-period and cumulative caps on reported queue lengths. These caps are unlikely to affect an agent following the prescribed strategy, but prevent arbitrary reports from making the simulations expensive. We also stop each simulation after a prescribed number of timesteps. This truncation makes the estimator slightly biased; we bound the resulting error and show that it preserves the utility guarantee up to the stated additive loss. Together, the coupling bound and these additional rules yield the improved runtimes even when agents report arbitrarily.

Finally, we believe that this approach extends beyond our specific
queuing model. As demonstrated by~\citet{csoka2023collusion}, pairwise externality compensation provides a way to protect individual utility in dynamic allocation problems; however its implementation requires accounting for the future effects of reports. Our results suggest that these effects can be estimated efficiently in other Markovian settings where the prescribed process can be
simulated and local mixing makes coupled trajectories agree quickly.
This offers a route to making similar robust guarantees computationally accessible in a broader range of non-monetary allocation problems.

\subsection{Related Work}

\noindent\textbf{Non-Monetary Repeated Resource Allocation:}
Prior work has investigated designing mechanisms in our setting in the special case when $\delta=1$. 
The problem was first introduced by \citet{guo2009competitive}. Subsequent work \cite{jackson,cavallo2014incentive,gorokh2021monetary} studied repeated non-monetary allocation through linked decisions; in particular, \citet{gorokh2021monetary} gave a general technique to convert the efficiency guarantee of any monetary mechanism using a non-monetary mechanism up to a sublinear error. However, unlike our work, these works assume the principal has full knowledge of the value distributions.

A more recent line studies distribution-agnostic mechanisms with
agent-specific guarantees relative to the ideal utility benchmark, introduced by \citet{gorokh2021remarkable}. For requests that must be served immediately (i.e., $\delta=1$ in our model), a sequence of works~\cite{gorokh2021remarkable,lin2025online,lin2026itcs} obtain progressively stronger
worst-case guarantees that hold regardless of other agents' behavior.
\citet{lin2026robust} obtain the optimal worst-case fraction at equilibrium,
together with a weaker guarantee against arbitrary behavior by other
agents. Beyond worst-case value distributions,
\citet{onyeze2025allocating} study equilibrium guarantees. Finally, recent work~\cite{csoka2025prior,csoka20261} on so-called `fair-floor guarantees' implies the optimal worst-case fraction
of ideal utility when $\delta=1$; we discuss these results and their
relationship to our mechanism below.

A related form of intertemporal demand arises when agents require the
resource for several consecutive periods. \citet{banerjee2023robust,fikioris2025beyond} study such reusable-resource
settings and obtain robust guarantees relative to ideal utility. This source of temporal dependence differs from ours: their requests can occupy the resource for multiple periods, whereas our jobs require one period of service but may wait in a queue before being served or expiring.

\noindent\textbf{Pairwise Externality Compensation and Guaranteed Utility:} Our mechanism builds on the general Guaranteed Utility Mechanism (GUM) of
\citet{csoka2023collusion}, which combines techniques from \citet{safronov2018coalition} and \citet{gorokh2021monetary} to build non-monetary mechanisms for dynamic settings. Pairwise externality
compensation accounts for the effects of one agent's reports on another
agent's anticipated payoff. Combined with techniques for replacing
monetary transfers by artificial currencies
\cite{gorokh2021monetary}, this yields non-monetary mechanisms with
individual utility guarantees. \citet{csoka2026guaranteed} surveys guaranteed utility equilibria and
their applications; these ideas have also found use in auction design~\cite{jagadeesan2026limits}.

In the immediate-expiration setting, \citet{csoka2025prior} uses a
prior-free version of the GUM framework to get robust guarantees relative to a new \emph{fair-floor} benchmark; \citet{csoka20261} obtains similar guarantees through a repeated virtual first-price auction.
The fair-floor benchmark is more complicated to define, but it is at
least a $\lambda(1,n,T)$ fraction of the ideal utility. Consequently, these
results already imply the optimal worst-case fraction of ideal
utility when $\delta=1$, up to a sublinear additive loss, regardless
of other agents' behavior. 

Our contribution is to characterize the optimal fraction of ideal
utility with persistent jobs and to attain it through computationally
efficient sampled compensation. Persistence makes a report affect
other agents' service opportunities over many future periods, which makes the externalities required by \pec complex.
Rather than evaluating these effects exactly, we estimate them
through coupled simulations of the high-valued job-count process.
This process can be simulated without knowing agents' value
distributions, and local mixing allows the simulations to stop once
their states agree. Controlling sampling fluctuations and truncation
bias, together with report caps that rarely affect the prescribed
strategy, preserves the robust utility guarantee while yielding
$O(n^2)$ expected amortized computation per period for fixed
$\delta>0$. Each agent secures its guarantee through a simple
strategy based on reporting high-valued job counts, without
requiring other agents to follow an equilibrium strategy.

\textbf{Strategic Queues, Decentralized Learning, and Fair Sharing:} There is a long literature on each of the classical topics from learning in queuing systems, dynamic matching with abandonment and fair sharing. Work on learning in queuing systems, such as \cite{DBLP:conf/nips/KrishnasamySJS16,DBLP:journals/ior/KrishnasamySJS21,gaitonde2023price,abel2025learning,fu2022stability,baudin2023strategic,freund2024efficient,sentenac2021decentralized} considers outcomes of no-regret learning and clarifies how coordination, information, and learning
affect stability and service capacity. Our objective is instead to guarantee each agent utility from
heterogeneous, privately valued jobs that may expire. 

Queueing and expiration also arise in dynamic matching. \citet{amanihamedani2025adaptive} study a continuous-time model in which
suppliers wait in queues and may abandon, while customers must be
matched on arrival by centralized system, highlighting the value of adapting
decisions to queue states when opportunities may expire. In contrast, we design a non-monetary mechanism for agents with
private values and queue states, with robust individual utility guarantees.

There is also a long literature of online mechanisms that aim to provide fair shares for all agents. For example recent work of \citet{DBLP:conf/sigmetrics/SalemIN23} studies repeated allocation decisions when utilities vary over time and the goal is to satisfy $\alpha$-fairness in the long run. 
In this paper, we consider ideal utility, an individual fair share benchmark, and develop a mechanism that guarantees each player a fraction of this benchmark independently of the behavior of other players.

\section{Preliminaries}
\subsection{Model}

A fixed set of $n$ agents compete for a shared resource over a horizon $T$. Time proceeds in timesteps $t=1,2,\dots, T$. Each agent maintains a queue of jobs, initially empty. In each timestep $t$, each agent~$i$ receives a job $p_i[t]$ in their queue.
The job has value $V[p_i[t]]$, where $V[p_i[t]]$ is drawn from some fixed agent-specific distribution $\mathcal F_i$ independently from the past history and the other agents' job values. We assume $\mathcal F_i$ is non-negative and bounded above by $1$. There is a single server that can serve one job per time period. Unserved jobs in agents' queues each expire in each period with probability $\delta\in(0,1]$, independently from the past history and other job expirations. Unexpired jobs remain in their respective queues for the next time period. The order of events within a timestep is: (1) agents receive jobs, (2) at most $1$ job gets served, (3) unserved jobs expire with probability $\delta$ each independently.

Agents aim to maximize the total sum of the value of their served jobs. Specifically, agent $i$'s utility is defined as
\[
    U_i = \sum_{p_i[t]:\text{$p_i[t]$ served before expiration}}V[p_i[t]].
\]

Agent $i$'s queue state is private to agent $i$. In particular, for an agent $i$, neither the principal nor the other agents $j\neq i$ know the job values $V[p_i[t]]$ nor the expiration events in agent $i$'s queue. Additionally, the principal does not have knowledge of the value distributions $\mathcal F_i$, but each agent knows their own distribution. In each round, the principal has the ability to grant a single agent access to the server. That agent can choose a job from their queue to serve or let the server idle. 

\noindent\textbf{Asymptotic Notation Convention:} There are three main parameters to our setting: $T$, $n$, and $\delta$. We treat $T$ as very large and $n$ as potentially large. We treat $1/\delta$ as a fixed constant bounded by a constant independent of $T$ and $n$. We shall use asymptotic notation (e.g., $O$, $\Omega$, etc.) to denote dependence on $T$ and $n$ only and treat $\delta$ as fixed. Furthermore, we assume we are in a parameter regime where $n\leq T$. We use $\tilde O$ and $\tilde\Theta$ to suppress polylogarithmic factors in $T$ and $n$. We also use the notation $O_n(\cdot), o_n(\cdot)$, etc., to suppress dependence on $n$. 

\subsection{Ideal Utility}
To measure how well an agent would ideally do in our setting given the existence of others, we use the \textit{ideal utility} benchmark used in previous work \cite{gorokh2021remarkable,banerjee2023robust,fikioris2025beyond,onyeze2025allocating,lin2025online,lin2026robust,lin2026itcs} in our setting.
\begin{definition}[Ideal Utility]
Let $F_i^{-1}$ denote the inverse CDF\footnote{Formally, $F_i^{-1}(q_i) = \inf\{v_i: \Pr_{V_i\sim\mathcal F_i}(V_i\leq v_i)\geq q_i\}$} of agent $i$'s value distribution $\mathcal F_i$. Let $Q_i\sim\Unif[0,1]$ and $V_i = F_i^{-1}(Q_i)$. The \textit{ideal utility} of agent $i$ is
\[
    v_i^\star = \E[V_i\mathbb 1_{\{Q_i\geq 1-1/n\}}].
\]
\end{definition}
In the above definition, $Q_i$ represents the quantile of the value  $V_i$ drawn from $\mathcal F_i$. Agent $i$'s ideal utility is then the integral of $V_i\sim\mathcal F_i$ over the event that $V_i$ is in the top $1/n$-quantile of $\mathcal F_i$.

The quantity $v_i^\star T$ is a natural upper bound on the utility agent $i$ should expect to receive in a fair and efficient mechanism. As an individual fair-share benchmark, we allow agent $i$ an expected $T/n$ units of service over the horizon and let it devote that capacity to its highest-valued arrivals. The quantity $v_i^\star T$ is thus the maximum expected utility agent $i$ can obtain under this average service cap.

\subsection{High-Valued Jobs and the Target Policy} \label{ssec:target_policy}
Fix $\delta$, $n$, and $T$. We shall now propose a policy, which we call the target policy, for a social planner who knows all the agents' private information that guarantees each agent the best achievable fraction of $v_i^\star T$ (i.e., the minimax fraction over all value distributions $\mathcal F_i$). Formally, for any $(\delta,n,T)$, the target policy will give each agent at least $\lambda(\delta,n,T)\cdot v_i^\star T$ expected utility for any value distributions~$\mathcal F_i$, while conversely, under any other policy, there exist value distributions $\mathcal F_i$ such that at least one agent gets at most $\lambda(\delta,n,T)\cdot v_i^\star T$ expected utility. We first need one definition:

\begin{definition}[High-Valued Job] \label{def:high_valued}
We call a job $p_i[t]$ \textit{high-valued} if its value $V[p_i[t]]$ is in the top $1/n$-quantile of the value distribution $\mathcal F_i$.\footnote{If $\mathcal F_i$ has atoms and the top $1/n$-quantile is not uniquely defined, we randomize appropriately at the cutoff to ensure that we satisfy that $\Pr[p_i[t]\mbox{ is high-valued}]=1/n$}.
\end{definition}
The target policy goes as follows. In each round, if there are no high-valued jobs, leave the server idle. Otherwise, select an agent $i$ among the set of agents who have at least $1$ high-valued job uniformly at random. Process the highest-valued high-valued job from agent $i$'s queue, breaking ties arbitrarily. 

\begin{restatable}{theorem}{thmTargetPolicyOptimality} \label{thm:target_policy_optimality}
For every $\delta$, $n$, and $T$, there exists $\lambda(\delta,n,T) \in [0, 1]$ independent of the value distributions $\mathcal F_i$ such that under the target policy, each agent obtains at least $\lambda(\delta,n,T)\cdot v_i^\star T$ expected utility. Specifically, $\lambda(\delta,n,T)\cdot T$ is the expected number of high-valued jobs across all agents served under the target policy.

Furthermore, if $\mathcal F_i=\Bern(1/n)$ for every agent $i$, under any policy, there exists an agent $i$ obtaining at most $\lambda(\delta,n,T)\cdot v_i^\star T$ expected utility.
\end{restatable}
\begin{proof}[Proof Sketch]
We set $\lambda(\delta,n,T)\cdot T$ to be the expected number of high-valued jobs served across all agents and timesteps by the target policy. By symmetry, each agent gets $\lambda(\delta,n,T)\cdot T/n$ high-valued jobs served in expectation. By definition of high-valued and ideal utility, the expected value of a job conditioned on being high-valued is $nv_i^\star$. If the target policy chose a uniformly random high-valued job from the selected agent $i$ given access, the target policy's selection process for which job gets served would be independent of the value of a job conditioned on whether or not it is high-valued. It would follow that each agent obtains $\lambda(\delta,n,T)\cdot v_i^\star T$ expected utility. The target policy selects the highest-valued high-valued job from the selected agent, which yields utility that is no worse than selecting a uniformly random high-valued job.

Now, assume $\mathcal F_i=\Bern(1/n)$ for every agent $i$. Then, $v_i^\star = 1/n$ and the utility of an agent is equal to the number of high-valued jobs they get served. Any policy, including the target policy, that serves an arbitrary high-valued job whenever one is available maximizes the expected number of high-valued jobs served across all agents and timesteps, because all high-valued jobs look identical at any given timestep. By symmetry and the definition of $\lambda(\delta,n,T)$, under any policy, there exists an agent $i$ obtaining at most $\lambda(\delta,n,T)\cdot v_i^\star T$ expected utility.

The complete proof is in \Cref{sec:target_policy_optimality_proof}.
\end{proof}

\begin{wrapfigure}[12]{R}{0.4\textwidth}
    \vspace{-13pt}
    \includegraphics[width=0.4\textwidth]{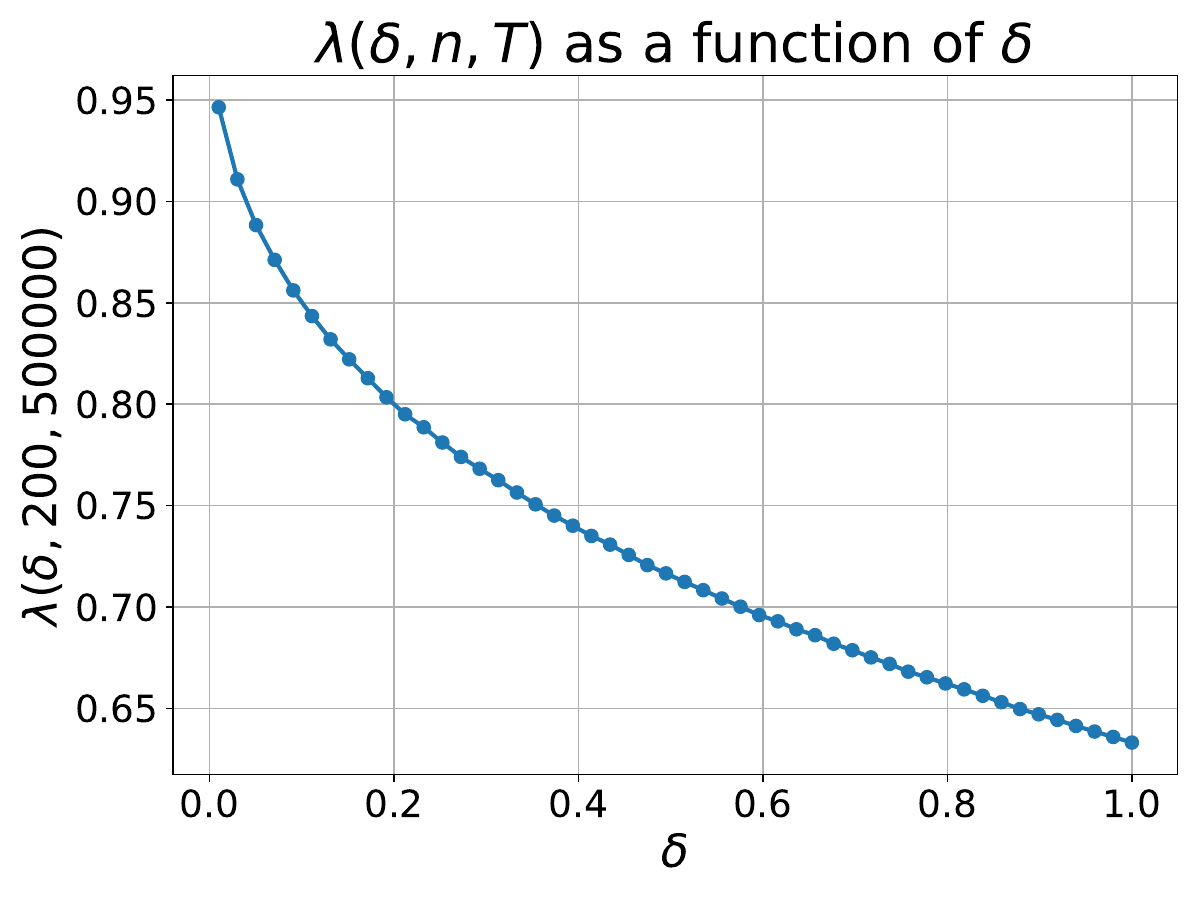}
    \vspace{-26pt}
    \caption{Empirical estimate of $\lambda(\delta,n,T)$ as a function of $\delta$ for $n=200$ and $T=500000$.
    }
    \Description{}
    \label{fig:lambda_vs_delta}
\end{wrapfigure}

For $\delta=1$, one can show that $\lambda(1,n,T)$ is exactly the $1-(1-1/n)^n$ bound as mentioned in the introduction using the $\Bern(1/n)$ value distributions, since the target policy always serves a job of value $1$ when one exists. In general, $\lambda(\delta,n,T)$ is nonincreasing in $\delta$ with $\lim_{\delta\to 0}\lim_{T\to\infty}\lambda(\delta,n,T)=1$ (\cref{prop:iterated_limit}). The core reason why $\lambda(1,n,T)$ is lower is that when multiple agents have high-valued jobs in a given round, only one can be served and the rest expire. As $\delta$ gets smaller, the fraction of high-valued jobs that can be served increases because they do not instantly expire when not served. We do not know of a closed-form expression for $\lambda(\delta,n,T)$, but we plot empirical estimates of $\lambda(\delta,n,T)$ as a function of $\delta$ to illustrate the effect of $\delta$ in \cref{fig:lambda_vs_delta}.

\begin{example}[Random Lottery Does Poorly]
The target policy always guarantees each agent at least a $\lambda(\delta,n,T)\geq \lambda(1,n,T) = 1-(1-1/n)^n \geq 1-1/e\approx 0.63$ fraction of their ideal utility. In this example, we show that the policy of giving a uniformly random agent access to the server each timestep, which is trivially implementable even if agents are strategic, can give each agent a small fraction of ideal utility even if $\delta$ is not $1$ but positive. Suppose agent $i$ has a $\Bern(1/n)$ value distribution. The fraction of ideal utility agent $i$ obtains is the fraction of value $1$ jobs served. By linearity of expectation, it suffices to bound the probability that a single job gets served. A job's lifetime is at most some $G\sim\Geom(\delta)$ independent of service. By union bounding over the events that a job gets served at some time step in its lifetime, the probability that the job gets served conditioned on $G$ is at most $G/n$ with the random lottery. Therefore, the probability it gets served is at most $\mathbb E[G/n] = 1/(n\delta)\to 0$ as $n\to\infty$.
In contrast, \cref{fig:lambda_vs_delta} is high for all $\delta$, even for $n = 200$.
\end{example}

\section{Pairwise Externality Compensations and Utility Guarantees} \label{sec:gum}

We now provide a brief summary of the steps we follow in the rest of the paper to prove our main result. We first define the pairwise virtual externality of an agent's report as the change it causes in another agent's anticipated future utility (\cref{subsec:externalities}). We then formally define the \pec mechanism (\cref{alg:gum_no_computation} in \cref{subsec:mech+GUM}), and prove that under this mechanism, agents have a strategy that guarantees them a worst-case optimal fraction of their ideal utility. 

This guarantee requires a $O(\text{polylog}(T))$ bound on the magnitude of each pairwise externality, to control the increments of the virtual-debt martingales and thus ensure an honest agent is rarely timed out; a priori, it is unclear if this is bound holds.
In \cref{subsec:externality_almost_sure_bound} we show that the externalities of such individual reports are bounded by a term that grows logarithmically with $T$. To do so, we couple two variants of our queuing process: one incorporates the agent's report at the current period, while the other samples that report from the target-policy transition kernel. We show that the two processes sync after $O(\log T)$ periods in expectation, yielding the externality bound.

The above coupling turns out to be critical for our subsequent results. 
In particular, in~\cref{sec:computation}, we use this coupling to construct a computationally efficient, slightly biased estimator of each pairwise externality (\cref{alg:compute_estimator}). We then bound the bias and combine the estimates with report caps to obtain our \spec mechanism (\cref{alg:gum_report_capping}).
For suitable parameters, we show that \spec has expected per-period runtime $O(n^2\log\log T)$ and expected amortized runtime $O(n^2)$ for fixed $\delta>0$, even under arbitrary reporting strategies.

\subsection{Description of the High-Valued Jobs Stochastic Process}

To define the externality charged to each report in \pec, we first describe a \emph{target-policy count} process. This is purely an accounting device: it lets us attribute changes in another agent's anticipated service to one report at a time. In particular, \pec processes reports in each period in the same order, and records externalities in pairwise virtual accounts. The underlying arrival, service, and expiration model remains the one specified in the preliminaries.

\paragraph{Splitting Each Timestep into Finer-Grained Updates}
In addition to considering stochastic processes with the timesteps $t=1,2,\dots, T$, we shall consider stochastic processes with more fine-grained \textit{updates}. We break down each timestep $t$ into $n+1$ updates $(t,1), (t,2), \dots, (t,n), (t,n+1)$ as follows. At each update $(t,i)$ for $i\in[n]$, job $p_i[t]$ arrives to agent $i$'s queue and the jobs in agent $i$'s queue from timestep $t-1$ expire with probability $\delta$ each independently. Agent $i$'s queue does not change at updates $(t,j)$ for $j\in[n]\setminus\{i\}$. At update $(t,n+1)$, the central server either chooses some job from some agent's queue to serve and removes that job or chooses no job to serve. Updates $(t,i)$ are ordered lexicographically; we let $\leq$ denote this lexicographic ordering on updates $(t,i)$. For convenience, we denote $(t,0) = (t-1,n+1)$ for $t>0$ and $(0,n+1)$ be an initial update at which every agent's queue is empty. In this way, $(t,i-1)$ is  the update before update $(t,i)$ for all $t>0$.

\begin{remark}
We think of the breakdown of each timestep $t$ into updates $(t,i)$ as part of how the algorithm runs and not how the physical job process operates. At the beginning of each time $t$, agents obtain their updated queues, and the algorithm then runs the updates $(t,i)$ for $1\leq i\leq n+1$. 
\end{remark}

\paragraph{States}
Let $\theta_i[t,j]$ denote the number of high-valued jobs that agent $i$ has at (right after) update $(t,j)$. We let $\theta[t,j]$ denote the vector of $\theta_i[t,j]$ for $i\in[n]$. For convenience, we let $\theta_i[t] = \theta_i[t,i]$, the number of high-valued jobs that agent $i$ has after the job $p_i[t]$ arrives and the expirations of jobs from timestep $t-1$ occur but before service, and $\theta[t]$ denotes the vector of the $\theta_i[t]$.

\paragraph{Markovian Stochastic Transition Kernel}

We define Markovian stochastic transition kernels $P_{t,j}:\mathbb Z_{\geq 0}^n\to\Delta(\mathbb Z_{\geq 0}^n)$ so that $P_{t,j}(\theta[t,j-1])$ is the law of $\theta[t,j]$ conditioned on the history up to and including update $(t,j-1)$ under the target policy. For $j\in[n]$, we define
\[
    (P_{t,j}(\bar\theta[t,j-1]))_j = \Binom(\bar\theta_j[t,j-1],1-\delta)\ast\Bern\left(\frac1n\right),
\]
since at update $(t,j)$, a $\Binom(\theta_j[t,j-1],1-\delta)$ number of high-valued jobs survive in agent $j$'s queue from the previous timestep, and with probability $1/n$, a new high-valued job arrives. For $i\neq j$, we define $(P_{t,j}(\bar\theta[t,j-1]))_i$ to be deterministically $\bar\theta_i[t,j-1]$ since only agent $j$'s queue changes at update $(t,j)$. For $j=n+1$, we define the distribution $P_{t,n+1}(\bar\theta[t,n])$ to be deterministically $0$ if $\bar\theta[t,n]=0$, and otherwise, $(P_{t,n+1}(\bar\theta[t,n]))_k = \bar\theta_k[t,n]-1$ for some $k$ chosen uniformly from $\{i:\bar\theta_i[t,n]>0\}$ and $(P_{t,n+1}(\bar\theta[t,n]))_i = \bar\theta_i[t,n]$ for $i\neq k$.

\paragraph{Histories}
Let $H_{t,j} = (\mathbb Z_{\geq 0}^n)^{\{(s,i):(s,i)\leq (t,j)\}}$ denote the set of sequences of states where there is one state for each update $(s,i)\leq (t,j)$. We call an element $h_{t,j}\in H_{t,j}$ a partial history up to update $(t,j)$. Given a partial history $h_{t,j}\in H_{t,j}$ and an update $(s,i)\geq (t,j)$, let $\mu_{s,i}(\cdot|h_{t,j})$ be the law of $h_{s,i}\in H_{s,i}$ obtained starting from the partial history $h_{t,j}$. Formally, define $\mu_{t,j}(\cdot|h_{t,j})$ at update $(t,j)$ to be the distribution that is deterministically $h_{t,j}$ and for $(s,i)>(t,j)$, inductively define $\mu_{s,i}(\cdot|h_{t,j})$ to be the law of $h_{s,i}$ where $h_{s,i}$ is obtained from $h_{s,i-1}\sim\mu_{s,i-1}(\cdot|h_{t,j})$ by applying $P_{s,i}$ to the last element of $h_{s,i-1}$ and appending it to $h_{s,i-1}$.

\subsection{Virtual Anticipated Payoffs and Externalities}
\label{subsec:externalities}
Define $\tilde U_j:H_{t,n+1}\to\mathbb Z$ by letting $\tilde U_j(h_{t,n+1})$ be the total number of high-valued jobs of agent $j$ served when the history of states is $h_{t,n+1}$. Formally, writing $h_{t,n+1} = (\bar\theta_{t,j})_{(t,j)}$,
\[
    \tilde U_j(h_{t,n+1}) = \sum_{s=1}^t (\bar\theta_j[s,n] - \bar\theta_j[s,n+1]).
\]

We now define the virtual anticipated payoffs and externalities used by \pec.
\begin{definition}[Virtual Anticipated Payoff] \label{def:virtual_anticipated_payoff}
The \textit{virtual anticipated payoff} of agent $j$ given a partial history $h_{t,i}\in H_{t,i}$ is
\[
    \Upsilon_j(h_{t,i}) = \E_{h_{T,n+1}\sim\mu_{T,n+1}(\cdot|h_{t,i})}[\tilde U_j(h_{T,n+1})].
\]
\end{definition}
Intuitively, $\Upsilon_j(h_{t,i})$ is the expected number of high-valued jobs that agent $j$ has served conditioned on the history up to update $(t,i)$ being $h_{t,i}$ if the target policy is used on updates after $(t,i)$. When $h_{t,i}$ has a nonzero probability of being an actual history up to update $(t,i)$, this is precisely what $\Upsilon_j(h_{t,i})$ is.

\begin{definition}[Virtual Externality] \label{def:virtual_externality}
Fix $i\in[n]$. Given a partial history $h_{t,i-1}\in H_{t,i-1}$, the \textit{externality} of agent $i$ on agent $j$ given a state $\bar\theta_i[t]\in\mathbb Z_{\geq 0}$ is
\[
    \gamma_{i\to j}(h_{t,i-1}, \bar\theta_i[t]) = \Upsilon_j(\bar h_{t,i}) - \Upsilon_j(h_{t,i-1})
\]
where $\bar h_{t,i}\in H_{t,i}$ is obtained by appending the last state of $h_{t,i-1}$ with coordinate $i$ replaced by $\bar\theta_i[t]$.
\end{definition}
One can think of $\gamma_{i\to j}(h_{t,i-1},\bar\theta_i[t])$ as the expected additional high-valued jobs that agent $j$ gets served conditioned on the history up to update $(t,i-1)$ being $h_{t,i-1}$ if agent $i$ has state $\bar\theta_i[t]$ at update $(t,i)$ instead of ``the expectation'' if the target policy is used on updates after $(t,i-1)$.

\subsection{The \pec Mechanism and its Guaranteed Utility Property}
\label{subsec:mech+GUM}

We now formally describe our \pec mechanism, which adapts the transfer-free version of the Guaranteed Utility Mechanism (GUM) of~\citet{csoka2023collusion} to our queueing setting. The high-level idea is to combine pairwise externality accounting~\cite{safronov2018coalition} with non-monetary enforcement via virtual accounts~\cite{gorokh2021monetary}. 
Our enforcement rule differs however: transfer-free GUM replaces all of an agent's subsequent reports with simulated reports once a debt limit is exceeded, whereas \pec forces zero reports only while the agent's virtual debt remains above the limit.
For completeness, \cref{appsec:gum_review} provides a brief overview of how pairwise virtual externality accounting works with transferable utilities (i.e., with money) in a simple setting. To adapt this to non-monetary settings, similar to transfer-free GUM, we use virtual debt accounts to limit how much an agent's reports can reduce another agent's anticipated payoff. 

In more detail, the \pec mechanism is parameterized by a \textit{virtual debt limit} $B$. \pec shall define \textit{virtual transfers} $p_{i\to j}[t]$ at each time $t$. For each time period $t=1,2,\dots, T$, go in order of agent number $i$ and do the following. If $\sum_{s=1}^{t-1}p_{i\to j}[s] > B$ for some $j\neq i$, set $\hat\theta_i[t]=0$. Otherwise, elicit a report for the number of high-valued jobs $\hat\theta_i[t]\in\{0,1,\dots,T\}$ from agent $i$. (We enforce that $\hat\theta_i[t]\leq T$.) Define the \textit{virtual externality} of agent $i$ on agent $j\neq i$ at time $t$ to be $\gamma_{i\to j}[t] := \gamma_{i\to j}(\hat h_{t,i-1}, \hat\theta_i[t])$ where $\hat h_{t,i-1}$ is the history containing the reports and server allocations right before agent $i$'s report $\hat\theta_i[t]$ (see \cref{alg:gum_no_computation} for a precise definition of $\hat h_{t,i-1}$). Set agent $i$'s \textit{virtual transfer} to each agent $j\neq i$ at time $t$ by $p_{i\to j}[t] := -\gamma_{i\to j}[t]$. In a time period $t$, after processing all agents $i$ in this fashion, if $\hat\theta_i[t] = 0$ for all $i$, leave the server idle; otherwise, give access to an agent $i$ uniformly at random chosen from agents with $\hat\theta_i[t]>0$. Pseudocode for the mechanism is given in \cref{alg:gum_no_computation}.

\begin{algorithm}
\caption{Pairwise Externality Compensation (\pec) Mechanism}
\label{alg:gum_no_computation}
\KwIn{virtual debt limit $B$}
Initialize $\hat\theta[0,n+1]=0$\;
\For{$t=1,2,\dots,T$}{
    \For{$i=1,2,\dots, n$}{
        \eIf{$\sum_{s=1}^{t-1}p_{i\to j}[s] > B$ for some $j\neq i$}{
            Set $\hat \theta_i[t]=0$\;
        }{
            Elicit a report $\hat\theta_i[t]\in\{0,1,\dots,T\}$ from agent $i$\;
        }
        Set $\hat\theta_i[t,i] = \hat\theta_i[t]$ and $\hat\theta_j[t,i]=\hat\theta_j[t,i-1]$ for $j\neq i$.\;
        Agent $i$ makes virtual transfers $p_{i\to j}[t]= -\gamma_{i\to j}[t]$ where $\gamma_{i\to j}[t] = \gamma_{i\to j}(\hat h_{t,i-1},\hat\theta_i[t])$ where $\hat h_{t,i-1} = (\hat\theta_{s,j})_{(s,j)\leq(t,i-1)}$\;
    }
    \eIf{$\hat\theta_i[t]=0$ for all $i$}{
        Let the server idle\;
        Set $\hat\theta[t,n+1]=\hat\theta[t,n]$\;
    }{
        Grant agent $i$ access to the server where $i\sim\Unif\{j:\hat\theta_j[t]>0\}$\;
        Set $\hat\theta_i[t,n+1] = \hat\theta_i[t]-1$ and $\hat\theta_j[t,n+1]=\hat\theta_j[t]$ for $j\neq i$\;
    }
}
\end{algorithm}

The analogue of honest play in this mechanism is always reporting $\hat\theta_i[t] = \theta_i[t]$.
Say that agent $i$ is \textit{timed out} at time $t$ if $\sum_{s=1}^{t-1}p_{i\to j}[s] > B$ for some $j\neq i$.
\begin{definition}[honest]
Agent $i$ is \textit{honest} if she reports $\hat\theta_i[t] = \theta_i[t]$ at every time $t$ that she is not timed out and always serves the highest-valued high-valued job in her queue each time she is granted access to the server (breaking ties arbitrarily). 
\end{definition}

Our main result for this mechanism is that each agent is guaranteed a utility level at least close to the benchmark $\lambda(\delta,n,T)v_i^\star T$ guaranteed by the target policy by being honest, regardless of the behavior of other agents.

\begin{definition}[Error]
A mechanism has \textit{error} at most $R$ if each agent $i$ has a strategy that obtains expected utility at least $\lambda(\delta,n,T)\cdot v_i^\star T - R$ regardless of the behavior of agents $j\neq i$.
\end{definition}

\begin{restatable}{theorem}{thmErrorNoComputation} \label{thm:error_no_computation}
There exists a virtual debt limit $B = \tilde O(\sqrt{T/n})$ such that \cref{alg:gum_no_computation} has error at most $\tilde O(\sqrt{nT})$.
\end{restatable}
We prove the theorem in \cref{sec:deferred_proofs_no_computation} and give intuition for why this theorem holds now.

Let $\Upsilon_j[t,k] = \Upsilon_j(\hat h_{t,k})$ where $\hat h_{t,k}$ is the reported history up to update $(t,k)$. Fix an honest agent $i$. Let $\tilde U_j$ be the number of high-valued jobs of agent $j$ that are served. We first consider a scenario in which the virtual debt limit $B$ is not enforced on agent $i$ and the utility of each agent $j$ were simply the number of high-valued jobs served offset by the virtual transfers (i.e., as if the virtual transfers were real money); now each agent $j$'s utility is
\begin{equation*}
    \tilde U_j' = \tilde U_j - \sum_{k\neq j}\sum_{t=1}^Tp_{j\to k}[t] + \sum_{k\neq j}\sum_{t=1}^T p_{k\to j}[t],
\end{equation*}
and show that if agent $i$ is honest, then $\E[\tilde U_i']=\Upsilon_i[0,n+1]$ regardless of the behavior of the other agents. The utility $\tilde U_j'$ can be thought of as the number of high-valued jobs that agent $j$ gets served minus compensations paid to other agents plus compensations paid by other agents. We make the following observations.
\begin{enumerate}[nosep,leftmargin=*]
\item The initial anticipated virtual payoff $\Upsilon_i[0,n+1]$ is the expected number of high-valued jobs served from agent $i$ under the target policy. Furthermore, the anticipated virtual payoff $\Upsilon_j[t,k]$ is the expected number of high-valued jobs served from agent $j$ given the history up to update $(t,k)$ if the target policy were used after update $(t,k)$.
\item By definition, $p_{k\to j}[t] = -\gamma_{k\to j}[t] = -(\Upsilon_j[t,k] - \Upsilon_j[t,k-1])$. Thus, the virtual transfer $p_{k\to j}[t]$ can be thought of as agent $k$ compensating agent $j$ for the change $\Upsilon_j[t,k] - \Upsilon_j[t,k-1]$ in agent $j$'s anticipated virtual payoff when agent $k$ reports $\hat\theta_k[t]$ at update $(t,k)$.
\item The honest agent $i$'s virtual transfer $p_{i\to j}[t]$ has expectation $0$ because agent $i$ reports $\hat\theta_i[t] = \theta_i[t]$, which is the exact state that the target policy updates the history with, so agent $j$'s anticipated virtual payoff does not change in expectation.
\item Similarly, $\E[\Upsilon_j[t,n+1] - \Upsilon_j[t,n]]=0$ for any agent $j$ because the server selection process in \cref{alg:gum_no_computation} is exactly the same as the target policy.
\end{enumerate}
Therefore, agent $i$ obtains the same utility in expectation as their anticipated payoff in the beginning. In particular, it doesn't matter what strategy the other agents take: agent $i$ is compensated by any harm done by others regardless of their strategy.
Formally, define $\mathcal H_{t,j} = \sigma(\hat h_{t,j})$, the filtration generated by the reported history up to update $(t,j)$.\footnote{Note that $\theta_k[t]$ is not $\mathcal H_{t,j}$-measurable for $k>j$ since agent $k$ makes his report after agent $j$, even though agent $k$ may know his true number of high-valued jobs $\theta_k[t]$ already at update $(t,j)$.} Define
\begin{align*}
    M_i[t,j] = \Upsilon_i[t,j] - \underbrace{\sum_{k\neq i}\left(\sum_{s=1}^{t-1}p_{i\to k}[s] + p_{i\to k}[t]\mathbb1_{j\geq i}\right)}_{\text{virtual transfers to others}} +\underbrace{\sum_{(s,k)\leq (t,j):k\neq i}p_{k\to i}[s]}_{\text{virtual transfers from others}}.
\end{align*}
One can argue that $(M_i[t,j])_{(t,j)=(0,n+1)}^{(T,n+1)}$ is an $\mathcal H_{t,j}$-martingale using the intuition above.

Notice that $\Upsilon_i[0,n+1]$, the number of high-valued jobs served from agent $i$ under the target policy, is exactly $\lambda(\delta,n,T)\cdot\frac Tn$ by definition. Putting the above claim together, we have
\begin{equation} \label{eq:virtual_utility}
    \E[\tilde U_i'] = \E[\tilde U_i] - \sum_{j\neq i}\E\left[\sum_{t=1}^Tp_{i\to j}[t]\right] + \sum_{j\neq i}\E\left[\sum_{t=1}^T p_{j\to i}[t]\right] \geq \lambda(\delta, n, T)\cdot \frac Tn
\end{equation}
regardless of agents' $j\neq i$ strategies. For any agent $j\neq i$, if $\hat\theta_j[t] = 0$, then $p_{j\to i}[t] \leq 0$ for all $i\neq j$, because reporting $0$ cannot produce a negative externality on the others. Additionally, assume we have an absolute bound $D$ on any externality on $i$ in that $|\gamma_{j\to i}[t]|\leq D$ for any $j\neq i$. Since we time out agent $j$ whenever their cumulative virtual transfers to agent $i$ exceeds $B$ and force them to report $0$, it follows that agent $j$'s cumulative virtual transfers to agent $i$ cannot exceed $B+D$. In particular, $\sum_{t=1}^T p_{j\to i}[t]\leq B+D$. Substituting this into \eqref{eq:virtual_utility}, we obtain
\begin{equation*}
    \E[\tilde U_i] - \sum_{j\neq i}\E\left[\sum_{t=1}^Tp_{i\to j}[t]\right] \geq \lambda(\delta, n, T)\cdot \frac Tn - (n-1)(B+D).
\end{equation*}
Let $\mathcal G_t$ denote the filtration generated by the history up to and including time $t$. For any time period $t$ in which agent $i$ is not timed out, $\E[p_{i\to j}[t]|\mathcal G_{t-1}]=0$ for any $j\neq i$ using the same reasoning that an honest agent has $0$ impact to others' anticipated virtual payoffs in expectation. Using the above display, we obtain
\begin{equation} \label{eq:virtual_utility_without_others_payments}
    \E[\tilde U_i] \geq \lambda(\delta, n, T)\cdot \frac Tn - (n-1)(B+D),
\end{equation}
i.e., the number of high-valued jobs utility without real transfers is lower bounded by the same value the transferable utility is lower bounded by (see \eqref{eq:virtual_utility}), up to the $(n-1)(B+D)$ error.

By a greedy argument, one can show that choosing the highest-valued high-valued job whenever granted access to the server is at least as good as choosing a random high-valued job when granted access. If agent $i$ is honest except she chooses a random high-valued job whenever granted access, the event that a high-valued job gets served is independent of its value conditioned on the job being high-valued, and the expected value of a job conditioned on being high-valued is $nv_i^\star$ by the definition of ideal utility. Thus, the above display implies the following lower bound on the expected utility:
\[
    \E[U_i] \geq \lambda(\delta,n,T)\cdot v_i^\star T - (n-1)(B+D) nv_i^\star.
\]
This is a utility guarantee for agent $i$ in the version of the mechanism where the virtual debt limit is not enforced on agent $i$. To obtain a utility guarantee for the actual version of the mechanism where the virtual debt limit is enforced on agent $i$, it suffices to show that the probability that the agent hits the virtual debt limit is small. Specifically, define the process $X_{i\to j}[t] = \sum_{s=1}^t p_{i\to j}[s]$ in the mechanism where the virtual debt limit is not enforced on agent $i$. By an appropriate coupling argument, in the actual mechanism,
\[
    \E[U_i] \geq \lambda(\delta,n,T)\cdot v_i^\star T - (n-1)(B+D) nv_i^\star - \Pr(\exists t,\ j\neq i: X_{i\to j}[t]>B)\cdot T.
\]
Observe that $X_{i\to j}[t]$ is a martingale by the previous observation that $\E[p_{i\to j}[t]|\mathcal G_{t-1}]=0$. In \cref{lem:as_externality_bound}, we give a value of $D = O(\log T)$ bounding the absolute value of any externality. This directly bounds the increments of $X_{i\to j}[t]$ almost surely, allowing us to use martingale concentration inequalities to show that $\Pr(\exists t,\ j\neq i: X_{i\to j}[t]>B)\leq 1/T$ for some $B = \tilde O(\sqrt{T/n})$, which we do in \cref{sec:deferred_proofs_no_computation}. Substituting these bounds into the above display, we obtain the $\tilde O(\sqrt{nT})$ error bound in \cref{thm:error_no_computation}. (A simple Azuma-Hoeffding bound lets us perform this martingale concentration argument with $B = \tilde O(\sqrt T)$ which yields the weaker error bound $\tilde O(n\sqrt{T})$; we use Freedman's inequality (\cref{thm:freedman}) for this scenario in \cref{sec:deferred_proofs_computation} to obtain the stronger result.)

\begin{remark} \label{rem:not_equilibrium}
We emphasize that \cref{thm:error_no_computation} gives a guaranteed utility property that by being honest, an agent is guaranteed a certain level of utility, as opposed to an equilibrium statement. In fact, every agent being honest might not be an equilibrium. For example, suppose agent $i$ has a $\Bern(1/(2n))$ value distribution and all other agents are honest. In this case, agent $i$ reporting the number of jobs with value $1$ makes the system less congested, clearing other agents' jobs faster, in turn helping agent $i$ process more value $1$ jobs.
\end{remark}

\subsection{Bounding Pairwise Externalities via Coupling}
\label{subsec:externality_almost_sure_bound}
The below lemma gives a bound on the externalities that is needed for the proof of \cref{thm:error_no_computation} as described at the end of the last subsection. The proof uses a coupling of two high-valued jobs stochastic processes. The same coupling is the basis for the efficient externality estimates used by \spec in~\cref{sec:computation}.

\begin{lemma} \label{lem:as_externality_bound}
For any agents $i,j$ and time $t_0$, and any report $\hat\theta_j[t_0]$,
\[
    |\gamma_{j\to i}[t_0]|\leq \begin{cases}\left\lceil \frac{\log(T+1)}{-\log(1-\delta)}\right\rceil + \frac1\delta & \text{if $\delta < 1$}\\1 & \text{if $\delta=1$}\end{cases}.
\]
\end{lemma}
\begin{proof}
The proof goes by coupling two high-valued jobs stochastic processes, which are as follows. Let the two processes be $(\theta^{\text{report}}[t,i])$ and $(\theta^{\text{random}}[t,i])$. The histories of both processes up to update $(t_0,j-1)$ are the same and are equal to the reported history $\hat h_{t_0,j-1}$. Let $\theta^{\text{report}}[t_0,j] = \hat\theta[t_0,j]$, and $\theta^{\text{random}}[t_0,j]$ be drawn from the high-valued jobs stochastic transition kernel. For shorthand, let $\{A,B\} = \{\text{random}, \text{report}\}$. The states in updates $(t,i)>(t_0,j)$ are determined by the stochastic transition kernel applied to both processes in the following coupled way.

\paragraph{Arrivals}
A high-valued job arrives at update $(t,i)$ in process $A$ if and only if it arrives at update $(t,i)$ in process $B$.

\paragraph{Expirations}
Fix an update $(t,i)$ with $i\in[n]$. Name the processes so that $\theta_i^A[t,i-1]\geq \theta_i^B[t,i-1]$. Condition on the history of both processes up to update $(t,i-1)$ in the following description. Denote the number of high-valued job expirations in agent $i$'s queue at update $(t,i)$ in process $B$ by $Z$, which has distribution $\Binom(\theta_i^B[t,i-1], \delta)$. The number of high-valued job expirations in agent $i$'s queue at update $(t,i)$ in process $A$ is $Z$ plus an independent $\Binom(\theta_i^A[t,i-1]- \theta_i^B[t,i-1], \delta)$ random variable.

\paragraph{Services}
At each time $t\geq t_0$, at the service update $(t,n+1)$, sample a permutation $\pi$ of $[n]$ uniformly at random. In both processes, we service the agent $i$ with the lowest value of $\pi(i)$ among agents $i$ with $\theta_i[t]>0$.

Let $\sigma = \inf\{t\geq t_0:\theta^A[t] = \theta^B[t]\}$ where $\inf\emptyset = T+1$. By the coupling, the two processes sync at time $\sigma$ and are identical before service at all $t\geq\sigma$. By the definition of virtual externality, $|\gamma_{j\to i}[t_0]|$ is the absolute expected difference in the number of high-valued jobs that agent $i$ serves between the two processes. The absolute difference in realized service counts is at most $\sigma-t_0$ since the two processes are identical right before service at times $t < t_0$ and $t \geq \sigma$. Hence, we aim to bound $\mathbb E[\sigma] - t_0$.

Without loss of generality, name the processes such that $\theta_j^A[t_0,j]\geq \theta_j^B[t_0,j]$. We shall show that $\theta^A[t,i]\geq \theta^B[t,i]$ coordinate-wise for all $(t,i)\geq (t_0,j)$ by induction on the update $(t,i)$. For the base case, $\theta^A[t_0,j]\geq \theta^B[t_0,j]$ holds: $\theta_i^A[t_0,j] = \theta_i^B[t_0,j]$ for $i\neq j$ since $A$ and $B$ were perfectly coupled up to update $(t_0,j-1)$, and $\theta_j^A[t_0,j] \geq \theta_j^B[t_0,j]$ by assumption. Now we consider updates to the states caused by arrivals, expirations, and services to the two processes and show they maintain the invariant that each agent has at least as many high-valued jobs in process $A$ as in process $B$.

\paragraph{Arrivals} Since arrivals are the same in both processes, the invariant is maintained.

\paragraph{Expirations} By the coupling rule, at update $(t,i)$, agent $i$ has at most $\theta_i^A[t,i-1] - \theta_i^B[t,i-1]$ more high-valued job expirations in process $A$ than process $B$, maintaining the invariant.

\paragraph{Services} If the same agent in both processes is served, the invariant is maintained. If no agent in process $B$ is served, then no agent has a high-valued job in process $B$, so the invariant is maintained. Otherwise, suppose agent $i$ is served in process $A$ and agent $k\neq i $ is served in process $B$. Denote the permutation used to determine service by $\pi$. Since $k$ was served in process $B$, $k$ had a non-zero number of high-valued jobs before service in process $B$. By the invariant, $k$ had a non-zero number of high-valued jobs before service in process $A$. Since $i$ was served instead of $k$ in process $A$, it must have been that $\pi(i) < \pi(k)$. Despite this, $k$ was served instead of $i$ in process $B$, so it must have been that $i$ had $0$ high-valued jobs in process $B$. Therefore, the service maintains the invariant.

For $t\geq t_0$, let $\Delta_t = \|\theta^A[t] - \theta^B[t]\|_1 := \sum_{i=1}^n (\theta_i^A[t] - \theta_i^B[t])$, the difference in the number of high-valued jobs across all agents in processes $A$ and $B$ right before service at time $t$. Let us analyze how $\Delta_t$ evolves over time for $t\geq t_0$. Arrivals do not change $\Delta_t$ since arrivals are the same for both processes $A$ and $B$. Services can only decrease $\Delta_t$ in that $\|\theta^A[t,n+1] - \theta^B[t,n+1]\|_1\leq \Delta_t$ because if a high-valued job in process $B$ gets served, then one gets served in $A$ too. By the way the expirations occur in the coupling, the difference in the number of high-valued job expirations between times $t$ and $t+1$ between processes $A$ and $B$ is distributed as $\Binom(\|\theta^A[t,n+1] - \theta^B[t,n+1]\|_1, \delta)$ conditioned on $\mathcal G_t$ where $\mathcal G_t$ denotes the history up to time $t$. This implies
\[
    \E[\Delta_{t+1}|\mathcal G_t] \leq (1-\delta)\|\theta^A[t,n+1] - \theta^B[t,n+1]\|_1 \leq (1-\delta)\Delta_t.
\]
Thus, $\E[\Delta_{t+1}|\Delta_t]\leq (1-\delta)\Delta_t$. By induction, $\E[\Delta_{t_0+s}|\Delta_{t_0}]\leq (1-\delta)^s \Delta_{t_0}$. For $s\leq T-t_0$, the event $\sigma > t_0+s$ is precisely the event that $\Delta_{t_0+s}\geq 1$, so
\[
    \Pr(\sigma > t_0+s|\Delta_{t_0}) = \Pr(\Delta_{t_0+s}\geq 1|\Delta_{t_0}) \leq \E[\Delta_{t_0+s}|\Delta_{t_0}] \leq (1-\delta)^s \Delta_{t_0}.
\]
Summing,
\begin{align*}
    \E[\sigma|\Delta_{t_0}] - t_0 & = \sum_{s=0}^{T-t_0} \Pr(\sigma > t_0+s|\Delta_{t_0})\leq \sum_{s=0}^\infty \min\{1,(1-\delta)^s \Delta_{t_0}\}.
\end{align*}
If $\Delta_{t_0}=0$, the above is equal to $0$. For $\delta=1$, the above is equal to $1$.

 For $\Delta_{t_0}>0$ and $\delta < 1$, the above is at most
\begin{align*}
    \left\lceil\frac{\log(\Delta_{t_0})}{-\log(1-\delta)}\right\rceil + & \sum_{s=\lceil \log(\Delta_{t_0})/-\log(1-\delta)\rceil}^\infty (1-\delta)^s\Delta_{t_0}\\
    & = \left\lceil\frac{\log(\Delta_{t_0})}{-\log(1-\delta)}\right\rceil + \frac{(1-\delta)^{\lceil \log(\Delta_{t_0})/-\log(1-\delta)\rceil}}{\delta}\cdot\Delta_{t_0}\\
    & \leq \left\lceil\frac{\log(\Delta_{t_0})}{-\log(1-\delta)}\right\rceil + \frac1\delta.
\end{align*}
Since $\theta^A[t_0]$ and $\theta^B[t_0]$ only differ in agent $j$, $\Delta_{t_0}\leq \theta_j^A[t_0] - \theta_j^B[t_0] \leq T+1$, where the last inequality comes from the fact that we never allow agents to report a value greater than $T$ in \cref{alg:gum_no_computation}, which implies that both $\theta_j^A[t_0]$ and $\theta_j^B[t_0]$ are at most $T+1$. The expression in the above display is then at most the expression in the lemma statement.
\end{proof}

\section{Efficiently Computing the Externalities} \label{sec:computation}
It is not clear how to efficiently implement \cref{alg:gum_no_computation} because we have not given an algorithm to compute the externalities $\gamma_{i\to j}[t]$. In this section, we first give a mechanism that efficiently computes estimators $\hat\gamma_{i\to j}[t]$ for all $j\neq i$ in time $O(n\log T)$ at update $(t,i)$ such that when we use the estimators $\hat\gamma_{i\to j}[t]$ instead of the actual externalities $\gamma_{i\to j}[t]$ in \cref{alg:gum_no_computation}, we obtain the same guarantees as in \cref{thm:error_no_computation}. We then show how to improve the algorithm's runtime by modifying the mechanism slightly.

\subsection{Computational Model} \label{ssec:computational_model}
In this paper, we assume a standard computational model (e.g., Word RAM with randomization) augmented with a sampling oracle that can draw from the distribution $\Binom(\theta,\delta)$ for $1\leq \theta\leq 2T$ in $O(1)$ time.

\subsection{Basic Algorithm}
The proof of \cref{lem:as_externality_bound} already gives an algorithm to compute an unbiased estimator of $\gamma_{j\to i}[t_0]$: we couple two simulated instances of the high-valued jobs stochastic process starting at update $(t_0,j)$ until the two instances sync, where one of the processes starts with a random state and the other starts with a state with the report $\hat\theta_j[t_0]$, and we output the difference in the number of high-valued jobs that agent $i$ serves between the two processes as the estimator $\hat\gamma_{j\to i}[t_0]$. This estimator is unbiased by the definition of externality. By the proof of \cref{lem:as_externality_bound}, the two coupled processes will sync in expected $O(\log T)$ timesteps. Since simulating each timestep in each process takes $O(n)$ computational time since there are $n$ updates per timestep, computing the estimator in this fashion takes $O(n\log T)$ expected computational time.

The proof of \cref{thm:error_no_computation} would work if we used an unbiased estimator $\hat\gamma_{i\to j}[t]$ at each update instead of the actual externalities $\gamma_{i\to j}[t]$ as long as the martingale $\hat X_{i\to j}[t] := -\sum_{s=1}^t\hat \gamma_{i\to j}[s]$ concentrates around $0$ for an honest agent $i$. The only problem with the algorithm described above to compute an estimator $\hat\gamma_{i\to j}[t]$ is that $|\hat\gamma_{i\to j}[t]|$ need not be almost surely bounded by a nontrivial constant since the two coupled instances need not sync until the end of time, which makes proving the martingale concentration difficult. To fix this, we modify the algorithm to stop the simulation after $L$ timesteps. Stopping before the two simulations sync introduces some bias into the estimator, but we shall show that we can make $L$ sufficiently large to reduce the bias to still obtain the guarantees of \cref{thm:error_no_computation}. Pseudocode for this algorithm is given in \cref{alg:compute_estimator}.

\begin{algorithm}
\caption{Compute Estimated Externalities via Coupled Samples} \label{alg:compute_estimator}
\KwIn{update $(t_0,j)$, report $\hat\theta_j[t_0]$, reported past states $\hat\theta[t_0,j-1]$, simulation limit $L$}
Simulate two coupled high-valued jobs processes $(\theta^{\text{report}}[(t,i)])$ and $(\theta^{\text{random}}[(t,i)])$ as described in the proof of \cref{lem:as_externality_bound} through timestep $\min\{\sigma, t_0+L-1,T\}$ where $\sigma = \inf\{t\geq t_0:\theta^{\text{report}}[t] = \theta^{\text{random}}[t]\}$\;
For each $i\neq j$, set $\hat\gamma_{j\to i}[t_0]$ to be the number of high-valued jobs from agent $i$ served in simulation $(\theta^{\text{report}}[(t,i)])$ minus the number of high-valued jobs from agent $i$ served in simulation $(\theta^{\text{random}}[(t,i)])$ during the timesteps that have been simulated\;
\KwOut{estimated externalities $\hat\gamma_{j\to i}[t_0]$ for $i\neq j$}
\end{algorithm}
Next, we define the \textit{predictable bias} of $\hat\gamma_{j\to i}[t_0]$ as
\[
    \beta_{j\to i}[t_0] := \E\left[\hat\gamma_{j\to i}[t_0]\middle|\hat\theta[t_0,j-1],\hat\theta[t_0,j]\right] - \gamma_{j\to i}[t_0].
\]

\begin{restatable}{lemma}{lemBiasBoundSimple} \label{lem:bias_bound_simple}
Given $\epsilon > 0$, by running \cref{alg:compute_estimator} with
\[
    L \geq \begin{cases}\frac{\log \frac{T+1}{\epsilon\delta}}{-\log(1-\delta)} & \text{if $\delta<1$}\\1 & \text{if $\delta=1$}\end{cases},
\]
the predictable bias $\beta_{j\to i}[t_0]$ is at most $\epsilon$.
\end{restatable}
\begin{proof}
As in the proof of \cref{lem:as_externality_bound}, if we define $\Delta_{t_0} = \|\theta^{\text{report}}[t_0] - \theta^{\text{random}}[t_0]\|_1$, then $\Pr(\sigma>t_0+s|\Delta_{t_0}) \leq (1-\delta)^s\Delta_{t_0}$ and $\Delta_{t_0}\leq T+1$ almost surely, yielding the bound $\Pr(\sigma>t_0+s)\leq (1-\delta)^s(T+1)$. By definition, the predictable bias $\beta_{j\to i}[t_0]$ is the expected difference between the output of \cref{alg:compute_estimator} when run with a simulation limit and the output when run without a simulation limit. This difference is at most $(\sigma-(t_0 + L))^+$. Using the tail sum formula,
\[
    \E[(\sigma-(t_0 + L))^+] \leq \sum_{s=L}^\infty \Pr(\sigma > t_0 + s) \leq (T + 1)\sum_{s=L}^\infty (1-\delta)^s = (T+1)\cdot\frac{(1-\delta)^L}{\delta}.
\]
For $L$ as chosen in the lemma statement, this is at most $\epsilon$.
\end{proof}

Replacing the exact externalities in \pec with the estimates computed by \cref{alg:compute_estimator} gives a basic version of \spec. We first prove its utility and runtime guarantees, then add report caps to obtain the improved runtimes. We prove the following guarantee for \spec.
\begin{restatable}{theorem}{thmBasicMechanismWithComputation} \label{thm:basic_mechanism_with_computation}
There exist a simulation limit and a virtual debt limit
\[
    L = \begin{cases}\Theta(\log T) & \text{if $\delta<1$},\\1 & \text{if $\delta=1$},\end{cases}
    \qquad B = \tilde\Theta\!\left(\sqrt{\frac{T}{n}}\right)
\]
such that \cref{alg:gum_no_computation}, using estimators $\hat\gamma_{i\to j}[t]$ computed by \cref{alg:compute_estimator} with simulation limit $L$ instead of the externalities $\gamma_{i\to j}[t]$, has error $\tilde O(\sqrt{nT})$.

Furthermore, the mechanism takes worst-case $O(n^2\log T)$ computational time per timestep $t$.
\end{restatable}
\begin{proof}[Proof Sketch]
First, let us prove the runtime bound. Our choice of $L$ satisfies $L\leq O(\log T)$, so each simulation takes at most $O(\log T)$ timesteps. Each timestep of a simulation takes $O(n)$ time since we must update $O(n)$ queues. Each timestep of the mechanism runs $O(n)$ coupled simulations, one for each agent's report. Thus, a timestep of the mechanism takes $O(n^2\log T)$ time.

For the error bound, the idea is to use \cref{lem:bias_bound_simple} with our choice of $\epsilon = 1/T$. With a predictable bias so small, one can modify the proof of \cref{thm:error_no_computation} to obtain the same error bound. In more detail, fix an honest agent $i$. Conditioned on the reported history up to an update $(t,i-1)$, the estimated externality $\hat\gamma_{i\to j}[t]$ has expectation $0$ for every other agent $j$ because both simulations in the execution of \cref{alg:compute_estimator} are identically distributed since the report $\hat\theta_i[t]$ and the simulated random state $\theta_i^{\text{random}}[t]$ are identically distributed conditioned on the reported history up to an update $(t,i-1)$. We can then use a similar martingale concentration argument and virtual debt limit $B$ as in \cref{thm:error_no_computation} to show that agent $i$ does not exceed any virtual debt limit with high probability. The virtual debt limit $B$ in \cref{thm:error_no_computation} represented a budget on how much of an externality one agent is allowed on another; now, it limits the cumulative estimated externality that one agent may impose on another. Since the estimated externality is off from the true externality by at most the predictable bias in expectation each time period, which is at most $\epsilon$, the amount of externality one agent can impose on another is bounded by about $B + \epsilon T$. But since $\epsilon=1/T$, the extra $\epsilon T$ term has no significant effect.
\end{proof}
The full proof is given in \cref{sec:deferred_proofs_computation}.

\subsection{Improved Algorithm}

By the proof of \cref{lem:as_externality_bound}, the expected number of timesteps until the two coupled simulations in \cref{alg:compute_estimator} at update $(t_0,i)$ sync conditioned on the initial difference $\Delta_{t_0}$ in the number of high-valued jobs agent $i$ has after update $(t_0,i)$ between the two simulations is at most $\lceil \log(\Delta_{t_0}) / -\log(1-\delta)\rceil + 1/\delta$ for $\delta < 1$ and $\Delta_{t_0}\geq 1$ (and is $0$ if $\Delta_{t_0}=0)$. In the proof of \cref{lem:as_externality_bound}, we bounded $\Delta_{t_0}\leq T+1$, giving us the $O(\log T)$ bound on the expected number of timesteps that we need to simulate. If agent $i$ is honest, this is a very pessimistic bound: the bound $\Delta_{t_0} \leq \theta_{i}[t_0-1]+1$, which happens since the number of high-valued jobs agent $i$ can have can only increase by $1$ per timestep, is a much better bound. One can show that with probability at least $1-1/T$, $\theta_{i}[t]$ is bounded by $O(\sqrt{\log T})$ (\cref{lem:honest_agent_exceed_per_round_report_cap_main_body}). Thus, if agent $i$ is honest, the expected number of timesteps that \cref{alg:compute_estimator} runs is $O(\log\log T)$. Furthermore, one can show that with probability at least $1-1/T$, $\sum_{t=1}^T \theta_{i}[t]$ is bounded by $O(T)$ (\cref{lem:occupation_bound_main_body}).
Then, the total expected number of timesteps that \cref{alg:compute_estimator} runs simulations on updates $(t,i)$ across time $t$ is bounded by $O(T)$. In this sense, the amortized expected number of timesteps spent on simulations to compute externalities $\gamma_{i\to j}[t]$ is $O(1)$ per timestep amortized across timesteps $t$.

The above analysis works if agent $i$ is honest. We can extend it to work under arbitrary behavior of agents by enforcing that agents' reports are not too large in the mechanism. Specifically, to obtain the expected number of simulated timesteps $O(\log\log T)$, we enforce that $\hat\theta_{i}[t]\leq\overline\theta$ for some $\overline\theta = \Theta(\sqrt{\log T})$, and to obtain the expected amortized number of simulated timesteps $O(1)$, we require that $\sum_{s=1}^t\hat\theta_{i}[s]\leq \overline\Sigma$ at each timestep $t$ for a suitable $\overline\Sigma=\tilde\Theta(T/n)$. For a careful choice of $\overline\theta$ and $\overline\Sigma$, the reports $\hat\theta_{i}[t]$ are enforced to be not too high so that we obtain improved runtime guarantees and that honest agents are not affected by the report capping with high probability so we obtain a similar error bound. We give pseudocode for the mechanism with this report capping using estimates $\hat\gamma_{j\to i}[t]$ computed by \cref{alg:compute_estimator} in \cref{alg:gum_report_capping}.

\begin{algorithm}[H]
\caption{Sampled Pairwise Externality Compensator (\spec) with Report Capping and Estimated Externalities}
\label{alg:gum_report_capping}
\KwIn{virtual debt limit $B$, per-round report cap $\overline\theta$, cumulative report cap $\overline\Sigma$, simulation limit $L$}
Initialize $\hat\theta[0,n+1]=0$\;
\For{$t=1,2,\dots,T$}{
    \For{$i=1,2,\dots, n$}{
        \eIf{$\sum_{s=1}^{t-1}p_{i\to j}[s] > B$ for some $j\neq i$}{
            Set $\hat \theta_i[t]=0$\;
        }{
            Elicit a report $\hat\theta_i[t]\in\{0,1,\dots,T\}$ from agent $i$\;
            Enforce report caps: $\hat\theta_i[t]\gets\min\left\{\hat\theta_i[t], \overline\theta, \overline\Sigma-\sum_{s=1}^{t-1}\hat\theta_i[s]\right\}$\;
        }
        Set $\hat\theta_i[t,i] = \hat\theta_i[t]$ and $\hat\theta_j[t,i]=\hat\theta_j[t,i-1]$ for $j\neq i$.\;
        Agent $i$ makes virtual transfers $p_{i\to j}[t]= -\hat\gamma_{i\to j}[t]$ where $\hat\gamma_{i\to j}[t]$ is computed by \cref{alg:compute_estimator} at update $(t,i)$ with report $\hat\theta_i[t]$, reported past states $\hat\theta[t,i-1]$, and simulation limit $L$\;
    }
    \eIf{$\hat\theta_i[t]=0$ for all $i$}{
       Let the server idle\;
       Set $\hat\theta[t,n+1]=\hat\theta[t,n]$\;
    }{
        Give agent $i$ access to the server where $i$ is chosen uniformly at random from $\{i:\hat\theta_i[t]>0\}$\;
        Set $\hat\theta_i[t,n+1] = \hat\theta_i[t]-1$ and $\hat\theta_j[t,n+1]=\hat\theta_j[t]$ for $j\neq i$\;
    }
}
\end{algorithm}

Our main results are listed formally below and are summarized in \cref{tab:mechanism_summary}. One of them uses a simulation limit $L=\Theta(\log T)$ with a worst-case per-timestep runtime of $O(n^2\log T)$ and an error bound of $\tilde O(\sqrt{nT})$, while the other uses a simulation limit $L=\Theta(\log\log T)$ and obtains a worst-case per-timestep runtime of $O(n^2\log\log T)$ and an error bound of $O(nT/\log T)$. Both results achieve an expected per-timestep runtime $O(n^2\log\log T)$ and an expected amortized runtime of $O(n^2)$. 
Adapting the proof of \cref{lem:bias_bound}, a simulation limit $L=\Omega\big(\log\big((\overline\theta+1)/\epsilon\big)\big)$ makes the predictable bias at most $\epsilon$. For fixed $0<\delta<1$ and $\overline\theta=\Theta(\sqrt{\log T})$, choosing $\epsilon=1/T$ permits $L=\Theta(\log T)$ and choosing $\epsilon=1/\log T$ permits $L=\Theta(\log\log T)$. The bias can accumulate over $T$ periods, contributing $O(n\epsilon T)$ to the error. We use $\epsilon=1/T$ to obtain \cref{thm:main_result_one} and $\epsilon=1/\ln T$ to obtain \cref{thm:main_result_two}.

\begin{restatable}[Main Result with $L=\Theta(\log T)$]{theorem}{thmMainResultOne} \label{thm:main_result_one}
There exist parameters
\begin{align*}
    \overline\theta &= \begin{cases}\Theta(\sqrt{\log T}) & \text{if $\delta<1$},\\1 & \text{if $\delta=1$},\end{cases}
    & L &= \begin{cases}\Theta(\log T) & \text{if $\delta<1$},\\1 & \text{if $\delta=1$},\end{cases}\\
    \overline\Sigma &= \tilde\Theta\!\left(\frac{T}{n}\right),
    & B &= \tilde\Theta\!\left(\sqrt{\frac{T}{n}}\right)
\end{align*}
such that \cref{alg:gum_report_capping} has error at most $\tilde O(\sqrt{nT})$, worst-case $O(n^2\log T)$ and expected $O(n^2\log\log T)$ per-step runtime, and expected total runtime spent across all time periods at most $O(n^2T)$.
\end{restatable}

\begin{restatable}[Main Result with $L=\Theta(\log\log T)$]{theorem}{thmMainResultTwo} \label{thm:main_result_two}
There exist parameters
\begin{align*}
    \overline\theta &= \begin{cases}\Theta(\sqrt{\log T}) & \text{if $\delta<1$},\\1 & \text{if $\delta=1$},\end{cases}
    & L &= \begin{cases}\Theta(\log\log T) & \text{if $\delta<1$},\\1 & \text{if $\delta=1$},\end{cases}\\
    \overline\Sigma &= \tilde\Theta\!\left(\frac{T}{n}\right),
    & B &= \tilde\Theta\!\left(\sqrt{\frac{T}{n}}\right)
\end{align*}
such that \cref{alg:gum_report_capping} has error at most $O(nT/\log T)$, worst-case $O(n^2\log\log T)$ per-step runtime, and expected total runtime spent across all time periods at most $O(n^2T)$.
\end{restatable}
\begin{table}
\centering
\begin{tabular}{@{}lcc@{}}
\hline
\textbf{$L$} & $\Theta(\log T)$ & $\Theta(\log\log T)$ \\
\hline
\textbf{Error} & $\tilde O(\sqrt{nT})$ & $O(nT/\log T)$ \\
\textbf{Worst-case} & $O(n^2\log T)$ & $O(n^2\log\log T)$ \\
\textbf{Expected} & $O(n^2\log\log T)$ & $O(n^2\log\log T)$ \\
\textbf{Expected amortized} & $O(n^2)$ & $O(n^2)$ \\
\hline
\end{tabular}
\caption{Guarantees of \cref{thm:main_result_one,thm:main_result_two}. The bottom three rows show the per-step runtime of the algorithm.}
\label{tab:mechanism_summary}
\end{table}
Both of these theorems are proved in \cref{sec:deferred_proofs_computation}. The only parts of the proofs that are significantly different than the proof of \cref{thm:basic_mechanism_with_computation} are the parts where we must prove that honest agents do not hit either the per-round or cumulative report cap with high probability. We give sketches for that part of the argument below.

\begin{lemma} \label{lem:honest_agent_exceed_per_round_report_cap_main_body}
Fix an agent $i$ and $\eta\in(0,1]$. If we set
\[
    \overline\theta \geq \begin{cases}\min\left\{T,\sqrt{\frac{2}{-\ln(1-\delta)}\left(\frac2\delta+\ln\frac T\eta\right)}+1\right\} & \text{if $\delta<1$}\\1 & \text{if $\delta=1$}\end{cases},
\]
then $\Pr(\exists t: \theta_i[t] > \overline\theta)\leq \eta$.
\end{lemma}
\begin{proof}
If $\delta =1$, then an agent's queue only contains $1$ job each time period, so the lemma statement holds because $\overline\theta\geq 1$. Otherwise, assume $\delta < 1$.

If $\overline\theta\geq T$, the claim holds deterministically because at most $T$ jobs arrive to an agent. Otherwise, assume $\overline\theta < T$. Fix a time $t$. Let $X_s\sim\Bern\left(\frac1n\right)$ be the indicator that a high-valued job arrived to agent $i$'s queue in timestep $t-s$, and let $Y_s\sim\Bern((1-\delta)^s)$ be the indicator that the job arriving at timestep $t-s$ would survive to timestep $t$ if not served. Set $Z_s = X_sY_s \sim\Bern\left(\frac1n\cdot(1-\delta)^s\right)$. Then, $\theta_i[t]\leq \sum_{s=0}^{t-1} Z_s$ almost surely. We apply a Chernoff bound to $\sum_{s=0}^{t-1}Z_s$. Specifically, apply \cref{lem:poisson_binomial_chernoff} with $p=1/n$, $q=1-\delta$, and $m=\overline\theta$ to obtain
\[
    \Pr(\theta_i[t] > \overline\theta) \leq e^{2/\delta}\left(\frac1n\right)^{\overline\theta}(1-\delta)^{\overline\theta(\overline\theta-1)/2}.
\]
The right-hand side can be bounded as
\[
    e^{2/\delta}\left(\frac1n\right)^{\overline\theta}(1-\delta)^{\overline\theta(\overline\theta-1)/2}\leq e^{2/\delta}(1-\delta)^{(\overline\theta-1)^2/2}\leq\frac\eta T
\]
for our choice of $\overline\theta$. By union bounding over all $T$ timesteps $t$, we obtain the lemma statement:
\[
    \Pr(\exists t:\theta_i[t]>\overline\theta)\leq\eta.
\]
\end{proof}

\begin{lemma} \label{lem:occupation_bound_main_body}
For any agent $i$, for any $x>0$, with probability at least $1-e^{-x}$,
\[
    \delta\sum_{t=1}^T \theta_i[t]\leq \frac Tn + \sqrt{\frac{32Tx}{n}} + 2x.
\]
\end{lemma}
\begin{proof}
Let $J_t\sim\Bern(1/n)$ be the indicator that a high-valued job arrives to agent $i$'s queue in timestep $t$, and let $G_t\sim\Geom(\delta)$ be the number of time periods that job would survive if not served and if we extended the time horizon to infinity. By the counting argument that $\sum_{t=1}^T\theta_i[t]$ is at most the total potential lifetime of high-valued jobs in agent $i$'s queue,
\[
    \sum_{t=1}^T\theta_i[t]\leq \sum_{t=1}^TJ_tG_t.
\]
The result follows from applying a Bernstein bound to the above sum; specifically, we apply \cref{lem:thinned_geometric_bernstein} with $H_s = \delta G_s$, $Z_s = J_sH_s$, and $p=1/n$.
\end{proof}

\printbibliography

@inproceedings{csoka2026guaranteed,
  title={Guaranteed Utility Equilibrium},
  author={Cs{\'o}ka, Endre and Pongr{\'a}cz, Andr{\'a}s and Rodivilov, Alexander},
  booktitle={Proceedings of the 27th ACM Conference on Economics and Computation},
  year={2026},
  note={To appear}
}

@article{safronov2018coalition,
  title={Coalition-proof full efficient implementation},
  author={Safronov, Mikhail},
  journal={Journal of Economic Theory},
  volume={177},
  pages={659--677},
  year={2018},
  publisher={Elsevier}
}

@article{csoka2025prior,
  title={Prior-free Collusion-proof Dynamic Mechanisms},
  author={Cs{\'o}ka, Endre},
  journal={arXiv preprint arXiv:2511.15727},
  year={2025}
}

@article{csoka2023collusion,
  title={A collusion-proof efficient dynamic mechanism},
  author={Cs{\'o}ka, Endre and Liu, Heng and Rodivilov, Alexander and Teytelboym, Alexander},
  journal={Available at SSRN 4419623},
  year={2023}
}

@article{gorokh2021monetary,
  title={From monetary to nonmonetary mechanism design via artificial currencies},
  author={Gorokh, Artur and Banerjee, Siddhartha and Iyer, Krishnamurthy},
  journal={Mathematics of Operations Research},
  volume={46},
  number={3},
  pages={835--855},
  year={2021},
  publisher={INFORMS}
}

@inproceedings{gorokh2021remarkable,
  title={The Remarkable Robustness of the Repeated Fisher Market},
  author={Gorokh, Artur and Banerjee, Siddhartha and Iyer, Krishnamurthy},
  booktitle={Proceedings of the 22nd ACM Conference on Economics and Computation},
  pages={562--562},
  year={2021}
}

@inproceedings{banerjee2023robust,
  title={Robust Pseudo-Markets for Reusable Public Resources},
  author={Banerjee, Siddhartha and Fikioris, Giannis and Tardos, Eva},
  booktitle={Proceedings of the 24th ACM Conference on Economics and Computation},
  pages={241--241},
  year={2023}
}

@article{onyeze2025allocating,
  title={Allocating public goods via dynamic max-min fairness: Long-run behavior and competitive equilibria},
  author={Onyeze, Chido and Banerjee, Siddhartha and Fikioris, Giannis and Tardos, {\'E}va},
  journal={Proceedings of the ACM on Measurement and Analysis of Computing Systems},
  volume={9},
  number={1},
  pages={1--45},
  year={2025},
  publisher={ACM New York, NY, USA}
}

@inproceedings{lin2025online,
  title={Online resource sharing: better robust guarantees via randomized strategies},
  author={Lin, David X and Hall, Daniel and Fikioris, Giannis and Banerjee, Siddhartha and Tardos, {\'E}va},
  booktitle={Proceedings of the Thirty-Fourth International Joint Conference on Artificial Intelligence},
  pages={179--186},
  year={2025}
}

@inproceedings{fikioris2025beyond,
author = {Fikioris, Giannis and Banerjee, Siddhartha and Tardos, Eva},
title = {Beyond Worst-Case Online Allocation via Dynamic Max-min Fairness},
year = {2025},
booktitle = {Proceedings of the 26th ACM Conference on Economics and Computation},
pages = {92},
}

@inproceedings{lin2026robust,
  title={Robust equilibria in shared resource allocation via strengthening border’s theorem},
  author={Lin, David X and Banerjee, Siddhartha and Fikioris, Giannis and Tardos, {\'E}va},
  booktitle={Proceedings of the 2026 Annual ACM-SIAM Symposium on Discrete Algorithms (SODA)},
  pages={6571--6612},
  year={2026},
  organization={SIAM}
}

@InProceedings{lin2026itcs,
  author =	{Lin, David X. and Fikioris, Giannis and Banerjee, Siddhartha and Tardos, \'{E}va},
  title =	{{Robust Resource Allocation via Competitive Subsidies}},
  booktitle =	{17th Innovations in Theoretical Computer Science Conference (ITCS 2026)},
  pages =	{96:1--96:15},
  series =	{Leibniz International Proceedings in Informatics (LIPIcs)},
  ISBN =	{978-3-95977-410-9},
  ISSN =	{1868-8969},
  year =	{2026},
  volume =	{362},
  editor =	{Saraf, Shubhangi},
  publisher =	{Schloss Dagstuhl -- Leibniz-Zentrum f{\"u}r Informatik},
  address =	{Dagstuhl, Germany},
  URL =		{https://drops.dagstuhl.de/entities/document/10.4230/LIPIcs.ITCS.2026.96},
  URN =		{urn:nbn:de:0030-drops-253835},
  doi =		{10.4230/LIPIcs.ITCS.2026.96}
}

@article{fan2015exponential,
  title={Exponential inequalities for martingales with applications},
  author={Fan, Xiequan and Grama, Ion and Liu, Quansheng},
  journal={Electron. J. Probab.},
  volume={20},
  pages={1-22}, 
  year={2015}
}

@article{freedman1975tail,
  title={On tail probabilities for martingales},
  author={Freedman, David A},
  journal={the Annals of Probability},
  pages={100--118},
  year={1975},
  publisher={JSTOR}
}

@article{csoka20261,
  title={A 1.283 Price-of-Anarchy Bound for the Repeated Virtual First-Price Auction},
  author={Cs{\'o}ka, Endre},
  journal={arXiv preprint arXiv:2609.05499},
  year={2026}
}

@inproceedings{guo2009competitive,
  title={Competitive repeated allocation without payments},
  author={Guo, Mingyu and Conitzer, Vincent and Reeves, Daniel M},
  booktitle={International Workshop on Internet and Network Economics},
  pages={244--255},
  year={2009},
  organization={Springer}
}

@inproceedings{cavallo2014incentive,
  title={Incentive compatible two-tiered resource allocation without money},
  author={Cavallo, Ruggiero},
  booktitle={Proceedings of the 2014 international conference on Autonomous agents and multi-agent systems},
  pages={1313--1320},
  year={2014}
}

@article{jackson,
  title     = {Overcoming incentive constraints by linking decisions},
  author    = {Jackson, Matthew O and Sonnenschein, Hugo F},
  journal   = {Econometrica},
  volume    = {75},
  number    = {1},
  pages     = {241--257},
  year      = {2007},
  publisher = {Wiley Online Library}
}

@article{jagadeesan2026limits,
  title={The Limits of Auctions under Ex-Ante Collusion},
  author={Jagadeesan, Ravi and Segal, Ilya R and Tordjman, Alex},
  journal={Available at SSRN},
  year={2026},
  doi={https://doi.org/10.2139/ssrn.6993038}
}

@book{10.1093/acprof:oso/9780199535255.001.0001,
    author = {Boucheron, Stéphane and Lugosi, Gábor and Massart, Pascal},
    title = {Concentration Inequalities: A Nonasymptotic Theory of Independence},
    publisher = {Oxford University Press},
    year = {2013},
    month = {02},
    isbn = {9780199535255},
    doi = {10.1093/acprof:oso/9780199535255.001.0001},
    url = {https://doi.org/10.1093/acprof:oso/9780199535255.001.0001},
}

@article{charalambides2005moments,
  title={Moments of a class of discrete q-distributions},
  author={Charalambides, Ch A},
  journal={Journal of statistical planning and inference},
  volume={135},
  number={1},
  pages={64--76},
  year={2005},
  publisher={Elsevier}
}

@inproceedings{amanihamedani2025adaptive,
  title={Adaptive approximation schemes for matching queues},
  author={AmaniHamedani, Alireza and Aouad, Ali and Saberi, Amin},
  booktitle={Proceedings of the 57th Annual ACM Symposium on Theory of Computing},
  pages={1454--1464},
  year={2025}
}

@article{gaitonde2023price,
  title={The price of anarchy of strategic queuing systems},
  author={Gaitonde, Jason and Tardos, {\'E}va},
  journal={Journal of the ACM},
  volume={70},
  number={3},
  pages={1--63},
  year={2023},
  publisher={ACM New York, NY}
}

@inproceedings{fu2022stability,
  title={Stability of decentralized queueing networks beyond complete bipartite cases},
  author={Fu, Hu and Hu, Qun and Lin, Jia’nan},
  booktitle={International Conference on Web and Internet Economics},
  pages={96--114},
  year={2022},
  organization={Springer}
}

@inproceedings{abel2025learning,
  title={Learning in strategic queuing systems with small buffers},
  author={Abel, Ariana and Kolumbus, Yoav and Duque, Jeronimo Martin and {Palma Foster}, Cristian and Tardos, Eva},
  booktitle={Proceedings of the 27th ACM Conference on Economics and Computation},
  year={2026},
  note={To appear}
}

@article{baudin2023strategic,
  title={Strategic behavior and no-regret learning in queueing systems},
  author={Baudin, Lucas and Scarsini, Marco and Venel, Xavier},
  journal={arXiv preprint arXiv:2302.03614},
  year={2023}
}

@article{sentenac2021decentralized,
  title={Decentralized learning in online queuing systems},
  author={Sentenac, Flore and Boursier, Etienne and Perchet, Vianney},
  journal={Advances in Neural Information Processing Systems},
  volume={34},
  pages={18501--18512},
  year={2021}
}

@article{freund2024efficient,
  title={Efficient decentralized multi-agent learning in asymmetric bipartite queueing systems},
  author={Freund, Daniel and Lykouris, Thodoris and Weng, Wentao},
  journal={Operations Research},
  volume={72},
  number={3},
  pages={1049--1070},
  year={2024},
  publisher={INFORMS}
}

@inproceedings{sivan2026quota,
  title={Quota Marketplace: Dynamic Pricing for Efficient Allocation of ML Training Resources},
  author={Sivan, Balasubramanian and Leme, Renato Paes and Tiuca, Mihai and McFarlane, Ian and Gkatzelis, Vasilis and Mehta, Nehal and Yeganeh, Soheil Hassas and Mirrokni, Vahab and Vahdat, Amin},
  booktitle={20th USENIX Symposium on Operating Systems Design and Implementation (OSDI 26)},
  pages={1223--1241},
  year={2026}
}

@inproceedings{DBLP:conf/sigmetrics/SalemIN23,
  author       = {Tareq Si Salem and
                  Georgios Iosifidis and
                  Giovanni Neglia},
  editor       = {Evgenia Smirni and
                  Konstantin Avrachenkov and
                  Phillipa Gill and
                  Bhuvan Urgaonkar},
  title        = {Enabling Long-term Fairness in Dynamic Resource Allocation},
  booktitle    = {Abstract Proceedings of the 2023 {ACM} {SIGMETRICS} International
                  Conference on Measurement and Modeling of Computer Systems, {SIGMETRICS}
                  2023, Orlando, FL, USA, June 19-23, 2023},
  pages        = {31--32},
  publisher    = {{ACM}},
  year         = {2023},
  url          = {https://doi.org/10.1145/3578338.3593541},
  doi          = {10.1145/3578338.3593541},
  bibsource    = {dblp computer science bibliography, https://dblp.org}
}

@article{DBLP:journals/ior/KrishnasamySJS21,
  author       = {Subhashini Krishnasamy and
                  Rajat Sen and
                  Ramesh Johari and
                  Sanjay Shakkottai},
  title        = {Learning Unknown Service Rates in Queues: {A} Multiarmed Bandit Approach},
  journal      = {Oper. Res.},
  volume       = {69},
  number       = {1},
  pages        = {315--330},
  year         = {2021},
  url          = {https://doi.org/10.1287/opre.2020.1995},
  doi          = {10.1287/opre.2020.1995},
  bibsource    = {dblp computer science bibliography, https://dblp.org}
}

@inproceedings{DBLP:conf/nips/KrishnasamySJS16,
  author    = {Subhashini Krishnasamy and
               Rajat Sen and
               Ramesh Johari and
               Sanjay Shakkottai},
  title     = {Regret of Queueing Bandits},
  booktitle = {Advances in Neural Information Processing Systems},
  volume = {29},
  pages     = {1669--1677},
  year      = {2016},
  publisher = {Curran Associates, Inc.},
}

\clearpage
\appendix
\crefalias{section}{appendix}
\section{Review of the Guaranteed Utility Mechanism (GUM)} 
\label{appsec:gum_review}

\subsection{GUM for Single Item Allocation with Transferable Utilities and Known Priors}
In this section, we review the mechanism of \cite{safronov2018coalition} or \cite{csoka2023collusion} in a simplified setting, which following \cite{csoka2023collusion}, we call the Guaranteed Utility Mechanism (GUM). This setting is simpler than the one in which we use GUM, but it illustrates the important ideas.

We consider a setting of allocating a single item to an agent among $n$ agents with transferable utilities. Each agent $i$ has a private value $V_i$ drawn from some distribution $\mathcal D_i$, independently across agents with values bounded in $[0,1]$. We assume utilities are transferable and that each agent $i$ can make a transfer $p_{i\to j}$ to another agent $j$. With such transfers, agent $i$ receives utility
\[
    U_i = \begin{cases}V_i - \sum_{j\neq i}p_{i\to j} + \sum_{j\neq i}p_{j\to i} & \text{if agent $i$ wins the item}\\- \sum_{j\neq i}p_{i\to j} + \sum_{j\neq i}p_{j\to i} & \text{otherwise}\end{cases}.
\]

The Guaranteed Utility Mechanism works as follows. It is parameterized by a desired allocation rule $\chi:[0,1]^n\to\Delta([n])$, a randomized rule that takes as input the value profile $(V_1, \dots, V_n)$ and outputs a winning agent to receive the item. Denote the probability that agent $i$ wins the item under $\chi$ with value profile $(V_1, \dots, V_n)$ by $\chi_i(V_1, \dots, V_n)$. In order of agent number $i$, do the following. Elicit a reported value $\hat V_i\in[0,1]$ from agent $i$. For each agent $j\in[n]$, define
\begin{equation} \label{eq:anticipated_payoff_single_period}
    \Upsilon_j[i] = \begin{cases}\E_{V_{i+1}\sim\mathcal D_{i+1}, \dots, V_n\sim\mathcal D_n}\left[\hat V_j\cdot\chi_j(\hat V_1, \dots, \hat V_i, V_{i+1}, \dots, V_n)\right] & \text{if $j\leq i$}\\\E_{V_{i+1}\sim\mathcal D_{i+1}, \dots, V_n\sim\mathcal D_n}\left[V_j\cdot\chi_j(\hat V_1, \dots, \hat V_i, V_{i+1}, \dots, V_n)\right] & \text{if $j>i$}\end{cases}.
\end{equation}
and
\begin{equation} \label{eq:externality_single_period}
    \gamma_{i\to j} = \Upsilon_j[i] - \Upsilon_j[i-1],
\end{equation}
where $\Upsilon_j[0]$ is still defined by \eqref{eq:anticipated_payoff_single_period}. Agent $i$ makes a transfer $p_{i\to j} = -\gamma_{i\to j}$ to each agent $j\neq i$. At the end, allocate the item according to the rule $\chi(\hat V_1, \dots, \hat V_n)$. Pseudocode for the mechanism is given in \cref{alg:gum_single_period}.

\begin{algorithm}
\caption{Guaranteed Utility Mechanism (GUM) for the allocation of a single item} \label{alg:gum_single_period}
\KwIn{allocation rule $\chi$}
\For{$i=1,2,\dots, n$}{
    Elicit a reported value $\hat V_i\in[0,1]$ from agent $i$\;
    Agent $i$ makes transfers $p_{i\to j}= -\gamma_{i\to j}$ where $\gamma_{i\to j}$ is defined by \eqref{eq:externality_single_period}\;
}
Allocate the item to agent $j$ with probability $\chi_j(\hat V_1, \dots, \hat V_n)$\;
\end{algorithm}

Say that agent $i$ is \text{guaranteed utility} $\underline u_i$ if by acting truthfully, i.e., reporting $\hat V_i = V_i$, agent $i$'s expected utility is at least $\underline u_i$ regardless of the behavior of the other agents. The main property of GUM is the following, proved in \cite{safronov2018coalition,csoka2023collusion}.
\begin{theorem} \label{thm:gup_single_period}
Under GUM, each agent $i$ is guaranteed utility $\E_{V_1\sim\mathcal D_1, \dots, V_n\sim\mathcal D_n}[V_{i}\cdot\chi_{i}(V_1, \dots, V_n)]$.

Additionally, if agent $i$ behaves truthfully, then $\E[\gamma_{i\to k}] = 0$.
\end{theorem}
To explain why this theorem is true, we explain the meanings of $\Upsilon_{i}[j]$ and $\gamma_{j\to i}$. We call $\Upsilon_{i}[j]$ agent $i$'s \textit{anticipated payoff} after agent $j$ announces $\hat V_{j}$. It is agent $i$'s expected utility under the mechanism conditioned on the reports $\hat V_1, \dots, \hat V_{j}$ from agent $j$ and before, assuming $\hat V_{i} = V_{i}$ if $i\leq j$ and that the reports of agents $k>j$ are unknown and drawn from $\mathcal D_k$ (which includes $k=i$ if $i>j$). We call $\gamma_{j\to i}$ agent $j$'s \textit{externality} on agent $i$. It is the difference between agent $i$'s expected payoff after agent $j$ announces $\hat V_{j}$ and agent $i$'s expected payoff before agent $j$ announces $\hat V_{j}$. Intuitively, this is the harm that agent $j$ places on agent $i$ by announcing $\hat V_{j}$ and deviating from expectation. The payment rule $p_{j\to i} = -\gamma_{j\to i}$ has a simple interpretation: agent $j$ pays agent $i$ for the harm done.

\cref{thm:gup_single_period} can be explained by following observations.
\begin{enumerate}
\item Every agent $i$ has anticipated payoff $\Upsilon_{i}[0] = \E_{V_1\sim\mathcal D_1, \dots, V_n\sim\mathcal D_n}[V_{i}\cdot\chi_{i}(V_1, \dots, V_n)]$ at the beginning.
\item Any changes in the anticipated payoff that occur because agents $j\neq i$ report values that change this expectation get compensated by those agents.
\item If agent $i$ reports $\hat V_{i} = V_{i}$, she does not change any other agents' $k\neq i$ anticipated payoffs in expectation, so $\E[\gamma_{i\to k}] = 0$.
\end{enumerate}
Therefore, if agent $i$ behaves truthfully, they obtain the same utility in expectation as their anticipated payoff in the beginning. In particular, it doesn't matter what strategy the other agents take: agent $i$ is compensated by any harm done by others regardless of their strategy. Formally, one can define $\mathcal H_{j} = \sigma(\hat V_1, \dots, \hat V_{j})$ to be filtration generated by the first $j$ reports and show that
\[
    M_{i}[j] := \Upsilon_{i}[j] - \sum_{k\neq i}p_{i\to k}\mathbb1_{j\geq i} + \sum_{k\leq j, k\neq i}p_{k\to i}
\]
is an $\mathcal H_{j}$-martingale for every truthful agent $i$, which implies \cref{thm:gup_single_period}.

\begin{example}
Suppose there are $2$ agents with value distributions both uniform on $\{0,1\}$, and suppose $\chi$ is the allocation rule that allocates to the agent with the highest value, allocating to agent $2$ in the case of a tie. At the beginning, agent $1$'s anticipated payoff is $1/4$ and agent $2$'s anticipated payoff is $1/2$. Consider what happens after agent $1$ reports $\hat V_1$. Agent $1$'s anticipated payoff updates to $\hat V_1/2$. Agent $2$'s anticipated payoff does not change, so agent $1$ makes $0$ payment to agent $2$. Now consider what happens when agent $2$ reports $\hat V_2$. Agent $2$'s anticipated payoff updates to $\hat V_2\mathbb1_{\hat V_2\geq\hat V_1}$. Agent $1$'s anticipated payoff updates to $\hat V_1\mathbb 1_{\hat V_1>\hat V_2}$, so agent $2$ pays agent $1$ the difference $\hat V_1/2- \hat V_1\mathbb1_{\hat V_1>\hat V_2}$.

If agent $1$ is truthful, then her expected utility is
\[
    \underbrace{\frac12\Pr\left(\hat V_1 > \hat V_2\middle|\hat V_1 =1\right)}_{\text{expected utility from the item}} + \underbrace{\frac14 - \frac12\Pr\left(\hat V_1 > \hat V_2\middle| \hat V_1=1\right)}_{\text{expected payment from agent $2$}} = \frac14.
\]
If agent $2$ is truthful, then her expected utility is
\[
    \underbrace{\frac12}_{\text{expected utility from the item}} - \underbrace{\left(\frac12\E[\hat V_1] - \frac12\E[\hat V_1\mathbb 1_{\hat V_1>0}]\right)}_{\text{expected payment to agent $1$}} = \frac12.
\]
\end{example}

\begin{remark}
\citet{safronov2018coalition} and \citet{csoka2023collusion} require the allocation rule $\chi$ be efficient in that it always allocates the item to an agent who has the highest value. This allows them to argue that truthful reporting is a Bayes-Nash equilibrium: if all agents $j\neq i$ are truthful, they are getting their guaranteed utility, so agent $i$ cannot get higher than their guaranteed utility since if they could, the resulting total surplus would violate the efficiency of $\chi$.

In contrast, we are not requiring $\chi$ to be efficient here. We still retain the guaranteed utility property of \cref{thm:gup_single_period}. However, it may not be that truthful reporting is a Bayes-Nash equilibrium. Agents may be able to deviate from truth-telling to obtain a utility level even higher than the guaranteed utility level.
\end{remark}

\begin{remark}
\citet{safronov2018coalition} and \citet{csoka2023collusion} also require the allocation rule $\chi$ be deterministic, whereas we allow $\chi$ to be randomized. Allowing $\chi$ to be randomized essentially makes no difference in the analysis.
\end{remark}
\section{Auxiliary Concentration Bounds}
We use the following inequality from \cite{freedman1975tail} recorded as Theorem A in \cite{fan2015exponential}.
\begin{theorem}[Freedman's Inequality] \label{thm:freedman}
Let $S_0, S_1,\dots$ be a $\mathcal G_t$-supermartingale with $S_0=0$. Define $\xi_t = S_t - S_{t-1}$ and assume there exists $c>0$ such that $\xi_t\leq c$ almost surely for all $t$. Define $\langle S\rangle_t = \sum_{s=1}^t \E[\xi_s^2|\mathcal G_{s-1}]$. Then, for any $a,v>0$,
\[
    \Pr(\exists t: S_t \geq a, \langle S\rangle_t \leq v) \leq \exp\left(-\frac{a^2}{2(v+ca)}\right).
\]
\end{theorem}

The below Bernstein bound for the sum of i.i.d. thinned geometric random variables is based on bounding the moments of the sum; this technique can be found in \cite{10.1093/acprof:oso/9780199535255.001.0001}.
\begin{lemma}[Bernstein bound for thinned geometric random variables]
\label{lem:thinned_geometric_bernstein}
Let \(J_1,\ldots,J_T\) be independent $\Bern(p)$ variables and let \(G_1,\ldots,G_T\) be independent geometric variables on \(\{1,2,\ldots\}\) with success parameter \(\delta\), independent of the \(J_s\).  Set \(H_s=\delta G_s\) and \(Z_s=J_sH_s\).  Then, for every \(x>0\),
\[
\Pr\left(
\sum_{s=1}^T(Z_s-p)
>
\sqrt{32Tpx}+2x
\right)
\le e^{-x}.
\]
\end{lemma}
\begin{proof}
For \(y\ge0\),
\[
\Pr(H_s>y)
=(1-\delta)^{\lfloor y/\delta\rfloor}
\le e^{-\delta\lfloor y/\delta\rfloor}
\le e^{1-y}.
\]
The tail-integral formula for a nonnegative random variable and the defining integral of the Gamma function now give, for every integer \(r\ge1\),
\begin{align*}
\mathbb E[H_s^r]
&=\int_0^\infty r y^{r-1}\Pr(H_s>y)dy\\
&\le e r\int_0^\infty y^{r-1}e^{-y}dy
=e r\Gamma(r)
=er!.
\end{align*}
Moreover, \(\mathbb E[H_s]=\delta\mathbb E[G_s]=1\), and hence \(\mathbb E[Z_s]=p\).  For every integer \(r\ge2\), the well-known inequality $|a+b|^r\leq 2^{r-1}(|a|^r+|b|^r)$ gives
\begin{align*}
\E[|Z_s-p|^r] \le2^{r-1}\bigl(\mathbb E[Z_s^r]+p^r\bigr) =2^{r-1}\bigl(p\mathbb E[H_s^r]+p^r\bigr) \le2^{r-1}p\bigl(e r!+1\bigr)\le2^{r-1}p(4r!)\\
=\frac{r!}{2}(16p)2^{r-2}.
\end{align*}
We used \(e+1/r!<4\) for \(r\ge2\) for the last inequality.

\pagebreak[3]
Let \(X_s=Z_s-p\).  For \(0<\vartheta<1/2\), the preceding Bernstein moment bound makes the following series absolutely convergent and gives
\begin{align*}
\mathbb E[e^{\vartheta X_s}] =1+\sum_{r=2}^\infty\frac{\vartheta^r\mathbb E[X_s^r]}{r!}\le1+\sum_{r=2}^\infty\frac{\vartheta^r\mathbb E[|X_s|^r]}{r!}\le1+\frac{16p}{2}\sum_{r=2}^\infty\vartheta^r2^{r-2}\\
=1+\frac{16p\vartheta^2}{2(1-2\vartheta)}\le\exp\left(\frac{16p\vartheta^2}{2(1-2\vartheta)}\right).
\end{align*}
By independence,
\[
    \E\left[\exp\left(\vartheta\sum_{s=1}^T X_s\right)\right] = \E\left[\prod_{s=1}^Te^{\vartheta X_s}\right] = \prod_{s=1}^T \E[e^{\vartheta X_s}] \le\exp\left(\frac{16Tp\vartheta^2}{2(1-2\vartheta)}\right).
\]
The case $p=0$ is immediate; otherwise put \(u=\sqrt{2x/(16Tp)}\) and choose
\[
\vartheta=\frac{u}{1+2u}\in(0,1/2).
\]
Writing \(a=\sqrt{32Tpx}+2x\), this choice satisfies
\[
1-2\vartheta=\frac{1}{1+2u},
\qquad
\frac{16Tp\vartheta^2}{2(1-2\vartheta)}
=\frac{x}{1+2u},
\qquad
\vartheta a=\frac{x(2+2u)}{1+2u}.
\]
Markov's inequality and the preceding MGF bound therefore yield
\begin{align*}
\Pr\left(\sum_{s=1}^T X_s > a\right)\le\exp\left(-\vartheta a+
\frac{16Tp\vartheta^2}{2(1-2\vartheta)}\right)=\exp\left(-\frac{x(2+2u)}{1+2u}
+\frac{x}{1+2u}\right)
=e^{-x}.
\end{align*}
This proves the claim.
\end{proof}

The below lemma is a tail bound for a particular Poisson-Binomial distribution based off a Chernoff bound; a similar bound can actually be derived from the bounds on the factorial moments of the Poisson-Binomial distribution found in \cite{charalambides2005moments}.
\begin{lemma}[Chernoff bound for a Poisson-Binomial distribution] \label{lem:poisson_binomial_chernoff}
Let $p\in(0,1]$, $q\in(0,1)$, and $Z_0, Z_1,\dots$ be independent random variables with $Z_t\sim\Bern(pq^t)$. For any integer $m\geq 1$, 
\[
    \Pr\left(\sum_{t=0}^\infty Z_t \geq m\right)\leq e^{2/(1-q)}p^mq^{m(m-1)/2}.
\]
\end{lemma}
\begin{proof}
For any $\vartheta>0$, Markov's inequality applied to the MGF of $\sum_{t=0}^T Z_t$ and independence give
\[
    \Pr\left(\sum_{t=0}^\infty Z_t \geq m\right) = \Pr\left(\prod_{t=0}^\infty e^{\vartheta Z_t} \geq e^{\vartheta m}\right) \leq e^{-\vartheta m}\prod_{t=0}^\infty \E[e^{\vartheta Z_t}].
\]
Take the logarithm of both sides and substitute $\E[e^{\vartheta Z_t}] = 1 - pq^t +  pq^te^{\vartheta}$ to obtain
\begin{equation} \label{eq:generic_chernoff_poisson_binomial}
    \ln\left(\Pr\left(\sum_{t=0}^\infty Z_t \geq m\right)\right)\leq -\vartheta m + \sum_{t=0}^\infty \ln(1 - pq^t +  pq^te^{\vartheta})\leq -\vartheta m + \sum_{t=0}^\infty \ln(1 + pq^te^{\vartheta}).
\end{equation}
Use the identities $\ln(1+x) \leq \ln(x) + 1/x$ and $\ln(1+x)\leq x$ to obtain
\[
    \ln(1 + pq^te^{\vartheta}) \leq \min\left\{\ln(pq^t) + \vartheta + \frac{1}{pq^te^{\vartheta}}, pq^te^{\vartheta}\right\}.
\]
Use the first inequality for $t < m$ and the second inequality for $t\geq m$ in \eqref{eq:generic_chernoff_poisson_binomial} to obtain
\begin{align*}
    \ln\left(\Pr\left(\sum_{t=0}^\infty Z_t \geq m\right)\right)& \leq -\vartheta m + \sum_{t=0}^{m-1}\left(\ln(pq^t) +  \vartheta + \frac{1}{pq^te^{\vartheta}}\right) + \sum_{t=m}^\infty  pq^te^{\vartheta}
\end{align*}
Choose $\vartheta = \ln(1/pq^m) > 0$. Substituting this value of $\vartheta$ above, we obtain
\begin{align*}
    \ln\left(\Pr\left(\sum_{t=0}^\infty Z_t \geq m\right)\right) & \leq m\ln(pq^m) + \sum_{t=0}^{m-1}\ln(q^{t-m}) + \sum_{t=0}^{m-1}q^{m-t} + \sum_{t=m}^\infty q^{t-m}\\
    & = m\ln p + m^2\ln q + \frac{m(m-1)}{2}\ln q - m^2\ln q + \sum_{t=0}^{m-1}q^{m-t} + \sum_{t=m}^\infty q^{t-m}\\
    & = m\ln p + \frac{m(m-1)}{2}\ln q + \sum_{t=0}^{m-1}q^{m-t} + \sum_{t=m}^\infty q^{t-m}\\
    & \leq m\ln p + \frac{m(m-1)}{2}\ln q + \frac{2}{1-q}.
\end{align*}
Exponentiating both sides gives the desired bound.
\end{proof}
\section{A Lemma for the Deferred Proofs: Converting Guarantees on the Number of High-Valued Jobs Served into Utility Guarantees}
Our main tool to give lower bounds on the utility of agents is to give lower bounds on the number of high-valued jobs served from that agent. Converting lower bounds on the number of high-valued jobs served into a lower bound on the utility is relatively straightforward if the agent served a random high-valued job each time they are granted access to the server. However, in our results, agents instead serve the highest-valued job each time they are granted access. Intuitively, this can only be better than serving a random high-valued job. We give the formal argument below.

For convenience, we assume that for agent $i$, there are i.i.d. $Q_i[t]\sim\Unif[0,1]$ such that $V[p_i[t]] = F_i^{-1}(Q_i[t])$ where $F_i^{-1}$ is the inverse CDF of the value distribution $\mathcal F_i$. Here, $Q_i[t]$ is the quantile of the value $V[p_i[t]]$. Formally, a job $p_i[t]$ is high-valued if and only if $Q_i[t]\geq 1-1/n$. This specifies what we mean by top $1/n$-quantile in \cref{def:high_valued}. The assumption that these quantiles $Q_i[t]$ exist is without loss of generality, because agent $i$ can produce such $Q_i[t]$ by setting $Q_i[t] = F_i(V[p_i[t]])$ if $F_i$ is continuous at $V[p_i[t]]$ and otherwise, sampling $Q_i[t]\sim\Unif(F_i(V[p_i[t]]-), F_i(V[p_i[t]])]$, where $F_i$ is the CDF of $\mathcal F_i$, i.e., $F_i(v_i) = \Pr_{V_i\sim\mathcal F_i}(V_i\leq v_i)$, and $F_i(v_i-) = \lim_{v\uparrow v_i}F_i(v_i)$.

\begin{lemma} \label{lem:utility_conversion}
Assume that when agent $i$ gets access to the server, she always chooses to serve her highest-valued job if she has a job, and when a tie among job values occurs, she always chooses to serve a high-valued job if she has one. Additionally, assume that the event that agent $i$ obtains access to the server at any given time $t$ is independent of the values in her queue conditioned on the events $\mathbb1_{Q_i[s]\geq 1-1/n}$ and $\theta_i[s]$ (the number of high-valued jobs in agent $i$'s queue at time $s$) for $s\leq t$.

Let $\tilde U_i^G$ denote the number of high-valued jobs served from agent $i$, and let $U_i^G$ denote the total utility of agent $i$. Then,
\[
    \E[U_i^G] \geq nv_i^\star \E[\tilde U_i^G].
\]
\end{lemma}
\begin{proof}
In the lemma statement, $G$ stands for ``greedy''. We couple this process with a process $R$, standing for ``random'', in which whenever agent $i$ gets access to the server and has at least one high-valued job in her queue, she processes a uniformly random high-valued job among the high-valued jobs in her queue, in the following way. At each timestep, the arriving jobs' values are the same in both processes. The principal grants server access to the same agent in both processes. If agent $i$ is granted access to the server, she chooses to serve her highest-valued job if she has a job in process $G$ as assumed, and in process $R$, agent $i$ chooses to serve a uniformly random high-valued job among the high-valued jobs in her queue if she has a high-valued job, and otherwise, chooses to serve her highest-valued (and not high-valued) job if she has one. Each unserved job of agent $i$ expires with probability $\delta$ between times $t$ and $t+1$, and we couple expirations such that the $r$th highest-valued job of agent $i$ expires in process $G$ if and only if the $r$th highest-valued job of agent $i$ expires in process $R$ (where $r$th highest is determined by the quantiles of the job values).

Observe that the coupled process $R$ is a valid job process in the sense that the arrival values and expirations have the correct law. Furthermore, observe that agent $i$ has both the same number of jobs and also the same number of high-valued jobs in the processes $G$ and $R$.

Denote the utility and number of high-valued jobs served from agent $i$ in process $R$ by $U_i^R$ and $\tilde U_i^R$, respectively. Write
\begin{equation} \label{eq:actual_utility_coupled_process}
\begin{split}
    U_i^R & = \sum_{t=1}^T V[p_i[t]]\mathbb 1_{\text{$p_i[t]$ served in $R$}}\\
    & \geq \sum_{t=1}^T V[p_i[t]]\mathbb 1_{\text{$p_i[t]$ served in $R$}}\mathbb 1_{\text{$Q_i[t]\geq1-1/n$}}\\
    & = \sum_{t=1}^T\sum_{s=t}^T V[p_i[t]]\mathbb 1_{\text{$Q_i[t]\geq1-1/n$}}\mathbb 1_{\text{$p_i[t]$ served in $R$ at time $s$}}.
\end{split}
\end{equation}
Let $\mathcal H_s$ denote the $\sigma$-field generated by the events $\{Q_i[u]\geq1-1/n\}$ and the random variables $\theta_i[u]$ for $u\leq s$.  By assumption of the lemma, the event that agent $i$ is served at time $s$ is independent of the job values in agent $i$'s queue at time $s$ when conditioned on $\mathcal H_s$ in the process $G$, by the coupling, the same holds in the process $R$. Furthermore, in the process $R$, since agent $i$ picks a uniformly random high-valued job to serve at each timestep when given access, the event that the particular job $p_i[t]$ gets served in $R$ at time $s$ is independent of its value $V[p_i[t]]$ when conditioned on $\mathcal H_s$. Thus,
\begin{align*}
    & \E\left[V[p_i[t]]\mathbb 1_{\text{$Q_i[t]\geq1-1/n$}}\mathbb 1_{\text{$p_i[t]$ served in $R$ at time $s$}}\right]\\
    & = \E\left[\E\left[V[p_i[t]]\mathbb 1_{\text{$Q_i[t]\geq1-1/n$}}\mathbb 1_{\text{$p_i[t]$ served in $R$ at time $s$}}\middle|\mathcal H_s\right]\right]\\
    & = \E\left[\E\left[V[p_i[t]]\mathbb1_{\text{$Q_i[t]\geq1-1/n$}}\middle|\mathcal H_s\right]\E\left[\mathbb1_{\text{$p_i[t]$ served in $R$ at time $s$}}\middle|\mathcal H_s\right]\right].
\end{align*}
Similarly, we can argue that $V[p_i[t]]$ is independent of $\mathcal H_s$ conditioned on $Q_i[t]\geq 1-1/n$ for any $s$ by induction on $s$. Clearly, $V[p_i[t]]$ is independent of $\mathcal H_s$ for $s<t$. The count $\theta_i[t]$ of high-valued jobs is independent of $V[p_i[t]]$ conditioned on $Q_i[t]$, so $V[p_i[t]]$ is independent of $\mathcal H_t$ conditioned on $Q_i[t]\geq1-1/n$. Now assume that $V[p_i[t]]$ is independent of $\mathcal H_s$ conditioned on $Q_i[t]\geq1-1/n$ for some $s\geq t$; we prove it for $s+1$. Conditioned on $\mathcal H_s$, the event that agent $i$ is served at time $s$ is independent of $V[p_i[t]]$ by the lemma's assumption. After that service, the expirations and arrivals of high-valued jobs are independent of $V[p_i[t]]$ conditioned on $\mathcal H_s$ by the model. It follows that $V[p_i[t]]$ is independent of $\mathcal H_{s+1}$ conditioned on $\mathcal H_s$. Combining with the inductive hypothesis that $V[p_i[t]]$ is independent of $\mathcal H_s$ conditioned on $Q_i[t]\geq1-1/n$, we conclude that $V[p_i[t]]$ is independent of $\mathcal H_{s+1}$ conditioned on $Q_i[t]\geq1-1/n$. Therefore,
\begin{align*}
    \E\left[V[p_i[t]]\mathbb1_{\text{$Q_i[t]\geq1-1/n$}}\middle|\mathcal H_s\right] = \E[V[p_i[t]]|Q_i[t]\geq1-1/n]\mathbb1_{Q_i[t]\geq1-1/n} = nv_i^\star\mathbb1_{Q_i[t]\geq1-1/n}.
\end{align*}
where the last equality follows from the definition of ideal utility. Combining the previous two displays,
\begin{align*}
    \E\left[V[p_i[t]]\mathbb 1_{\text{$p_i[t]$ served in $R$}}\mathbb 1_{\text{$Q_i[t]\geq1-1/n$}}\right] & = nv_i^\star\sum_{s=t}^T \E[\mathbb1_{Q_i[t]\geq1-1/n}\mathbb1_{\text{$p_i[t]$ served in $R$ at time $s$}}]\\
    & = nv_i^\star\E[\mathbb 1_{\text{$Q_i[t]\geq1-1/n$}}\mathbb 1_{\text{$p_i[t]$ served in $R$}}].
\end{align*}
Substitute the above into \eqref{eq:actual_utility_coupled_process} to obtain
\begin{equation} \label{eq:utility_transfer_random}
    \E[U_i^R] \geq nv_i^\star\sum_{t=1}^T \E[\mathbb 1_{\text{$Q_i[t]\geq1-1/n$}}\mathbb 1_{\text{$p_i[t]$ served in $R$}}] \geq nv_i^\star\cdot\E[\tilde U_i^R].
\end{equation}

Now we use a greedy argument to show that the same holds in the process $G$. At each time $t$, immediately after service and before expirations, suppose there are $l[t]$ jobs in agent $i$'s queue. Write their values as $v_1^G[t]\leq \dots\leq v_{l[t]}^G[t]$ in process $G$ and $v_1^R[t]\leq\dots\leq v_{l[t]}^R[t]$ in process $R$. We shall show that $v_r^G[t]\leq v_r^R[t]$ for all $r\in[l[t]]$ is maintained by induction on $t$. At $t=0$, agent $i$'s queue is empty, so the invariant vacuously holds. Since arrival values are the same, arrivals maintain the invariant. Since the $r$th highest-valued job in process $G$ expires if and only if the $r$th highest-valued job in process $R$ expires, expirations preserve the invariant. One can reason that the greedy rule of processing the highest-valued job in process $G$ makes the services also preserve the invariant.

A job goes unserved if it either expires or is unserved at time $T$. For $t<T$, let $D_r[t]$ indicate that the $r$th smallest job expires in time $t$; set $D_r[T]=1$ for every unserved job after the final service. One can see that for either process $P\in\{G,R\}$,
\[
    U_i^P = \sum_{t=1}^TV[p_i[t]] - \sum_{t=1}^T\sum_{r=1}^{l[t]}v_r^P[t]D_r[t],
\]
the total value of all jobs minus the value of the unserved jobs. Since $v_r^G[t]\leq v_r^R[t]$ for all $t$ and all $r$, the above expression implies that $U_i^G\geq U_i^R$. Therefore, by \eqref{eq:utility_transfer_random},
\[
    \E[U_i^G] \geq nv_i^\star\cdot\E[\tilde U_i^R] = nv_i^\star\cdot\E[\tilde U_i^G]
\]
where the last equality follows from the fact that $\tilde U_i^R = \tilde U_i^G$ almost surely since a high-valued job gets served in process $G$ if and only if a high-valued job gets served in process $R$ by the coupling.
\end{proof}

\section{Deferred Proofs from \cref{ssec:target_policy}} \label{sec:target_policy_optimality_proof}

\thmTargetPolicyOptimality*
\begin{proof}
We set $\lambda(\delta,n,T)$ such that $\lambda(\delta,n,T)\cdot T$ is the expected number of high-valued jobs served across all agents and timesteps by the target policy. By symmetry, each agent gets $\lambda(\delta,n,T)\cdot T/n$ high-valued jobs served in expectation. \cref{lem:utility_conversion} implies that each agent obtains at least $\lambda(\delta,n,T)\cdot v_i^\star T$ in expected utility. Indeed the conditions of \cref{lem:utility_conversion} are satisfied: the target policy always serves the highest-valued job of an agent and the target policy only depends on the high-valued jobs process.

Now, assume $\mathcal F_i=\Bern(1/n)$ for every agent $i$. Then, high-valued jobs are precisely the jobs of value $1$ and $v_i^\star = 1/n$. The utility of an agent is equal to the number of high-valued jobs they get served. In particular, by the definition of $\lambda(\delta,n,T)$, the total utility under the target policy, summed across agents, is $\lambda(\delta,n,T)\cdot T$ in expectation. We shall show that no other policy can guarantee a higher expected total utility, which implies that there is always an agent $i$ obtaining at most $\lambda(\delta,n,T)\cdot T/n = \lambda(\delta,n,T)\cdot v_i^\star T$ expected utility by symmetry. Take an arbitrary policy $\pi$ and couple two of the job processes, one using the target policy and one using the policy $\pi$, in the following way. All arriving jobs have the same values in both processes. After service at each time $t$, let $m[t]$ be the number of total unserved jobs with value $1$ across all queues in the process with the target policy and $m'[t]$ be the number total unserved jobs with value $1$ across all queues in the process with the policy $\pi$. We shall inductively define how the expirations are coupled to maintain the invariant that $m[t]\leq m'[t]$ for each $t$, and the number of expirations of jobs of value $1$ in the process with the target policy is at most that of the process with the policy $\pi$. We have $m[1]\leq m'[1]$: the target policy always serves a job of value $1$ if possible and the number of jobs of value $1$ arriving at time $1$ are the same across both processes by the coupling. Fix a time $t$, and suppose we have defined how the expirations in periods before $t$ were defined such that $m[t]\leq m'[t]$. Sample $D_1\sim \Binom(m[t],\delta)$ and $D_2\sim\Binom(m'[t]-m[t],\delta)$ independently. Each job with value $0$ expires independently with probability $\delta$, arbitrarily coupled across both processes. Choose a uniformly random subset of size $D_1$ of the $m[t]$ jobs with value $1$ in the process with the target policy to expire, and choose a uniformly random subset of size $D_1 + D_2$ of the $m'[t]$ jobs with value $1$ in the process with the policy $\pi$ to expire, independently of the expirations of the value-$0$ jobs. If $m'[t]\geq m[t]$, then $D_1$ and $D_2$ are well-defined and each job in a given process expires independently with probability $\delta$. Furthermore, if $t<T$, in time $t+1$ before service, the number of jobs of value $1$ in the process with the target policy is at most that of the process with policy $\pi$. The same remains true after service since the target policy serves a job of value $1$ if one exists, establishing $m[t+1]\leq m'[t+1]$. By induction, we have coupled these two processes such that $m[t]\leq m'[t]$ at each timestep $t$, and the number of expirations of jobs of value $1$ in the target policy process is at most that of the $\pi$ policy process. The total number of jobs of value $1$ served in either process is the total number of jobs arriving with value $1$ minus the number of jobs of value $1$ that expire before time $T$, minus the number left immediately after service at time $T$. Since the arrivals are the same across process, but both the expirations before $T$ and the terminal remainder ($m[T]\leq m'[T]$) are at least as big under $\pi$, the total number of jobs of value $1$ served is at least as high under the target policy, establishing the claim.
\end{proof}

\begin{proposition} \label{prop:iterated_limit}
For each fixed $\delta$ and $n$, the limit $\lim_{T\to\infty}\lambda(\delta,n,T)$ exists. Additionally, for each fixed $n$,
\[
    \lim_{\delta\to0}\lim_{T\to\infty}\lambda(\delta,n,T)=1.
\]
\end{proposition}
\begin{proof}
By \cref{thm:target_policy_optimality}, $\lambda(\delta,n,T)\cdot T$ is the expected number of high-valued jobs served across all agents and timesteps by the target policy.

First, let us handle the case when $n=1$. In that case, every job is a high-valued job and every job gets served, so $\lambda(\delta,n,T)\cdot T=T$. This implies that $\lambda(\delta,n,T)=1$, establishing all claims in the proposition for $n=1$.

In the case $\delta=1$, then $\lambda(1,n,T)=1-(1-1/n)^n$ for every $T$, so $\lim_{T\to\infty}\lambda(\delta,n,T)$ exists.

Now assume $n\geq2$ and fix $\delta < 1$. As in the notation defined in \cref{sec:gum}, let $\theta_i[t]$ be the number of high-valued jobs in agent $i$'s queue at timestep $t$ after arrivals and before service. One can reason that under the target policy, the vector $\theta[t] = (\theta_1[t],\dots,\theta_n[t])$ evolves as a Markov chain on $\mathbb Z_{\geq0}^n$. Specifically, the services, expirations, and arrivals of high-valued jobs between timesteps $t$ and $t+1$ only depend on the state $\theta[t]$ and independent randomness. A high-valued job gets served at time $t$ whenever $\theta[t]\neq 0$, so we can write
\[
    \lambda(\delta,n,T) = \E\left[\frac1T\sum_{t=1}^T \mathbb1_{\theta[t]\neq 0}\right]. 
\]
Thus, to show that $\lim_{T\to\infty}\lambda(\delta,n,T)$ exists, we shall show that the Markov chain $(\theta[t])_t$ is irreducible and positive recurrent. Then, we can apply the Ergodic theorem to see that $\frac1T\sum_{t=1}^T\mathbb 1_{\theta[t]\neq0}$ converges almost surely, and then conclude $\lim_{T\to\infty}\lambda(\delta,n,T)$ exists by the bounded convergence theorem.

We first show that $(\theta[t])_t$ is irreducible. From any state $\underline\theta\in\mathbb Z_{\geq 0}^n$, it is possible to get to $0$ if all the high-valued jobs expire and no new high-valued jobs arrive. Starting from $0$, it is possible to get to any state $\underline\theta\in\mathbb Z_{\geq 0}^n$ since up to $n\geq 2$ high-valued jobs can arrive in each period, and only at most $1$ is served in each period, and expirations need not necessarily happen. Now let us argue positive recurrence. Consider the Lyapunov function $\|\cdot\|_1$ on states $\underline\theta\in\mathbb Z_{\geq 0}^n$ given by $\|\underline\theta\|_1 = \sum_{i=1}^n \underline\theta_i$. The drifts $\|\theta[t+1]\|_1 - \|\theta[t]\|_1$ are bounded almost surely: they can increase by at most $n$ since there can only be $n$ high-valued job arrivals. When $\|\theta[t]\|_1\geq2$,
\[
    \E[\|\theta[t+1]\|_1|\theta[t]] - \|\theta[t]\|_1 = -1-\delta(\|\theta[t]\|_1-1)+1 = -\delta(\|\theta[t]\|_1-1)\leq-\delta.
\]
because there is $1$ service, an expected number $\delta(\|\theta[t]\|_1-1)$ then expire, and $1$ high-valued job arrives in expectation (each agent has a high-valued job arrive with probability $1/n$). The last inequality uses $\|\theta[t]\|_1\geq 2$. By Foster's criterion, $(\theta[t])_t$ is positive recurrent. This establishes that we can use the Ergodic theorem to show that $\lim_{T\to\infty}\lambda(\delta,n,T)$ exists.

Now fix $n\geq 2$ and we shall show $\lim_{\delta\to0}\lim_{T\to\infty}\lambda(\delta,n,T)=1$. First, let us define an alterative policy used by a social planner to select jobs. Fix a window size $w$, and divide the first $\lfloor T/w\rfloor\cdot w$ time periods into $\lfloor T/w\rfloor$ windows of size $w$. The alternative policy serves no jobs in the $T -\lfloor T/w\rfloor\cdot w$ time periods outside one of these windows. At each timestep in a given window, the alternative policy picks an agent uniformly at random among agents have an unserved high-valued job that arrived in that window, regardless of if that high-valued job has expired, and serves one of such jobs. In other words, the alternative policy is the same as the target policy, except it ignores high-valued jobs that arrived before the current window, and it also ignores expirations and might serve an expired job.

We shall first analyze the number of high-valued jobs, expired or not, served by the alternative policy, and then use this to analyze the number of unexpired high-valued jobs served. Fix a window, and without loss of generality, number the time periods in this window by $t=1,2,\dots,w$. Let $A_t$ be the number of high-valued jobs that arrive at time $t$, and let $S_t$ be the indicator of whether a high-valued job, expired or not, is served by the alternative policy at time $t$. Now, we claim and shall show that
\begin{equation} \label{eq:convert_to_maximal_inequality}
    \sum_{t=1}^w S_t \geq w - \max_{0\leq t\leq w}\left(t - \sum_{s=1}^t A_s\right).
\end{equation}
Let $t_0$ be the last time period $t$ in the window in which $S_t=0$. If there is no such $t$, then $\sum_{t=1}^wS_t=w$ and there is nothing to prove. Since no job is served at time $t_0$, it must be that every high-valued job up to time $t_0$ has been served, so $\sum_{t=1}^{t_0}S_t = \sum_{s=1}^{t_0}A_s$. By definition of $t_0$, in every time period after $t_0$, a service happens, so $\sum_{t=t_0+1}^w S_t = w - t_0$. Write
\[
    \sum_{t=1}^w S_t = \sum_{t=1}^{t_0}S_t + \sum_{t=t_0+1}^w S_t = \sum_{s=1}^{t_0}A_s + w - t_0 = w - \left(t_0 - \sum_{s=1}^{t_0} A_s\right)\geq w - \max_{0\leq t\leq w}\left(t-\sum_{s=1}^t A_s\right).
\]
Write the last term above as
\[
    \max_{0\leq t\leq w}\left(t - \sum_{s=1}^t A_s\right) = \max_{0\leq t\leq w}\left(\sum_{s=1}^t (1-A_s)\right).
\]
Each $1-A_s\sim1-\Binom(n,1/n)$, which has mean $\E[1-A_s] = 0$ and variance $\Var[1-A_s] = n\cdot 1/n\cdot (1-1/n)\leq 1$, and the $1-A_s$ are independent. By Doob's $L^2$ inequality,
\[
    \E\left[\left(\max_{0\leq t\leq w}\left(\sum_{s=1}^t (1-A_s)\right)\right)^2\right]\leq 4\E\left[\left(\sum_{t=1}^w (1-A_t)\right)^2\right] = 4\sum_{t=1}^w \Var(1-A_t)\leq 4w.
\]
By Jensen's inequality,
\[
    \E\left[\max_{0\leq t\leq w}\left(\sum_{s=1}^t (1-A_s)\right)\right] \leq \sqrt{\E\left[\left(\max_{0\leq t\leq w}\left(\sum_{s=1}^t (1-A_s)\right)\right)^2\right]} \leq 2\sqrt w.
\]
It now follows from \eqref{eq:convert_to_maximal_inequality} that
\[
    \E\left[\sum_{t=1}^w S_t\right] \geq w - 2\sqrt w.
\]
This lower bounds the number of high-valued jobs served in the window, but that might include expired jobs. To lower bound the number of unexpired jobs served, we note that the expected number of jobs that arrive in the window and expire before its end is at most $\delta w^2$, because $w$ high-valued jobs arrive in expectation, and each has a probability of at most $\delta w$ in the whole window of size $w$. Thus, the expected number of unexpired high-valued jobs served by the alternative policy is at least $w - 2\sqrt w - \delta w^2$. Sum this lower bound over all $\lfloor T/w\rfloor$ windows to see that over all $T$ time periods, the expected number of unexpired high-valued jobs served by the alternative policy is at least
\[
    \left\lfloor \frac Tw\right\rfloor(w - 2\sqrt w - \delta w^2).
\]
By the proof of \cref{thm:target_policy_optimality}, the target policy maximizes the number of unexpired high-valued jobs served, so $\lambda(\delta,n,T)\cdot T$ is at least the above display. (Alternatively, we can argue without invoking the proof of \cref{thm:target_policy_optimality} as follows. If every agent had a $\mathcal F_i=\Bern(1/n)$ value distribution, then a job is high-valued if and only if it has value $1$, the ideal utility $v_i^\star=1/n$, and the utility of an agent is the number of unexpired high-valued jobs they have served. By the second part \cref{thm:target_policy_optimality}, at least one agent obtains at most $\lambda(\delta,n,T)\cdot v_i^\star T$ utility, so by symmetry of the alternative policy and since $v_i^\star=1/n$, the expected total utility is at most $\lambda(\delta,n,T)\cdot T$ under the alternative policy. By the first part of \cref{thm:target_policy_optimality}, under the target policy, each agent obtains at least $\lambda(\delta,n,T)\cdot v_i^\star T$ utility, so the total utility is at least $\lambda(\delta,n,T)\cdot T$ under the target policy. Thus, the total number of high-valued jobs served by the target policy is at least that of the alternative policy).

Now, we conclude
\begin{align*}
    \lim_{\delta\to0}\lim_{T\to\infty}\lambda(\delta,n,T) &\geq \lim_{\delta\to0}\lim_{T\to\infty}\frac1T\left\lfloor \frac Tw\right\rfloor(w - 2\sqrt w - \delta w^2)\\
    &= \lim_{\delta\to 0}\left(1-\frac{2}{\sqrt w} - \delta w\right) = 1-\frac{2}{\sqrt w}.
\end{align*}
Since $w$ was arbitrary, we can send $w\to\infty$ to see that the above limit is at least $1$, and therefore equal to $1$ since $\lambda(\delta,n,T)\leq 1$.
\end{proof}
\section{Deferred Proofs from \cref{sec:computation}} \label{sec:deferred_proofs_computation}
\subsection{Lemmas about Individual Runs of \cref{alg:compute_estimator}}
In this subsection, we prove lemmas about an individual run of \cref{alg:compute_estimator}. Throughout, we consider a run of \cref{alg:compute_estimator} at update $(t_0,i)$, report $\hat\theta_{i}[t_0]$, reported past states $\hat\theta[t_0,i-1]$, and simulation limit $L$. Unless stated otherwise, in this subsection, we consider the reported past states not as random but rather fixed, and probabilistic statements condition on the reported past states.

The algorithm runs two coupled simulations as described in the proof of \cref{lem:as_externality_bound}, and we mostly keep the notation from there. The two coupled processes are $(\theta^{\text{report}}[t,j])$ and $(\theta^{\text{random}}[t,j])$. We also name the processes $A$ and $B$ such that $\theta_{i}^A[t_0,i]\geq \theta_{i}^B[t_0,i]$. The algorithm stops the coupled processes after timestep $r:=\min\{\sigma, t_0+L-1,T\}$ where $\sigma = \inf\{t\geq t_0:\theta^A[t]=\theta^B[t]\}$ where $\inf\emptyset = T+1$. At that time, the algorithm outputs $\hat\gamma_{i\to j}[t_0]$ for $j\neq i$, which is the difference in the number of high-valued jobs from agent $j$ served between the simulation $(\theta^{\text{report}}[(t,j)])$ and $(\theta^{\text{random}}[(t,j)])$ in timesteps $t\in [t_0, r]$. We denote $\Delta_t = \|\theta^A[t] - \theta^B[t]\|_1$.

\begin{lemma} \label{lem:leftover_facts_from_main_body}
For all $(t,j)\geq (t_0,i)$, $\theta^A[t,j] \geq \theta^B[t,j]$.

Additionally, for any $s\geq 0$,
\[
    \Pr(\sigma > t_0 + s|\Delta_{t_0}) \leq (1-\delta)^s\Delta_{t_0}
\]
and
\[
    \E[\sigma|\Delta_{t_0}] - t_0 \leq \begin{cases}0 & \text{if $\Delta_{t_0}=0$}\\ \left\lceil \frac{\log(\Delta_{t_0})}{-\log(1-\delta)}\right\rceil + 1/\delta & \text{if $\Delta_{t_0}\geq1$ and $\delta<1$}\\1 & \text{if $\Delta_{t_0}\geq1$ and $\delta=1$}\end{cases}.
\]
\end{lemma}
\begin{proof}
These were proven as intermediate statements in the proof of \cref{lem:as_externality_bound}.
\end{proof}

\begin{corollary} \label{cor:expected_simulated_timesteps}
Given $\overline\theta$ and $\epsilon>0$, if $\hat\theta_{i}[t_0-1], \hat\theta_{i}[t_0]\leq\overline\theta$, then
\[
    \E[\sigma|\Delta_{t_0}] - t_0 \leq \begin{cases}\left\lceil \frac{\log(\overline\theta+1)}{-\log(1-\delta)}\right\rceil + 1/\delta & \text{if $\delta<1$}\\1 & \text{if $\delta=1$}\end{cases}.
\]
\end{corollary}
\begin{proof}
Since an agent gains at most $1$ high-valued job per time period, $\theta_{i}^{\text{random}}[t_0]\leq \hat\theta_{i}[t_0-1] + 1 \leq \overline\theta + 1$. Since $\hat\theta_{i}[t_0]\leq\overline\theta$, $\Delta_{t_0} \leq \overline\theta+1$. Substitute this bound into the bound on $\E[\sigma|\Delta_{t_0}]$ in \cref{lem:leftover_facts_from_main_body}.
\end{proof}

\begin{lemma} \label{lem:ex_ante_unbiased_for_honest_agents}
If $\hat\theta_{i}[t_0]$ is drawn randomly according to the stochastic transition kernel applied to $\hat\theta[t_0,i-1]$ (and independently of $\theta_i^{\text{random}}[t_0,i]$), then $\E[\hat\gamma_{i\to j}[t_0]]=0$ for any $j\neq i$.
\end{lemma}
\begin{proof}
If $\hat\theta_{i}[t_0]$ is drawn randomly according to the stochastic transition kernel applied to $\hat\theta[t_0,i-1]$, then the processes $(\theta^{\text{report}}[(t,j)])$ and $(\theta^{\text{random}}[(t,j)])$ have the same law. Since $\hat\gamma_{i\to j}[t_0]$ is the difference in the number of high-valued jobs served from agent $j$ between these two processes equal in law, it has expectation $0$.
\end{proof}
\begin{lemma} \label{lem:token_tracking}
For any agent $j\neq i$, the sign of $\hat\gamma_{i\to j}[t_0]$ is opposite that of $\hat\theta_{i}[t_0]-\theta_{i}^{\text{random}}[t_0]$ in that $\hat\gamma_{i\to j}[t_0](\hat\theta_{i}[t_0]-\theta_{i}^{\text{random}}[t_0])\leq 0$, and $|\hat\gamma_{i\to j}[t_0]|\leq \Delta_{t_0}$ almost surely.
\end{lemma}
\begin{proof}
For each process $P\in\{A,B\}$, denote the number of high-valued jobs arriving to an arbitrary agent $k$ during updates after $(t_0,i)$ and not after timestep $r:=\min\{\sigma, t_0+L-1,T\}$ by $\text{Arrivals}_k^P$. Denote the number of agent $k$'s high-valued jobs that expire and the number of services of high-valued jobs of agent $k$ during the same update interval by $\text{Expirations}_k^P$ and $\text{Services}_k^P$, respectively. Then, by a flow conservation argument,
\[
    \theta_k^P[r,n+1] - \theta_k^P[t_0,i] = \text{Arrivals}_k^P - \text{Expirations}_k^P - \text{Services}_k^P.
\]
By the coupling, $\text{Arrivals}_k^A = \text{Arrivals}_k^B$, so by rearranging the above equation for $P=A,B$ and equating the arrivals, we obtain
\begin{equation} \label{eq:flow_conservation}
\begin{split}
    \left((\theta_k^{A}[r,n+1] - \theta_k^A[t_0,i]) - (\theta_k^B[r,n+1] - \theta_k^B[t_0,i])\right) + \left(\text{Expirations}_k^A -\text{Expirations}_k^B\right)\\
     + \left(\text{Services}_k^A - \text{Services}_k^B\right) = 0.
\end{split}
\end{equation}
As established in \cref{lem:leftover_facts_from_main_body}, every agent $k$ has at least as many high-valued jobs in process $A$ as in process $B$ at all updates after $(t_0,i)$. By the coupling, this implies $\theta_k^A[r,n+1] \geq \theta_k^B[r,n+1]$ and $\text{Expirations}_k^A \geq \text{Expirations}_k^B$. The above display then implies
\begin{equation}
    -(\theta_k^A[t_0,i] - \theta_k^B[t_0,i]) + \left(\text{Services}_k^A - \text{Services}_k^B\right) \leq 0.
\end{equation}
Fix an agent $j\neq i$, and observe $\theta_{j}^A[t_0,i] = \theta_{j}^B[t_0,i]$. \cref{eq:flow_conservation} with $k=j$ says
\[
    \text{Services}_{j}^A - \text{Services}_{j}^B \leq 0.
\]
By how we labeled $A$ and $B$, $\hat\gamma_{i\to j}[t_0] = \text{Services}_{j}^A - \text{Services}_{j}^B$ if $\hat\theta_{i}[t_0] \geq \theta_{i}^{\text{random}}[t_0]$ and $\hat\gamma_{i\to j}[t_0] = -(\text{Services}_{j}^A - \text{Services}_{j}^B)$ if $\hat\theta_{i}[t_0] < \theta_{i}^{\text{random}}[t_0]$. This observation and the previous display imply that $\hat\gamma_{i\to j}[t_0]$ has sign opposite that of $\hat\theta_{i}[t_0] - \theta_{i}^{\text{random}}[t_0]$.

This also establishes that
\[
    |\hat\gamma_{i\to j}[t_0]| = -(\text{Services}_{j}^A - \text{Services}_{j}^B).
\]
Substitute this into \eqref{eq:flow_conservation} with $k=j$ and the observation that $\theta_{j}^A[t_0,i] = \theta_{j}^B[t_0,i]$ to obtain
\[
    |\hat\gamma_{i\to j}[t_0]| = \left(\theta_{j}^A[r,n+1] - \theta_{j}^B[r,n+1]\right) + \left(\text{Expirations}_{j}^A -\text{Expirations}_{j}^B\right)
\]
By definition, $\Delta_{t_0} = \theta_{i}^A[t_0,i] - \theta_{i}^B[t_0,i]$. Use this, the previous display, and sum \eqref{eq:flow_conservation} across all $k\in[n]$ to obtain
\begin{equation} \label{eq:flow_conservation_summed}
\begin{split}
    -\Delta_{t_0} + \left(\theta_{i}^A[r,n+1] - \theta_{i}^B[r,n+1]\right) + \left(\text{Expirations}_{i}^A - \text{Expirations}_{i}^B\right) + \sum_{j\neq i}|\hat\gamma_{i\to j}[t_0]|\\
    + \sum_{k\in[n]}\left(\text{Services}_k^A - \text{Services}_k^B\right) = 0.
\end{split}
\end{equation}
It was established earlier that both $\theta_{i}^A[r,n+1] - \theta_{i}^B[r,n+1]\geq 0$ and $\text{Expirations}_{i}^A - \text{Expirations}_{i}^B\geq 0$. Since the server in process $B$ idles whenever the server in process $A$ idles by the coupling,
\[
    \sum_{k\in[n]}\left(\text{Services}_k^A - \text{Services}_k^B\right)\geq 0.
\]
\cref{eq:flow_conservation_summed} then implies
\[
    \sum_{j\neq i}|\hat\gamma_{i\to j}[t_0]|\leq \Delta_{t_0}.
\]
Therefore, each individual term in the sum is at most $\Delta_{t_0}$.
\end{proof}

\begin{corollary} \label{cor:positive_externality_from_timed_out_agents}
If $\hat\theta_{i}[t_0] = 0$, then $\hat\gamma_{i\to j}[t_0] \geq 0$ for all $j\neq i$.
\end{corollary}
\begin{proof}
If $\hat\theta_{i}[t_0]=0$, then $\hat\theta_{i}[t_0]-\theta_{i}^{\text{random}}[t_0]\leq 0$, so $\hat\gamma_{i\to j}[t_0]\geq 0$ by \cref{lem:token_tracking}.
\end{proof}

\begin{corollary} \label{cor:as_estimator_bound}
If $\hat\theta_{i}[t_0-1], \hat\theta_{i}[t_0]\leq \overline\theta$, then $|\hat\gamma_{i\to j}[t_0]|\leq \min\{L,\overline\theta+1\}$ almost surely.
\end{corollary}
\begin{proof}
The absolute count $|\hat\gamma_{i\to j}[t_0]|$ is at most the number of simulation steps run, so $|\hat\gamma_{i\to j}[t_0]|\leq L$.

Since an agent gains at most $1$ high-valued job per time period, $\theta_{i}^{\text{random}}[t_0]\leq \hat\theta_{i}[t_0-1] + 1 \leq \overline\theta + 1$. Since $\hat\theta_{i}[t_0]\leq\overline\theta$, $\Delta_{t_0} \leq \overline\theta+1$. Then, $|\hat\gamma_{i\to j}[t_0]|\leq \overline\theta+1$ by \cref{lem:token_tracking}.
\end{proof}

Denote the predictable bias of $\hat\gamma_{i\to j}[t_0]$ by
\[
    \beta_{i\to j}[t_0] = \E[\hat\gamma_{i\to j}[t_0]] - \gamma_{i\to j}[t_0] .
\]
\begin{lemma} \label{lem:bias_bound}
Given $\overline\theta$ and $\epsilon>0$, if $\hat\theta_{i}[t_0-1], \hat\theta_{i}[t_0]\leq\overline\theta$ and
\[
    L \geq \begin{cases}\frac{\log \frac{\overline\theta+1}{\epsilon\delta}}{-\log(1-\delta)} & \text{if $\delta<1$}\\1 & \text{if $\delta=1$}\end{cases},
\]
then $\beta_{i\to j}[t_0] \leq\epsilon$ for all $j\neq i$.
\end{lemma}
\begin{proof}
Since an agent gains at most $1$ high-valued job per time period, $\theta_{i}^{\text{random}}[t_0]\leq \hat\theta_{i}[t_0-1] + 1 \leq \overline\theta + 1$. Since $\hat\theta_{i}[t_0]\leq\overline\theta$, $\Delta_{t_0} \leq \overline\theta+1$. Substitute this bound into the bound on $\Pr(\sigma>t_0+s|\Delta_{t_0})$ in \cref{lem:leftover_facts_from_main_body}
\[
    \Pr(\sigma > t_0 + s) \leq (1-\delta)^s(\overline\theta + 1).
\]
By definition, the predictable bias $\beta_{i\to j}[t_0]$ is the expected difference between the output of \cref{alg:compute_estimator} when run with a simulation limit and the output when run without a simulation limit. This difference is at most $(\sigma-(t_0 + L))^+$. Using the tail sum formula,
\[
    \E[(\sigma-(t_0 + L))^+] \leq \sum_{s=L}^\infty \Pr(\sigma > t_0 + s) \leq (\overline\theta + 1)\sum_{s=L}^\infty (1-\delta)^s = (\overline\theta+1)\cdot\frac{(1-\delta)^L}{\delta}.
\]
For $L$ as chosen in the lemma statement, this is at most $\epsilon$.

\end{proof}

\begin{lemma} \label{lem:quadratic_variation_bound}
If $\hat\theta_{i}[t_0]$ is drawn randomly and independently from $\theta_i^{\text{random}}[t_0]$ according to the stochastic transition kernel applied to $\hat\theta[t_0-1,n+1]$, then for any $j\neq i$,
\[
    \E[\hat\gamma_{i\to j}[t_0]^2]\leq 2\delta\hat\theta_{i}[t_0-1] + \frac2n.
\]
\end{lemma}
\begin{proof}
By \cref{lem:token_tracking}, $|\hat\gamma_{i\to j}[t_0]|\leq \Delta_{t_0}$ almost surely. Observe that with the assumption that $\hat\theta_{i}[t_0]$ is drawn randomly,
\[
    \hat\theta_{i}[t_0], \theta_i^{\text{random}}[t_0]\sim\Binom(\hat\theta_{i}[t_0-1,n+1],1-\delta)\ast\Bern\left(\frac1n\right)
\]
because a $\Binom(\hat\theta_{i}[t_0-1,n+1],1-\delta)$ number of high-valued jobs survive from the previous period and with probability $1/n$, a new one arrives at time $t_0$. Hence,
\begin{align*}
    \E[\hat\gamma_{i\to j}[t_0]^2]\leq \E[\Delta_{t_0}^2] & = \E[(\hat\theta_{i}[t_0] - \theta_{i}^{\text{random}}[t_0])^2]\\
    & = \Var[\hat\theta_{i}[t_0]-\theta_{i}^{\text{random}}[t_0]] = 2\Var[\hat\theta_{i}[t_0]]\\
    & = 2\Var[\Binom(\hat\theta_{i}[t_0-1,n+1],1-\delta)] + 2\Var\left[\Bern\left(\frac1n\right)\right]\\
    & = 2\hat\theta_{i}[t_0-1,n+1](1-\delta)\delta + 2\left(\frac1n\right)\left(1-\frac1n\right)\\
    & \leq 2\delta\hat\theta_i[t_0-1,n+1] + \frac2n\leq 2\delta\hat\theta_i[t_0-1] + \frac2n.
\end{align*}
\end{proof}

\begin{lemma} \label{lem:naive_bound_difference}
Almost surely,
\[
    \Delta_{t_0}\leq \hat\theta_{i}[t_0-1] + \hat\theta_{i}[t_0] + 1.
\]
\end{lemma}
\begin{proof}
Since an agent gains at most $1$ high-valued job per time period, $\theta_{i}^{\text{random}}[t_0]\leq \hat\theta_{i}[t_0-1] + 1$. Thus,
\[
    \Delta_{t_0} = \left|\theta_{i}^{\text{random}}[t_0]-\hat\theta_{i}[t_0]\right|\leq \hat\theta_{i}[t_0-1] + \hat\theta_{i}[t_0] + 1.
\]
\end{proof}
\subsection{Deferred Proofs about \cref{alg:gum_report_capping}}
In this section, we prove theorems about \cref{alg:gum_report_capping}.

Let us first define some notation used throughout. We shall set the algorithm parameters $B$, $\overline\theta$, $\overline\Sigma$, and $L$ later.

Let $\mathcal H_{t,i}$ be the filtration generated by the history of all reports and server allocations recorded by the mechanism up to update $(t,i)$ and the estimated externalities $\hat\gamma_{j\to k}[s]$ used up to update $(t,i)$. (It doesn't matter for our results whether the estimated externalities are public.) Let $\mathcal G_t$ be the filtration generated by the history of all reports and server allocations recorded by the mechanism up to timestep $t$ and the estimated externalities $\hat\gamma_{j\to k}[s]$ for $s\leq t$, i.e., $\mathcal G_t = \mathcal H_{t,n+1}$.

Fix an honest agent $i$. Define an auxiliary mechanism that is the same as the actual mechanism except the virtual debt limit and both the per-round and cumulative report caps are not enforced on the fixed honest agent $i$; all restrictions on agents $j\neq i$ remain enforced in the auxiliary mechanism. We denote quantities in the auxiliary mechanism with the superscript $\text{aux}$. Since there is no virtual debt limit or report cap enforcement on agent $i$, we always have $\hat\theta_i^{\text{aux}}[t] = \theta_i^{\text{aux}}[t]$. Couple the auxiliary mechanism with the real mechanism in which all restrictions are enforced such that the mechanisms agree until agent $i$ makes a report that gets modified by one of the report caps or virtual debt limit.

\subsubsection{Proving that the two coupled processes never differ with high probability}

\begin{lemma} \label{lem:occupation_bound}
For any agent $i$, for any $x>0$, with probability at least $1-e^{-x}$,
\[
    \delta\sum_{t=1}^T \theta_i^{\text{aux}}[t]\leq \frac Tn + \sqrt{\frac{32Tx}{n}} + 2x.
\]
\end{lemma}
\begin{proof}
This is the same statement as \cref{lem:occupation_bound_main_body} except for the auxiliary mechanism, and the same proof applies.
\end{proof}

\begin{lemma} \label{lem:honest_agent_exceed_budget}
Fix an honest agent $i$. For $j\neq i$, define
\[
    \hat X_{i\to j}^{\text{aux}}[t] = -\sum_{s=1}^t\hat \gamma_{i\to j}^{\text{aux}}[s].
\]
Fix $\eta\in(0,1]$. Set $x = \ln(2n/\eta)$ and $v = 4T/n + 2\sqrt{32Tx/n} + 4x$. If we set $B\geq Lx + \sqrt{x(L^2x + 2v)}$, then $\Pr(\exists t, j\neq i: \hat X_{i\to j}^{\text{aux}}[t] > B) \leq\eta$.
\end{lemma}
\begin{proof}
For a fixed $j\neq i$, set $\xi_{i\to j}^{\text{aux}}[t] = \hat X_{i\to j}^{\text{aux}}[t] - \hat X_{i\to j}^{\text{aux}}[t-1]$. By \cref{lem:quadratic_variation_bound},
\[
    \E[\xi_{i\to j}^{\text{aux}}[t]^2|\mathcal G_{t-1}^{\text{aux}}] \leq 2\delta\theta_i^{\text{aux}}[t-1] + \frac{2}{n}.
\]
Summing over $t$,
\[
    \sum_{t=1}^T \E[\xi_{i\to j}^{\text{aux}}[t]^2|\mathcal G_{t-1}^{\text{aux}}] \leq 2\delta\sum_{t=1}^T\theta_i^{\text{aux}}[t-1]+\frac{2T}{n}.
\]
\cref{lem:occupation_bound} implies that $\delta\sum_{t=1}^T \theta_i^{\text{aux}}[t-1]\leq T/n + \sqrt{32Tx/n} + 2x$ with probability at least $1-e^{-x}$, so the above display implies
\[
    \Pr\left(\sum_{t=1}^T \E[\xi_{i\to j}^{\text{aux}}[t]^2|\mathcal G_{t-1}^{\text{aux}}] > \frac{4T}{n} + 2\sqrt{\frac{32Tx}{n}}+4x\right)\leq e^{-x}.
\]
Observe that $\hat X_{i\to j}^{\text{aux}}[t]$ is a $\mathcal G_t^{\text{aux}}$-martingale by \cref{lem:ex_ante_unbiased_for_honest_agents}. Furthermore, by \cref{cor:as_estimator_bound}, its increments are almost surely bounded by $L$. By \cref{thm:freedman} with $S_t = \hat X_{i\to j}^{\text{aux}}[t]$, $\xi_t = \xi_{i\to j}^{\text{aux}}[t]$, $c=L$, and $a=B$,
\[
    \Pr\left(\exists t: \hat X_{i\to j}^{\text{aux}}[t]\geq B, \sum_{s=1}^t \E[\xi_{i\to j}^{\text{aux}}[s]^2|\mathcal G_{s-1}^{\text{aux}}]\leq v\right)\leq \exp\left(-\frac{B^2}{2(v + LB)}\right).
\]
Combine the previous two displays to obtain
\[
    \Pr\left(\exists t: \hat X_{i\to j}^{\text{aux}}[t]\geq B\right)\leq \exp\left(-\frac{B^2}{2(v + LB)}\right) + e^{-x}.
\]
By a union bound over $j\neq i$,
\[
    \Pr\left(\exists t, j\neq i: \hat X_{i\to j}^{\text{aux}}[t]\geq B\right)\leq n\exp\left(-\frac{B^2}{2(v + LB)}\right) + ne^{-x}.
\]
Expanding $B \geq Lx + \sqrt{x(L^2x+2v)}$ gives $B^2 \geq 2x(v + LB)$, so
\[
    n\exp\left(-\frac{B^2}{2(v + LB)}\right) \leq ne^{-x}.
\]
Therefore,
\[
    \Pr\left(\exists t, j\neq i: \hat X_{i\to j}^{\text{aux}}[t]\geq B\right)\leq 2ne^{-x} = \eta.
\]
\end{proof}

\begin{lemma} \label{lem:honest_agent_exceed_per_round_report_cap}
Fix an agent $i$ and $\eta\in(0,1]$. If we set
\[
    \overline\theta \geq \begin{cases}\min\left\{T,\sqrt{\frac{2}{-\ln(1-\delta)}\left(\frac2\delta+\ln\frac T\eta\right)}+1\right\} & \text{if $\delta<1$}\\1 & \text{if $\delta=1$}\end{cases},
\]
then $\Pr(\exists t: \theta_i^{\text{aux}}[t] > \overline\theta)\leq \eta$.
\end{lemma}
\begin{proof}
This is the same statement as \cref{lem:honest_agent_exceed_per_round_report_cap_main_body} except for the auxiliary mechanism, and the same proof applies.
\end{proof}

\begin{lemma} \label{lem:honest_agent_exceed_cumulative_report_cap}
Fix an agent $i$ and $\eta\in(0,1]$. If we set
\[
    \overline\Sigma \geq \frac1\delta\left(\frac Tn + \sqrt{-\frac{32T}{n}\ln\eta} - 2\ln\eta\right),
\]
then $\Pr\left(\sum_{t=1}^T \theta_i^{\text{aux}}[t] > \overline\Sigma\right)\leq \eta$.
\end{lemma}
\begin{proof}
This follows from \cref{lem:occupation_bound} with $x=-\ln\eta$.
\end{proof}

\begin{corollary} \label{cor:honest_agent_ban_probability}

Let $\eta = 1/T$, $x = \ln(2n/\eta)$ and $v = 4T/n + 2\sqrt{32Tx/n} + 4x$. If we set
\begin{align*}
    B \geq Lx + \sqrt{x(L^2x + 2v)}& , T\geq \overline\theta \geq \begin{cases}\min\left\{T,\sqrt{\frac{2}{-\ln(1-\delta)}\left(\frac2\delta+\ln\frac T\eta\right)}+1\right\} & \text{if $\delta<1$}\\1 & \text{if $\delta=1$}\end{cases},\\
    &\text{and}\quad \overline\Sigma \geq \frac1\delta\left(\frac Tn + \sqrt{-\frac{32T}{n}\ln\eta} - 2\ln\eta\right),
\end{align*}
then the real mechanism and the auxiliary mechanism differ with probability at most $3/T$.
\end{corollary}
\begin{proof}
Fix an honest agent $i$. By \cref{lem:honest_agent_exceed_budget,lem:honest_agent_exceed_per_round_report_cap,lem:honest_agent_exceed_cumulative_report_cap}, with our parameter choices, with $\hat X_{i\to j}^{\text{aux}}[t]$ defined as in \cref{lem:honest_agent_exceed_budget},
\[
    \Pr(\exists t, j\neq i:\hat X_{i\to j}^{\text{aux}}[t]>B)\leq\eta, \Pr(\exists t:\theta_i^{\text{aux}}[t]>\overline\theta)\leq\eta,\Pr\left(\sum_{t=1}^T\theta_i^{\text{aux}}[t]>\overline\Sigma\right)\leq\eta.
\]
If none of the conditions occur in the above display, then neither the virtual debt limit nor either report cap alters agent $i$\textquotesingle s report in the real mechanism, so by the coupling, the real mechanism and auxiliary mechanism never differ. The result now follows from a union bound over the events in the above display.
\end{proof}

\subsubsection{Using \pec/\spec properties to prove low error bounds}
For each agent $k$, write $\Upsilon_k[t,j] := \Upsilon_k(\hat h_{t,j})$ for agent $i$'s anticipated payoff at update $(t,j)$, where $\hat h_{t,j}$ is the reported history through update $(t,j)$ (recall $\Upsilon_k(\cdot)$ is defined at \cref{def:virtual_anticipated_payoff})).
\begin{lemma} \label{lem:gum_martingale}
Fix an honest agent $i$. Define
\begin{align*}
    M_{i}^{\text{aux}}[t,j] = \Upsilon_{i}^{\text{aux}}[t,j] + \underbrace{\sum_{k\neq i}\left(\sum_{s=1}^{t-1}\gamma_{i\to k}^{\text{aux}}[s] + \gamma_{i\to k}^{\text{aux}}[t]\mathbb1_{j\geq i}\right)}_{\text{externality on agents $k\neq i$ by $i$}} \underbrace{-\sum_{(s,k)\leq (t,j):k\neq i}\gamma_{k\to i}^{\text{aux}}[s]}_{\text{externality on agent $i$}}.
\end{align*}
Then, $M_{i}^{\text{aux}}[t,j]$ is an $\mathcal H_{t,j}^{\text{aux}}$-martingale starting at $M_i^{\text{aux}}[0,n+1] = \lambda(\delta,n,T)\cdot T/n$.
\end{lemma}
\begin{proof}
It is readily checked that $M_{i}^{\text{aux}}[(t,j)]$ is $\mathcal H_{t,j}^{\text{aux}}$-measurable.
Fix some update $(t,j)$. To show that the increments of $M_{i}^{\text{aux}}[(t,j)]$ have expectation $0$ conditioned on the history, we break down into $3$ cases: $j=n+1$, $j=i$, and $j\in[n]\setminus\{i\}$.

If $j=n+1$, then $M_{i}^{\text{aux}}[t,j] - M_{i}^{\text{aux}}[t,j-1]= \Upsilon_{i}^{\text{aux}}[t,n+1] - \Upsilon_{i}^{\text{aux}}[t,n]$. Conditioned on $\mathcal H_{t,j-1}^{\text{aux}} = \mathcal H_{t,n}^{\text{aux}}$, this difference has expectation $0$ by the tower law: $\Upsilon_{i}^{\text{aux}}[t,n]$ is the expected value of $\tilde U_{i}(h^{\text{aux}}_{T,n+1})$ starting at the reported history $\hat h_{t,n}^{\text{aux}}$ up to update $(t,n)$, and $\Upsilon_{i}^{\text{aux}}[t,n+1]$ is the same thing but with the random service at time $t$ realized; in particular, note that the state $\hat\theta^{\text{aux}}[t,n+1]$ is drawn from the same distribution as the target policy when the reported history is $\hat h_{t,n}^{\text{aux}}$ as in \cref{alg:gum_report_capping}.

Now consider the case when $j=i$, in which case
\[
    M_{i}^{\text{aux}}[t,j] - M_{i}^{\text{aux}}[t,j-1] = \Upsilon_{i}^{\text{aux}}[t,i] - \Upsilon_{i}^{\text{aux}}[t,i-1] + \sum_{k\neq i}\gamma_{i\to k}^{\text{aux}}[t].
\]
By a similar argument as in the previous case,
\[
    \E[\Upsilon_{i}^{\text{aux}}[t,i] - \Upsilon_{i}^{\text{aux}}[t,i-1]|\mathcal H_{t,i-1}^{\text{aux}}]=0
\]
by the tower law. Specifically, $\Upsilon_{i}^{\text{aux}}[t,i-1]$ is the expected value of $\tilde U_{i}(h_{T,n+1}^{\text{aux}})$ starting at the reported partial history $\hat h_{t,i-1}^{\text{aux}}$ up to update $(t,i-1)$ with truthful random reports thereafter, and $\Upsilon_{i}^{\text{aux}}[t,i]$ is the same thing but starting at the reported partial history $\hat h_{t,i}^{\text{aux}}$. Since agent $i$ is honest, $\hat h_{t,i}^{\text{aux}}$ is $\hat h_{t,i-1}^{\text{aux}}$ with the last state vector with its $i$th coordinate replaced by $\theta_{i}^{\text{aux}}[t]$ appended. By the tower law, $\Upsilon_{i}^{\text{aux}}[t,i-1]$ and $\Upsilon_{i}^{\text{aux}}[t,i]$ have equal expectations conditioned on $\mathcal H_{t,i-1}^{\text{aux}}$. We can then apply \cref{lem:ex_ante_unbiased_for_honest_agents} to see that $\E[\gamma_{i\to k}^{\text{aux}}[t]|\mathcal H_{t,i-1}^{\text{aux}}]= 0$ for each $k\neq i$. Thus, $\E[M_{i}^{\text{aux}}[t,j] - M_{i}^{\text{aux}}[t,j-1]|\mathcal H_{t,j-1}^{\text{aux}}]=0$.

Now we consider the final case when $j\in[n]\setminus\{i\}$. In this case,
\[
    M_{i}^{\text{aux}}[t,j] - M_{i}^{\text{aux}}[t,j-1] = \Upsilon_{i}^{\text{aux}}[t,j] - \Upsilon_{i}^{\text{aux}}[t,j-1] - \gamma_{j\to i}^{\text{aux}}[t].
\]
By definition, $\gamma_{j\to i}^{\text{aux}}[t] = \Upsilon_{i}^{\text{aux}}[t,j] - \Upsilon_{i}^{\text{aux}}[t,j-1]$, so the above is $0$ almost surely, establishing that $M_{i}^{\text{aux}}[t,j]$ is an $\mathcal H_{t,j}^{\text{aux}}$-martingale. By definition, $\Upsilon_i^{\text{aux}}[0,n+1]$ is the expected number of high-valued jobs served from agent $i$ under the target policy described in \cref{ssec:target_policy}. By \cref{thm:target_policy_optimality} and symmetry, $\Upsilon_i^{\text{aux}}[0,n+1] = \lambda(\delta,n,T)\cdot T/n$.
\end{proof}

Denote
\[
    \beta_{j\to i}^{\text{aux}}[t] = \E[\hat\gamma_{j\to i}^{\text{aux}}[t]|\mathcal H_{t,j-1}^{\text{aux}}, \hat\theta_{j}^{\text{aux}}[t]] - \gamma_{j\to i}^{\text{aux}}[t] .
\]
\begin{lemma} \label{lem:generic_error}
Fix an honest agent $i$. Suppose $\beta_{j\to i}^{\text{aux}}[t]\leq \epsilon$ almost surely for all $j\neq i$ and all $t$. Additionally, suppose the probability that the real mechanism and auxiliary mechanism differ is bounded above by $q$. Define $D = \min\{L,\overline\theta + 1\}$. Then, 
\[
    \E[U_{i}] \geq \lambda(\delta,n,T)\cdot v_{i}^\star T  - \left((n-1)(B+D + \epsilon T) + q\cdot T\right).
\]
\end{lemma}
\begin{proof}
We first lower-bound agent $i$'s utility in the auxiliary mechanism.
Taking $M_{i}^{\text{aux}}[t,j]$ as defined in \cref{lem:gum_martingale}, we have $M_{i}^{\text{aux}}[0,n+1] = \Upsilon_{i}^{\text{aux}}[0,n+1]$. Because $M_{i}^{\text{aux}}[(t,j)]$ is a martingale starting at $\lambda(\delta,n,T)\cdot T/n$ as established in \cref{lem:gum_martingale},
\begin{equation} \label{eq:gum_martingale_stop}
    \E[M_{i}^{\text{aux}}[T,n+1]] = \lambda(\delta,n,T)\cdot \frac Tn
\end{equation}
Observe that the true externality $\gamma_{i\to k}^{\text{aux}}[s]$ is the conditional expectation of the output of \cref{alg:compute_estimator} when run at update $(s,i)$ with the reported history therebefore and $L=\infty$. We can then apply \cref{lem:ex_ante_unbiased_for_honest_agents} to see that $\E[\gamma_{i\to k}^{\text{aux}}[s]] = 0$ for every agent $k\neq i$ and timestep $s$. It follows from the definition of $M_{i}^{\text{aux}}[t,j]$ and \eqref{eq:gum_martingale_stop} that
\begin{equation} \label{eq:gum_martingale_consequence}
    \lambda(\delta,n,T)\cdot \frac Tn = \E[\Upsilon_{i}^{\text{aux}}[T,n+1]] - \sum_{k\neq i}\sum_{s=1}^T\E[\gamma_{k\to i}^{\text{aux}}[s]].
\end{equation}
Write
\[
    \E[\gamma_{k\to i}^{\text{aux}}[s]] = \E[\hat\gamma_{k\to i}^{\text{aux}}[s]] - \E[\beta_{k\to i}^{\text{aux}}[s]] = -\E[p_{k\to i}^{\text{aux}}[s]] - \E[\beta_{k\to i}^{\text{aux}}[s]].
\]
Then,
\[
    - \sum_{k\neq i}\sum_{s=1}^T\E[\gamma_{k\to i}^{\text{aux}}[s]] = \sum_{k\neq i}\sum_{s=1}^T(\E[p_{k\to i}^{\text{aux}}[s]] + \E[\beta_{k\to i}^{\text{aux}}[s]]).
\]
At any time $t$, if $\sum_{s=1}^{t-1}p_{k\to i}^{\text{aux}}[s] > B$, then the mechanism puts $\hat\theta_k^{\text{aux}}[t] = 0$. It follows from \cref{cor:positive_externality_from_timed_out_agents,cor:as_estimator_bound} that $\sum_{s=1}^t p_{k\to i}^{\text{aux}}[s]\leq B+D$ at all times $t$. Substituting this bound and the bound $\E[\beta_{j\to i}^{\text{aux}}[s]]\leq\epsilon$ into the above display, we obtain
\[
    - \sum_{k\neq i}\sum_{s=1}^T\E[\gamma_{k\to i}^{\text{aux}}[s]] \leq (n-1)(B+D + \epsilon T).
\]
Substitute into \eqref{eq:gum_martingale_consequence} to obtain
\[
    \lambda(\delta,n,T)\cdot \frac Tn \leq \E[\Upsilon_{i}^{\text{aux}}[T,n+1]] + (n-1)(B+D + \epsilon T).
\]
Let $\tilde U_{i}^{\text{aux}}$ be the total number of high-valued jobs of agent $i$ served. Observe that $\tilde U_{i}^{\text{aux}} = \Upsilon_{i}^{\text{aux}}[T,n+1]$. Substituting this into the above display and rearranging,
\begin{equation} \label{eq:virtual_utility_lower_bound}
    \E[\tilde U_{i}^{\text{aux}}] \geq \lambda(\delta,n,T)\cdot\frac Tn  - (n-1)(B+D + \epsilon T).
\end{equation}
We shall apply \cref{lem:utility_conversion} to obtain
\[
    \E[U_{i}^{\text{aux}}] \geq nv_{i}^\star\cdot\E[\tilde U_{i}^{\text{aux}}].
\]
Indeed, the conditions of \cref{lem:utility_conversion} are satisfied: agent $i$ is honest so she always chooses to serve her highest-valued job, the auxiliary mechanism grants agents access to the server only depending on the reports of the numbers of high-valued jobs, and the other agents' strategies do not depend on the job values of agent $i$ nor the high-valued arrivals of agent $i$ when conditioned on the reports of the numbers of high-valued jobs. Substitute \eqref{eq:virtual_utility_lower_bound} into the above to obtain
\[
    \E[U_{i}^{\text{aux}}] \geq \lambda(\delta,n,T)\cdot v_{i}^\star T  - (n-1)(B+D + \epsilon T)nv_{i}^\star.
\]
By the assumption that values are bounded in $[0,1]$, by the definition of ideal utility,
\[
    v_{i}^\star \leq \Pr_{Q_{i}^{\text{aux}}\sim\Unif[0,1]}(Q_{i}^{\text{aux}}\geq 1-1/n) = 1/n,
\]
so
\[
    \E[U_{i}^{\text{aux}}] \geq \lambda(\delta,n,T)\cdot v_{i}^\star T  - (n-1)(B+D + \epsilon T).
\]
Since $U_i^{\text{aux}}\leq T$ almost surely, using the fact that $U_{i}^{\text{aux}} = U_{i}$ with probability at least $1-q$ since the real mechanism and auxiliary mechanism agree with probability at least $1-q$,
\[
    \E[U_{i}] \geq \lambda(\delta,n,T)\cdot v_{i}^\star T  - (n-1)(B+D + \epsilon T) - q\cdot T.
\]
\end{proof}

\subsubsection{Main Results}
\begin{theorem}[Error Bound Result 1] \label{thm:error_result_one}
Set $\epsilon = 1/T$. Assume we set the algorithm's parameters so that
\[
    L = \begin{cases}\left\lceil\frac{\log \frac{\overline\theta+1}{\epsilon\delta}}{-\log(1-\delta)}\right\rceil & \text{if $\delta<1$}\\1 & \text{if $\delta=1$}\end{cases},
\]
and
\begin{equation} \label{eq:explicit_budget_one}
    B = L \ln(2Tn) + \sqrt{\left(L^2 \ln(2Tn) + \frac{8T}{n} + 16\sqrt{\frac{2T \ln(2Tn)}{n}} + 8\ln(2Tn)\right) \ln(2Tn)}
\end{equation}
where $\overline\theta$ and $\overline\Sigma$ satisfy the conditions of \cref{cor:honest_agent_ban_probability}. Then, \cref{alg:gum_report_capping} has error at most
\[
    O\left(\sqrt{nT\log T} + n\log^2 T\right).
\]
\end{theorem}
\begin{proof}
Substituting in the variables that the lower bound for $B$ in the condition of \cref{cor:honest_agent_ban_probability}, one can check that condition says
\[
    B \geq L \ln(2Tn) + \sqrt{\left(L^2 \ln(2Tn) + \frac{8T}{n} + 16\sqrt{\frac{2T \ln(2Tn)}{n}} + 8\ln(2Tn)\right) \ln(2Tn)}.
\]
This is exactly the budget in \eqref{eq:explicit_budget_one}. It is easy to see that $L$ satisfies the condition of \cref{cor:honest_agent_ban_probability}, and $\overline\theta$ and $\overline\Sigma$ satisfy the conditions of \cref{cor:honest_agent_ban_probability} by assumption. Thus, all our parameter choices satisfy the conditions of \cref{cor:honest_agent_ban_probability}.

For $\delta < 1$, substitute $\epsilon=1/T$ and $\overline\theta\leq T$ to obtain
\[
    L \leq \left\lceil\frac{\log\frac{T(T+1)}{\delta}}{-\log(1-\delta)}\right\rceil = O(\log T).
\]
By substituting the bound $D\leq L$ with $L=1$ for $\delta=1$ and the above bound on $L$ for $\delta < 1$ into \eqref{eq:explicit_budget_one}, one can find that
\[
    B \leq O\left(\sqrt{\frac{T\log T}{n}} + \log^2T\right).
\]
(Recall that we are in the regime where $n\leq T$.)

Using \cref{lem:generic_error} with the above bounds and the $q\leq3/T$ bound from \cref{cor:honest_agent_ban_probability} for honest $i$, the error is at most $O\left(\sqrt{nT\log T} + n\log^2T\right)$.
\end{proof}

\begin{theorem}[Error Bound Result 2] \label{thm:error_result_two}
Set $\epsilon = 1/\ln T$ and $\eta=1/T$. Assume we set the algorithm's parameters so that
\[
    L = \begin{cases}\left\lceil\frac{\log \frac{\overline\theta+1}{\epsilon\delta}}{-\log(1-\delta)}\right\rceil & \text{if $\delta<1$}\\1 & \text{if $\delta=1$}\end{cases},\overline\theta = \begin{cases}\min\left\{T,\left\lceil\sqrt{\frac{2}{-\ln(1-\delta)}\left(\frac2\delta+\ln\frac T\eta\right)}\right\rceil+1\right\} & \text{if $\delta<1$}\\1 & \text{if $\delta=1$}\end{cases}.
\]
and
\begin{equation} \label{eq:explicit_budget_two}
    B = L \ln(2Tn) + \sqrt{\left(L^2 \ln(2Tn) + \frac{8T}{n} + 16\sqrt{\frac{2T \ln(2Tn)}{n}} + 8\ln(2Tn)\right) \ln(2Tn)}
\end{equation}
where $\overline\Sigma$ satisfies the condition of \cref{cor:honest_agent_ban_probability}. Then, \cref{alg:gum_report_capping} has error at most
\[
    O\left(\frac{nT}{\log T}\right).
\]
\end{theorem}
\begin{proof}
Substituting in the variables that the lower bound for $B$ in the condition of \cref{cor:honest_agent_ban_probability}, one can check that condition says
\[
    B \geq L \ln(2Tn) + \sqrt{\left(L^2 \ln(2Tn) + \frac{8T}{n} + 16\sqrt{\frac{2T \ln(2Tn)}{n}} + 8\ln(2Tn)\right) \ln(2Tn)}.
\]
This is exactly the budget in \eqref{eq:explicit_budget_two}. It is easy to see that $L$ and $\overline\theta$ satisfy the conditions of \cref{cor:honest_agent_ban_probability}, and $\overline\Sigma$ satisfies the condition of \cref{cor:honest_agent_ban_probability} by assumption. Thus, all our parameter choices satisfy the conditions of \cref{cor:honest_agent_ban_probability}.

Substituting in the variables that define $L$, one can check that for $\delta < 1$,
\[
    L \leq \left\lceil\frac{\log(\ln T) - \log(\delta) + \log(\sqrt{\frac{2}{-\ln(1-\delta)}\left(\frac2\delta+2\ln T\right)}+3)}{-\log(1-\delta)}\right\rceil\leq O(\log\log T).
\]
By substituting the bound $D\leq L$ with $L=1$ for $\delta=1$ and the above bound on $L$ for $\delta<1$ into \eqref{eq:explicit_budget_two}, one can find that
\[
    B \leq O\left(\sqrt{\frac{T\log T}{n}} + \log(T)\log\log(T)\right).
\]
Using \cref{lem:generic_error} with the above bounds and the $q\leq3/T$ bound from \cref{cor:honest_agent_ban_probability} for honest $i$, the error is at most $O\left(\frac{nT}{\log T}\right)$.
\end{proof}

\begin{theorem}[Runtime Result] \label{thm:runtime_result}
To run one timestep of \cref{alg:gum_report_capping}, it takes $O(n^2L)$ worst-case and expected $O\left(n^2\left(1+\log(\overline\theta+1)\right)\right)$ computational time. To run all $T$ timesteps of $\cref{alg:gum_report_capping}$, it takes $O(n^2(\overline\Sigma + T))$ expected computational time.
\end{theorem}
\begin{proof}
The runtime is dominated by the amount of time it takes to compute the estimators $\hat\gamma_{j\to i}[t]$ in the calls to \cref{alg:compute_estimator}. Specifically, suppose we run $\sigma_{t,j}$ timesteps in the coupled simulation in \cref{alg:compute_estimator}. It takes $O(n\sigma_{t,j})$ to run the coupled simulation, since in each simulated timestep, one has to update $n$ queues, each can be done in $O(1)$ time as per the discussion in \cref{ssec:computational_model}. Thus, it takes $O\left(n\sum_{j=1}^n \sigma_{t,j}\right)$ to simulate each timestep. To bound this, we give uniform bounds on $\sigma_{t,j}$.

The per-timestep worst-case runtime bound $O(n^2L)$ follows from the fact that each $\sigma_{t,j}\leq L$ almost surely by definition. The per-timestep expected runtime bound $O\left(n^2\left(1+\log(\overline\theta+1)\right)\right)$ follows from \cref{cor:expected_simulated_timesteps}. By summing \cref{lem:naive_bound_difference} and using \cref{cor:expected_simulated_timesteps}, we obtain
\[
    \E\left[\sum_{t=1}^T \sigma_{t,j}\right]\leq O\left(\E\left[\sum_{t=1}^T\hat\theta_j[t] + \sum_{t=1}^T \hat\theta_j[t-1]+ T\right]\right) \leq O(\overline\Sigma + T).
\]
The total expected computational time bound follows.
\end{proof}

\thmBasicMechanismWithComputation*
\begin{proof}
Set $\epsilon = 1/T$ and
\[
    L = \begin{cases}\left\lceil\frac{\log \frac{T+1}{\epsilon\delta}}{-\log(1-\delta)}\right\rceil & \text{if $\delta<1$}\\1 & \text{if $\delta=1$}\end{cases}.
\]
Choose the virtual debt limit
\[
    B = L \ln(2Tn) + \sqrt{\left(L^2 \ln(2Tn) + \frac{8T}{n} + 16\sqrt{\frac{2T \ln(2Tn)}{n}} + 8\ln(2Tn)\right) \ln(2Tn)}
\]
For fixed $\delta$ and $n\leq T$, these choices have the orders as in the theorem statement.

The mechanism described in the theorem statement is \cref{alg:gum_report_capping} with $\overline\theta = T$ and $\overline\Sigma=\infty$. The choice of $L$ and $B$ is the same as in \cref{thm:error_result_one}. The error bound of $\tilde O(\sqrt{nT})$ comes from that theorem. For the runtime result, observe that $L\leq O(\log T)$ and apply \cref{thm:runtime_result}.
\end{proof}

\thmMainResultOne*
\begin{proof}
Choose the parameters explicitly as 
\begin{align*}
    &\overline\theta = \begin{cases}\min\left\{T,\left\lceil\sqrt{\frac{2}{-\ln(1-\delta)}\left(\frac2\delta+2\ln T\right)}\right\rceil+1\right\} & \text{if $\delta<1$}\\1 & \text{if $\delta=1$}\end{cases}, L = \begin{cases}\left\lceil \frac{\log\frac{(\overline\theta+1)T}{\delta}}{-\log(1-\delta)}\right\rceil & \text{if $\delta < 1$}\\1 & \text{if $\delta=1$}\end{cases},\\
     &\overline\Sigma = \left\lceil\frac1\delta\left(\frac Tn + \sqrt{\frac{32T}{n}\ln T} + 2\ln T\right)\right\rceil,\quad\text{and}\\
     & B = L \ln(2Tn) + \sqrt{\left(L^2 \ln(2Tn) + \frac{8T}{n} + 16\sqrt{\frac{2T \ln(2Tn)}{n}} + 8\ln(2Tn)\right) \ln(2Tn)}.
\end{align*}
For fixed $\delta$ and $n\leq T$, these choices have the orders stated in the theorem statement.

For the error bound, one can check that these parameter choices satisfy the conditions of \cref{thm:error_result_one} and apply that theorem. For the runtime bound, one can check that $\overline\theta\leq O(\sqrt{\log T})$, $L\leq O(\log T)$, and $\overline\Sigma\leq O(T)$ and apply \cref{thm:runtime_result}.
\end{proof}

\thmMainResultTwo*
\begin{proof}
Choose the parameters explicitly as
\begin{align*}
    &\overline\theta = \begin{cases}\min\left\{T,\left\lceil\sqrt{\frac{2}{-\ln(1-\delta)}\left(\frac2\delta+2\ln T\right)}\right\rceil+1\right\} & \text{if $\delta<1$}\\1 & \text{if $\delta=1$}\end{cases}, L = \begin{cases}\left\lceil \frac{\log\frac{(\overline\theta+1)\ln T}{\delta}}{-\log(1-\delta)}\right\rceil & \text{if $\delta < 1$}\\1 & \text{if $\delta=1$}\end{cases},\\
     &\overline\Sigma = \left\lceil\frac1\delta\left(\frac Tn + \sqrt{\frac{32T}{n}\ln T} + 2\ln T\right)\right\rceil,\quad\text{and}\\
     & B = L \ln(2Tn) + \sqrt{\left(L^2 \ln(2Tn) + \frac{8T}{n} + 16\sqrt{\frac{2T \ln(2Tn)}{n}} + 8\ln(2Tn)\right) \ln(2Tn)}.
\end{align*}
For fixed $\delta$ and $n\leq T$, these choices have the orders stated as in the theorem statement.

For the error bound, one can check that these parameter choices satisfy the conditions of \cref{thm:error_result_two} and apply that theorem. For the runtime bound, one can check that $L\leq O(\log\log T)$ and $\overline\Sigma\leq O(T)$ and apply \cref{thm:runtime_result}.
\end{proof}
\section{Deferred Proof of \cref{thm:error_no_computation}} \label{sec:deferred_proofs_no_computation}
In this section, we prove \cref{thm:error_no_computation}. The proof is a simpler version of the proofs in \cref{sec:deferred_proofs_computation}, and we reuse some of the logic from there. The difference is that we reason about \cref{alg:gum_no_computation} instead of \cref{alg:gum_report_capping}, which uses the actual externalities $\gamma_{i\to j}[t]$ instead of the estimated externalities $\hat\gamma_{i\to j}[t]$ for the virtual transfers and has no per-round nor cumulative report cap.

Let us first define some notation used throughout. We shall set the budget $B$ later. Let $\mathcal H_{t,i}$ be the filtration generated by the history of all reports and server allocations recorded by the mechanism up to update $(t,i)$. Let $\mathcal G_t$ be the filtration generated by the history of all reports and server allocations recorded by the mechanism up to timestep $t$, i.e., $\mathcal G_t = \mathcal H_{t,n+1}$.

The below lemma is analogous to \cref{lem:ex_ante_unbiased_for_honest_agents}.
\begin{lemma} \label{lem:no_computation_ex_ante_unbiased_for_honest_agents}
If $\hat\theta_{i}[t_0]$ is drawn randomly according to the stochastic transition kernel applied to $\hat\theta[t_0,i-1]$ (and independently of $\theta_i^{\text{random}}[t_0,i]$), then $\E[\gamma_{i\to j}[t_0]|\mathcal H_{t_0,i-1}]=0$ for any $j\neq i$.
\end{lemma}
\begin{proof}
Observe that the true externality $\gamma_{i\to j}[t_0]$ is the conditional expectation of the output $\hat\gamma_{i\to j}[t_0]$ of \cref{alg:compute_estimator} when run at update $(t_0,i)$ with the reported history therebefore and $L=\infty$. For any $j\neq i$, by \cref{lem:ex_ante_unbiased_for_honest_agents}, the expectation of $\hat\gamma_{i\to j}[t_0]$ conditioned on the reported state at $(t_0,i-1)$ is $0$, which implies $\E[\gamma_{i\to j}[t_0]|\mathcal H_{t_0,i-1}]= 0$.
\end{proof}

The below corollary is analogous to \cref{cor:positive_externality_from_timed_out_agents}.
\begin{corollary} \label{cor:no_computation_positive_externality_from_timed_out_agents}
If $\hat\theta_j[t_0] = 0$, then $\gamma_{j\to k}[t_0] \geq 0$ for all $k\neq j$.
\end{corollary}
\begin{proof}
Observe that the true externality $\gamma_{j\to k}[t_0]$ is the conditional expectation of the output $\hat\gamma_{j\to k}[t_0]$ of \cref{alg:compute_estimator} when run at update $(t_0,j)$ with the reported history therebefore and $L=\infty$. For any $k\neq j$, by \cref{cor:positive_externality_from_timed_out_agents}, $\hat\gamma_{j\to k}[t_0]\geq 0$, so it follows that $\gamma_{j\to k}[t_0] \geq 0$.
\end{proof}

The below lemma is analogous to \cref{lem:quadratic_variation_bound}.
\begin{lemma} \label{lem:no_computation_quadratic_variation_bound}
If $\hat\theta_{i}[t_0]$ is drawn randomly and independently from $\theta_i^{\text{random}}[t_0]$ according to the stochastic transition kernel applied to $\hat\theta[t_0-1,n+1]$, then for any $j\neq i$,
\[
    \E[\gamma_{i\to j}[t_0]^2|\mathcal H_{t_0,i-1}]\leq 2\delta\hat\theta_{i}[t_0-1] + \frac2n.
\]
\end{lemma}
\begin{proof}
Observe that the true externality $\gamma_{i\to j}[t_0]$ is the conditional expectation of the output $\hat\gamma_{i\to j}[t_0]$ of \cref{alg:compute_estimator} when run at update $(t_0,i)$ with the reported history therebefore and $L=\infty$. For any $j\neq i$, by \cref{lem:quadratic_variation_bound}, the expectation of $\hat\gamma_{i\to j}[t_0]^2$ conditioned on the reported state at $(t_0,i-1)$ is at most $2\delta\hat\theta_i[t_0-1]+2/n$, which by conditional Jensen\textquotesingle s inequality, implies
\[
    \E[\gamma_{i\to j}[t_0]^2|\mathcal H_{t_0,i-1}] \leq \E[\hat\gamma_{i\to j}[t_0]^2|\mathcal H_{t_0,i-1}]\leq 2\delta\hat\theta_i[t_0-1]+2/n.
\]
\end{proof}

As in \cref{sec:deferred_proofs_computation}, fix an honest agent $i$, and define an auxiliary mechanism that is the same as the actual mechanism except the virtual debt limit is not enforced on the fixed agent $i$; the virtual debt limit remains enforced on agents $j\neq i$ in the auxiliary mechanism. We denote quantities in the auxiliary mechanism with the superscript $\text{aux}$. Since there is no virtual debt limit enforcement on agent $i$, we always have $\hat\theta_i^{\text{aux}}=\theta_i^{\text{aux}}$. Couple the auxiliary mechanism with the real mechanism in which the virtual debt limit is actually enforced such that the mechanisms agree until agent $i$ makes a report that gets modified by the virtual debt limit.

Throughout, we denote
\[
    D = \begin{cases}\left\lceil \frac{\log(T+1)}{-\log(1-\delta)}\right\rceil + \frac1\delta & \text{if $\delta < 1$}\\1 & \text{if $\delta=1$}\end{cases},
\]
the almost sure bound on the externalities given in \cref{lem:as_externality_bound}. 

The below lemma is analogous to \cref{lem:honest_agent_exceed_budget}. Its proof below is basically the same as the proof of \cref{lem:honest_agent_exceed_budget} with $\gamma_{i\to j}^{\text{aux}}$ replacing $\hat\gamma_{i\to j}^{\text{aux}}$ and $D$ replacing $L$.
\begin{lemma} \label{lem:no_computation_honest_agent_exceed_budget}
Fix an honest agent $i$. For $j\neq i$, define
\[
    X_{i\to j}^{\text{aux}}[t] = -\sum_{s=1}^t\gamma_{i\to j}^{\text{aux}}[s].
\]
Fix $\eta\in(0,1]$. Set $x = \ln(2n/\eta)$ and $v = 4T/n + 2\sqrt{32Tx/n} + 4x$. If we set $B\geq Dx + \sqrt{x(D^2x + 2v)}$, then $\Pr(\exists t, j\neq i: X_{i\to j}^{\text{aux}}[t] > B) \leq\eta$.
\end{lemma}
\begin{proof}

For a fixed $j\neq i$, set $\xi_{i\to j}^{\text{aux}}[t] = X_{i\to j}^{\text{aux}}[t] - X_{i\to j}^{\text{aux}}[t-1]$. By \cref{lem:no_computation_quadratic_variation_bound},
\[
    \E[\xi_{i\to j}^{\text{aux}}[t]^2|\mathcal G_{t-1}^{\text{aux}}] \leq 2\delta\theta_i^{\text{aux}}[t-1] + \frac{2}{n}.
\]
Summing over $t$,
\[
    \sum_{t=1}^T \E[\xi_{i\to j}^{\text{aux}}[t]^2|\mathcal G_{t-1}^{\text{aux}}] \leq 2\delta\sum_{t=1}^T\theta_i^{\text{aux}}[t-1]+\frac{2T}{n}.
\]
The auxiliary mechanisms to \cref{alg:gum_no_computation} and \cref{alg:gum_report_capping} only differ in the virtual payments and the report caps. The proof of \cref{lem:honest_agent_exceed_budget} uses \cref{lem:occupation_bound}, whose proof does not use anything about the virtual payments, which does not use the virtual payments or report caps of \cref{alg:gum_report_capping}. Thus, \cref{lem:occupation_bound} applies here even though we are reasoning about the auxiliary mechanism to \cref{alg:gum_no_computation} instead of \cref{alg:gum_report_capping}. \cref{lem:occupation_bound} implies that $\delta\sum_{t=1}^T \theta_i^{\text{aux}}[t-1]\leq T/n + \sqrt{32Tx/n} + 2x$ with probability at least $1-e^{-x}$, so the above display implies
\[
    \Pr\left(\sum_{t=1}^T \E[\xi_{i\to j}^{\text{aux}}[t]^2|\mathcal G_{t-1}^{\text{aux}}] > \frac{4T}{n} + 2\sqrt{\frac{32Tx}{n}}+4x\right)\leq e^{-x}.
\]
Observe that $X_{i\to j}^{\text{aux}}[t]$ is a $\mathcal G_t^{\text{aux}}$-martingale by \cref{lem:no_computation_ex_ante_unbiased_for_honest_agents}. Furthermore, by \cref{lem:as_externality_bound}, its increments are almost surely bounded by $D$. By \cref{thm:freedman} with $S_t = X_{i\to j}^{\text{aux}}[t]$, $\xi_t = \xi_{i\to j}^{\text{aux}}[t]$, $c=D$, and $a=B$,
\[
    \Pr\left(\exists t: X_{i\to j}^{\text{aux}}[t]\geq B, \sum_{s=1}^t \E[\xi_{i\to j}^{\text{aux}}[s]^2|\mathcal G_{s-1}^{\text{aux}}]\leq v\right)\leq \exp\left(-\frac{B^2}{2(v + DB)}\right).
\]
Combine the previous two displays to obtain
\[
    \Pr\left(\exists t: X_{i\to j}^{\text{aux}}[t]\geq B\right)\leq \exp\left(-\frac{B^2}{2(v + DB)}\right) + e^{-x}.
\]
By a union bound over $j\neq i$,
\[
    \Pr\left(\exists t, j\neq i: X_{i\to j}^{\text{aux}}[t]\geq B\right)\leq n\exp\left(-\frac{B^2}{2(v + DB)}\right) + ne^{-x}.
\]
Expanding $B \geq Dx + \sqrt{x(D^2x+2v)}$ gives $B^2 \geq 2x(v + DB)$, so
\[
    n\exp\left(-\frac{B^2}{2(v + DB)}\right) \leq ne^{-x}.
\]
Therefore,
\[
    \Pr\left(\exists t, j\neq i: X_{i\to j}^{\text{aux}}[t]\geq B\right)\leq 2ne^{-x} = \eta.
\]

\end{proof}

The below lemma is analogous to \cref{lem:generic_error}.
\begin{lemma} \label{lem:no_computation_generic_error}
Fix an honest agent $i$. Suppose the probability that the real mechanism and auxiliary mechanism differ is bounded above by $q$. Then, 
\[
    \E[U_{i}] \geq \lambda(\delta,n,T)\cdot v_{i}^\star T  - \left((n-1)(B+D) + q\cdot T\right).
\]
\end{lemma}
\begin{proof}
The reasoning to reach \eqref{eq:gum_martingale_consequence} for any $k\neq i$ in the proof of \cref{lem:generic_error}, including the entire statement and proof of \cref{lem:gum_martingale} which is used in the proof of \cref{lem:generic_error}, holds in the auxiliary mechanism to \cref{alg:gum_no_computation} since it does not involve the virtual payments of the mechanism nor the report caps, which are the only ways the auxiliary mechanisms to \cref{alg:gum_no_computation} and \cref{alg:gum_report_capping} differ. At any time $t$, if $\sum_{s=1}^{t-1}p_{k\to i}^{\text{aux}}[s]>B$, then the mechanism puts $\hat\theta_k^{\text{aux}}=0$. It follows from \cref{cor:no_computation_positive_externality_from_timed_out_agents,lem:as_externality_bound} that $\sum_{s=1}^t p_{k\to i}^{\text{aux}}\leq B+D$ at all times $t$. Substituting this bound and the fact that $p_{k\to i}^{\text{aux}} = -\gamma_{k\to i}^{\text{aux}}$ into \eqref{eq:gum_martingale_consequence}, we obtain
\[
    \lambda(\delta,n,T)\cdot\frac Tn\leq \E[\Upsilon_i^{\text{aux}}[T,n+1]] + (n-1)(B+D).
\]
We shall apply \cref{lem:utility_conversion} to obtain
\[
    \E[U_{i}^{\text{aux}}] \geq nv_{i}^\star\cdot\E[\tilde U_{i}^{\text{aux}}].
\]
Indeed, the conditions of \cref{lem:utility_conversion} are satisfied: agent $i$ is honest so she always chooses to serve her highest-valued job, the auxiliary mechanism grants agents access to the server only depending on the reports of the numbers of high-valued jobs, and the other agents' strategies do not depend on the job values of agent $i$ nor the high-valued arrivals of agent $i$ when conditioned on the reports of the numbers of high-valued jobs. Combine the previous two displays with the identity $\Upsilon_i^{\text{aux}}[T,n+1] = \tilde U_i^{\text{aux}}$ to obtain
\[
    \E[U_{i}^{\text{aux}}] \geq \lambda(\delta,n,T)\cdot v_{i}^\star T  - (n-1)(B+D)nv_{i}^\star.
\]
By the assumption that values are bounded in $[0,1]$, by the definition of ideal utility,
\[
    v_{i}^\star \leq \Pr_{Q_{i}^{\text{aux}}\sim\Unif[0,1]}(Q_{i}^{\text{aux}}\geq 1-1/n) = 1/n,
\]
so
\[
    \E[U_{i}^{\text{aux}}] \geq \lambda(\delta,n,T)\cdot v_{i}^\star T  - (n-1)(B+D).
\]
Since $U_i^{\text{aux}}\leq T$ almost surely, using the fact that $U_{i}^{\text{aux}} = U_{i}$ with probability at least $1-q$ since the real mechanism and auxiliary mechanism agree with probability at least $1-q$,
\[
    \E[U_{i}] \geq \lambda(\delta,n,T)\cdot v_{i}^\star T  - (n-1)(B+D) - q\cdot T.
\]
\end{proof}

\thmErrorNoComputation*
\begin{proof}
Set the virtual debt limit $B$ as 
\begin{align*}
    B = & \left( 1 + \frac{1}{\delta} - \frac{\log(T+1)}{\log(1-\delta)} \right) \ln(2Tn)\\
    & + \sqrt{ \left( 1 + \frac{1}{\delta} - \frac{\log(T+1)}{\log(1-\delta)} \right)^2 \ln^2(2Tn) + \frac{8T \ln(2Tn)}{n} + 16\sqrt{\frac{2T \ln(2Tn)}{n}} \ln(2Tn) + 8\ln^2(2Tn) }
\end{align*}
if $\delta < 1$ and
\begin{align*}
    B = \ln(2Tn) + \sqrt{ \frac{8T \ln(2Tn)}{n} + 16\sqrt{\frac{2T \ln(2Tn)}{n}} \ln(2Tn) + 9\ln^2(2Tn) }
\end{align*}
if $\delta=1$.
For fixed $\delta$ and $n\leq T$, indeed $B = \tilde\Theta(\sqrt{T/n})$.

One can check that our choice of $B$ satisfies the condition on \cref{lem:no_computation_honest_agent_exceed_budget} with $\eta=1/T$. If $X_{i\to j}^{\text{aux}}[t]$ defined as in that lemma never exceeds $B$, then the virtual debt limit never alters agent $i$'s report in the real mechanism, so by the coupling, the real mechanism and the auxiliary mechanism never differ. Thus, the two coupled mechanisms differ with probability at most $1/T$. By \cref{lem:no_computation_generic_error} with $q=1/T$, the mechanism's error is at most $(n-1)(B+D) + 1$, which one can check to be at most $\tilde O(\sqrt{nT})$.
\end{proof}

\section{AI and LLM Usage Disclosure} \label{sec:ai_disclosure}

We used the LLMs Google Gemini in the Google Antigravity application and OpenAI ChatGPT in the OpenAI Codex extension in Visual Studio Code in the process of developing some of the techniques in this paper.

We used the ChatGPT 6 Astra (medium) to fill in routine mathematical details and computations and rewrote its proofs to be more stylistically in line with our writing.

After writing the paper, we used ChatGPT 6 Astra (medium) to check for typos and mathematical errors. Only very minor errors were found and the corrections needed were small.

Beyond checking for errors and filling in routine mathematical details, our usage of AI and LLMs was limited to developing the computational aspects of the mechanism. We developed all the modelling details and results of this paper, ignoring computational concerns, without any AI or LLM use. In particular, we came up with the general idea of the paper, the specific model, the idea to use the ideal utility benchmark, and the idea to use the Guaranteed Utility Mechanism all without any AI or LLM use.

As for the computational aspects of our paper, almost all high-level ideas came from us with the LLMs only helping to speed up the process of filling in technical details and giving first drafts of formal proofs. Specifically, our research process to come up with first drafts of formal proofs can be summarized as follows.
\begin{enumerate}
\item We proposed to the Gemini 3.1 Pro (High) model our general idea of simulating the future timesteps of the mechanism at each GUM update until the Markov chain mixes many times to efficiently compute the externality payments. Gemini 3.1 Pro formalized this idea as simulating two coupled Markov chains until they sync and wrote initial drafts of proofs.
\item With the help of the ChatGPT 5.6 Luna (High) model, we spotted multiple errors. We asked Gemini to fix its errors by following the following multi-agent setup with two nested loops. In the inner loop, the main Gemini agent would fix its errors and have two ``mathematical reviewer'' subagents review its work and find errors repeatedly until both ``mathematical reviewer'' subagents agreed there were no more errors. In the outer loop, a ``writing reviewer'' subagent would review the proof writing quality and comment on issues. Pseudocode for this multi-agent setup is given in \cref{alg:multi_agent_setup}.

\begin{algorithm}
\caption{Multi-Agent Setup for an Initial Algorithm}\label{alg:multi_agent_setup}
\While{the ``writing reviewer'' subagent does not approve}{
    \While{either ``mathematical reviewer'' subagent does not approve}{
        The main agent rewrites the proofs\;
        The main agent calls the two mathematical reviewers to evaluate\;
    }
    The main agent calls the writing reviewers to evaluate\;
}
\end{algorithm}

We repeated this multi-agent setup a few times, pointing out errors along the way that both we and independent ChatGPT 5.6 Luna (High) and ChatGPT 5.6 Luna (Very High) chats found. In this way, we quickly wrote an initial draft of our mechanism that is close to the initial mechanism that we presented with the $\tilde O_n(\sqrt T)$ utility error bound and $\tilde O_n(\log T)$ runtime at each step. Throughout this process, the LLMs also found that doing multiple simulations at each step is not helpful. Finally, we asked ChatGPT 5.6 Terra (Very High) to completely rewrite the proofs to be more readable and then reviewed the proofs carefully for correctness.
\item After the initial algorithm, we had the following improvement ideas.
\begin{enumerate}
\item Improving the runtime to $O_n(\log\log T)$ by capping per-round reports.
\item Improving the utility error bound to $\tilde O_n(\sqrt T)$ while maintaining $O_n(\log\log T)$ expected runtime at the cost of an $O_n(\log T)$ worst-case runtime.
\item Improving the sublinear utility error bounds by using a tighter exponential probability tail bound for the accumulation of virtual payments for honest agents. (We were initially using Azuma-Hoeffding for this instead of the tighter Freedman inequality that we use in the paper.)
\item Giving an $O_n(1)$ amortized runtime bound while retaining the previous guarantees by capping the cumulative reports.
\end{enumerate}
We developed these ideas in the order that they were listed above. For each of them, we did the following. We came up with the idea by reading the current draft of results and proofs without further LLM usage. Then, we asked Gemini 3.1 Pro (High) to draft a short ``proposal'' Markdown document that outlined the proposed improvement. We sent the proposal document to ChatGPT 5.6 Sol (Ultra) to implement in the draft. In addition, we used ChatGPT 5.6 Sol (High) to find the related work using similar probability concentration bounds as used for the above improvements.
\end{enumerate}

We did not directly use the draft of results and proofs that came out of the process described above; instead, we rewrote them.

\end{document}